\documentclass[%
 reprint,
 amsmath,amssymb,
 aps,
 prd,
]{revtex4-2}

\usepackage{graphicx}
\usepackage[caption=false]{subfig}

\usepackage{dcolumn}
\usepackage{bm}

\usepackage{afterpage}
\usepackage{floatpag}

\usepackage[colorlinks=true,allcolors=blue]{hyperref}
\usepackage{breakurl}
\usepackage{cleveref}

\usepackage{xcolor}

\begin{document}
\preprint{APS/123-QED}

\title{Extracting non-Gaussian Features in Gravitational Wave Observation Data Using Self-Supervised Learning}

\author{Yu-Chiung Lin}
\email{yuchiung.lin@mx.nthu.edu.tw}
\author{Albert K.~H.~Kong}%
 \email{akong@gapp.nthu.edu.tw}
\affiliation{%
Institute of Astronomy, National Tsing Hua University, Hsinchu 30013, Taiwan
}%

\begin{abstract}
    We propose a self-supervised learning model to denoise gravitational wave (GW) signals in the time series strain data without relying on waveform information. Denoising GW data is a crucial intermediate process for machine-learning-based data analysis techniques, as it can simplify the model for downstream tasks such as detections and parameter estimations. We use the blind-spot neural network and train it with whitened strain data with GW signals injected as both input data and target. Under the assumption of a Gaussian noise model, our model successfully denoises $38$\% of GW signals from binary black hole mergers in H1 data and $49$\% of signals in L1 data detected in the O1, O2, and O3 observation runs with an overlap greater than $0.5$. We also test the model's potential to extract glitch features, loud inspiral compact binary coalescence signals a few seconds before the merger, and unseen CCSN signals during training.
\end{abstract}

\maketitle

\section{\label{intro}Introduction}
The research on gravitational wave (GW) astronomy has been booming since the first detection of the GW signal from a binary black hole (BBH) coalescence (GW150914) \cite{PhysRevLett.116.061102} made by the advanced Laser Interferometer Gravitational-Wave Observatory (LIGO)~\cite{LIGOScientific:2014pky} in 2015. In addition, in 2017, the LIGO-Virgo Collaboration~\cite{LIGOScientific:2014pky, Acernese_2014} detected the first neutron star binary coalescence (GW170817)~\cite{TheLIGOScientific:2017qsa} with electromagnetic follow-up observations~\cite{Troja:2017nqp, Evans:2017mmy, 2017Sci...358.1579H, LIGOScientific:2017ync, LIGOScientific:2017zic, Mooley:2017enz}, which is the first multi-messenger event containing GW emissions in history.
In the first three observation runs from 2015 to 2020, the LIGO and Virgo Collaboration have reported $93$ confidently-detected GW events from compact binary coalescence (CBC) events~\cite{LIGOScientific:2018mvr, LIGOScientific:2020ibl, 2021arXiv210801045T, KAGRA:2021vkt}, including GWs from two binary neutron star (BNS) merger events~\cite{TheLIGOScientific:2017qsa, LIGOScientific:2020aai} and three possible candidates from neutron star-black hole binaries (NSBH)~\cite{KAGRA:2021vkt}. Moreover, in the first half of the fourth observation run (O4a), $81$ significant GW candidates from CBC events were uploaded to the Gravitational-Wave Candidate Event Database~\footnote{https://gracedb.ligo.org/superevents/public/O4/} (GraceDB) and public alerts were sent out.
On the other hand, the Kamioka Gravitational Wave Detector (KAGRA)~\cite{Akutsu2019, 10.1093/ptep/ptaa120} in Japan started the joint observation with the GEO600 detector~\cite{Dooley:2015fpa, Lough:2020xft} in 2020 (O3GK) for two weeks but did not find any events~\cite{KAGRA:2022fgc, KAGRA:2022twx}. 
With the addition of the KAGRA detector with the LIGO and Virgo forming the LIGO-Virgo-KAGRA (LVK) network, it is expected to detect more GW candidates from CBCs in O4b. Also, searches for GW signals from different sources such as subsolar-mass black hole binaries~\cite{LIGOScientific:2021job, LIGOScientific:2022hai}, bursts~\cite{LIGOScientific:2019ryq, KAGRA:2021tnv, KAGRA:2021bhs, LIGOScientific:2022sts, LIGOScientific:2022jpr, LIGOScientific:2021iyk} and continuous waves from spinning neutron stars~\cite{LIGOScientific:2020qhb, KAGRA:2021una, LIGOScientific:2021hvc, KAGRA:2022osp, LIGOScientific:2021ozr} and low-mass X-ray binaries~\cite{LIGOScientific:2022enz} are ongoing.

Most of the current GW search pipelines~\cite{2019arXiv190108580S, 2016CQGra..33u5004U, Hooper:2011rb, Chu:2020pjv, Aubin:2020goo} are based on the matched-filtering search method, which uses a set of known GW waveforms as templates to correlate the strain data and search for the maximum response~\cite{Sathyaprakash:2009xs, LIGOScientific:2019hgc}. 
The matched-filtering method is optimal for searching the GW signal if the background noise is Gaussian~\cite{Sathyaprakash:2009xs}. However, as the type of source and mass range increase, more GW templates that span wide frequency ranges are needed, which will require many computational resources during the search~\cite{George:2016hay}.
To tackle this challenge, the machine learning (ML) technique is a prominent alternative to the matched-filtering method for its short inference time, which is suitable for low-latency searches. So far the literature has explored the possibility of ML approaches to GW detection~\cite{George:2016hay, George:2017pmj, Gebhard:2019ldz, Gabbard:2017lja, Alvares:2020bjg, Huerta:2020xyq, Chan:2019fuz, 2020arXiv200914611S, 2020arXiv200100279I, Krastev:2020skk, 2021PhRvD.103f3034L, 2022ApJ...927..232A, Iess:2023quq}, glitch classification~\cite{2019CQGra..36g5005L, Yan:2022spw},  sky localization~\cite{Chatterjee:2019gqr, Kolmus:2021buf, Chatterjee:2022ggk}, parameter estimation~\cite{George:2016hay, Alvares:2020bjg, Krastev:2020skk} and denoising~\cite{2016PhRvD..94l4040T, 2020PhLB..80035081W, 2019arXiv190303105S, Ormiston:2020ele, Mogushi:2021cpw, Chatterjee:2021lit, Bacon:2022lsm, 2022arXiv221214283R}.

GW signal denoising is essential for enhancing the signal-to-noise ratio (SNR) and providing robust astrophysical parameter estimation~\cite{2020PhLB..80035081W}. 
Current works have used dictionary learning~\cite{2016PhRvD..94l4040T}, supervised learning~\cite{2020PhLB..80035081W}, and denoising auto-encoder (DAE)~\cite{2019arXiv190303105S, Ormiston:2020ele, Chatterjee:2021lit, Bacon:2022lsm, 2022arXiv221214283R} to recover the burst and the BBH waveforms from noisy strain data. However, these works rely on the clean waveform as the reference, which makes them effective only for denoising well-modeled waveforms such as CBC signals and a specific type of burst signals.
On the other hand, a neural network model was proposed to remove glitches overlapping with the signals by training a model to reconstruct the gated portion of strain data with injection~\cite{Mogushi:2021cpw}. This method does not use clean signals as references but can only reconstruct CBC signals with large SNR in a gated portion of a few hundred milliseconds before the merger.
Meanwhile, a total-variation method was employed to remove Gaussian noise in the strain data without relying on information from clean signals~\cite{Torres:2014zoa, Torres-Forne:2018yvv}. This method successfully denoised loud core-collapse supernova (CCSN) signals and BBH signals. However, achieving an optimal denoising result requires searching for a regularization term using an additional artificial neural network (ANN) model. 

We consider ML-based denoising as a feature-extraction task since it can remove noise and emphasize the important GW features in the strain data. Such a feature extractor can serve as an intermediate process for downstream tasks such as detection, parameter estimation, and glitch classification. It also simplifies the complexity of ML models that perform these tasks. 
We aim to build a general-purpose feature extractor in the ML-based GW processing pipeline, which can extract as many interesting features 
as possible in addition to CBC signals from the strain data. One can directly use our model to extract features for downstream tasks or fine-tune the specific tasks to focus on certain features for which the task is needed.
In recent years, the self-supervised learning (SSL) technique has become a popular research field in computer vision~\cite{2015arXiv150505192D, 2016arXiv160407379P, 2018arXiv180307728G, 2018arXiv181110980K, 2021arXiv210402057C} and natural language processing~\cite{2018arXiv181004805D, 2023arXiv230308774O}. By training with specific pre-text tasks and supervisory loss functions, the SSL model can learn to extract essential representations in the data without the need for human annotations~\cite{2016arXiv161109842Z, 2019arXiv190206162J}.

The SSL denoiser predicts the clean data point from the surrounding noisy data and is currently an active research field in computer vision due to the difficulty of obtaining clean ground-truth images in reality~\cite{2015arXiv150505192D, laine2019high, 2021arXiv210607009K, 2020arXiv201011971X}. With a suitable noise model as the supervisory loss function, the SSL denoiser can be trained using the original noisy image as both the input and the target. Training of the SSL denoiser can be seen as extracting the essential features in the noisy data under the given noise assumption. 
When using the noisy image as both the input and the target, traditional neural network models can only learn the identity since the pixels in the input image are directly mapped to the original image. 
Interestingly, in their work, Krull et al.~\cite{2018arXiv181110980K} provided a remedy by proposing the NOISE2VOID training technique based on the blind-spot neural network where the receptive field of the network does not include the center pixel. Consequently, the model is forced to predict the center clean pixel using the surrounding noisy data. The model can be trained with only noisy images, assuming the corruption is zero-mean and independent between pixels. However, they created the blind spot by applying the masking scheme to the loss function, reducing the training efficiency. Laine et al.~\cite{laine2019high} proposed a blind-spot structure that combines multiple branches in which the receptive field is restricted in different directions. The restriction in the receptive field can be realized by shifting the output of a normal convolution so that only pixels synthesized before the current pixel are allowed in the receptive field~\cite{2016arXiv160605328V, 2016arXiv160106759V, 2017arXiv170105517S}.

This work uses the SSL denoiser to extract GW features in the strain data without providing clean waveforms as references. Assuming the Gaussian noise background with zero means, we use the blind-spot method proposed in Ref.~\cite{laine2019high} to create the blind spot in the WaveNet model~\cite{2016arXiv160903499V} and train the model using the whitened strain data with CBC signals injected as both input and target. Our goal is to build a general-purpose feature extractor in the machine-learning-based GW processing pipeline, which can extract features from different types of GW sources such as CBCs, CCSNe, and continuous waves from spinning neutron stars. The extracted features can then be used for downstream tasks such as detection, parameter estimation, glitch classification, and sky localization.  

The rest of this paper is organized as follows: In Section~\ref{sec:ssl_denoise}, we give an overview of the self-supervised learning denoiser and the blind-spot neural network and introduce the method used in this work, including the noise model, the neural network architecture, the process of generating training and testing data, and the detail of the training process. In Section~\ref{sec:results}, we present the model performance results on the mock data and the actual events. This section also discusses the potential for extracting glitch features, CBC inspiral signals, and CCSN signals that have not been added to the training dataset. Finally, we conclude our work in Section~\ref{sec:conclusion}.

\section{\label{sec:ssl_denoise}Self-Supervised Learning Gravitational Wave Denoiser}

\subsection{\label{subsec:Gaussian_model}The Gaussian Noise Model}
The detector strain data $\textbf{d}(t)$ can be seen as a combination of detector noise $\textbf{n}(t)$ and the gravitational wave signals $\textbf{h}(t)$ such that
\begin{equation}
    \textbf{d}(t) = \textbf{n}(t) + \textbf{h}(t).
\end{equation}
If the noise is stationary and Gaussian, then we can model the noise of a single detector using the likelihood function \cite{LIGOScientific:2019hgc}:
\begin{equation}\label{eq:noise_likelihood}
    \begin{split}
        p(\textbf{d}&|\textbf{h}) = \\ 
        &\exp\left(-\frac{1}{2}\left[(\textbf{d}-\textbf{h}|\textbf{d}-\textbf{h})+\int\ln(S_n(f))df\right]\right),
    \end{split}
\end{equation}
where $S_n(f)$ is the noise power spectral density estimation for the detector,  and $(\textbf{a}|\textbf{b})$ is the noise-weighted inner product
\begin{equation}\label{eq:noise_weighted_inner_product}
    (\textbf{a}|\textbf{b}) = 2\int_0^\infty\frac{\tilde{a}(f)\tilde{b}^*(f) + \tilde{a}^*(f)\tilde{b}(f)}{S_n(f)}df.
\end{equation}

With the Gaussian noise likelihood described in Equation~\eqref{eq:noise_likelihood}, we can train a model to predict the clean signal $\hat{\textbf{h}}$ from the noisy data $\textbf{d}$, using the loss function
\begin{align}\label{eq:gaussian_loss}
    \mathcal{L} &= -\log p(\textbf{d}|\hat{\textbf{h}})\nonumber \\
    &= \frac{1}{2}\left[(\textbf{r}|\textbf{r})+\int\ln(S_n(f))df\right],
\end{align}
where $\textbf{r} = \textbf{d}-\hat{\textbf{h}}$ is the residual between the noisy data and the predicted clean signal.

Usually, the detector noise is several orders of magnitude greater than the GW signals, which makes it hard for the neural network model to extract GW features in such noisy data. To suppress the noise magnitude and emphasize the GW features, one has to whiten the time series data by dividing its Fourier coefficients by the noise amplitude spectral density estimation $\sqrt{S_n(f)}$ \cite{LIGOScientific:2019hgc}. Equation \eqref{eq:noise_weighted_inner_product} can therefore be seen as the correlation between two vector $\textbf{a}_w$ and $\textbf{b}_w$ whitened with the noise power spectral density $S_n(f)$:
\begin{equation}
    (\textbf{a}|\textbf{b}) = 2\int_0^\infty\left(\tilde{a}_w(f)\tilde{b}_w^*(f) + \tilde{a}_w^*(f)\tilde{b}_w(f)\right)df,
\end{equation}
where $\tilde{a}_w(f)$ and $\tilde{b}_w(f)$ are the Fourier transform of $\textbf{a}$ and $\textbf{b}$, respectively. In addition, we assume that $S(n)=1$ for whitened strain, and then the second term in Equation~\eqref{eq:gaussian_loss} becomes zero. Therefore, the loss function can be simplified to
\begin{equation}\label{eq:whitened_gaussian_loss}
    \mathcal{L} = 2\int_0^\infty\left|\tilde{r}_w(f)\right|^2df,
\end{equation}
where $\tilde{r}_w(f) = \tilde{d}_w(f) - \tilde{h}_w(f)$ is the whitened residual between the noisy data and the predicted clean signal. Note that the loss function~\eqref{eq:whitened_gaussian_loss} is the mean squared error in the Fourier space.

In reality, the detector data is usually neither stationary nor Gaussian, and the cleaned signal $\hat{\textbf{h}}(t)$ may contain non-stationary or non-Gaussian noises such as lines or glitches. However, we can also expect that the model will be able to probe the GW signals from different sources, such as CCSN and continuous waves from spinning neutron stars. Also, there is a possibility to have overlapping signals in the strain data. Our model can help remove the noise so that the other signal separation model can better split the overlapped signals.
Therefore, our model can be seen as a general-purpose feature extractor for GW data and simplify the complexity of the model performing downstream tasks such as classification, parameter estimation, and signal separation. 

\subsection{The Blind-Spot WaveNet Model}
\label{subsec:blind_spot_neural_network}
\begin{figure*}[tbp]
    \hspace*{-1cm}
    \centering
    \includegraphics[width=0.99\linewidth]{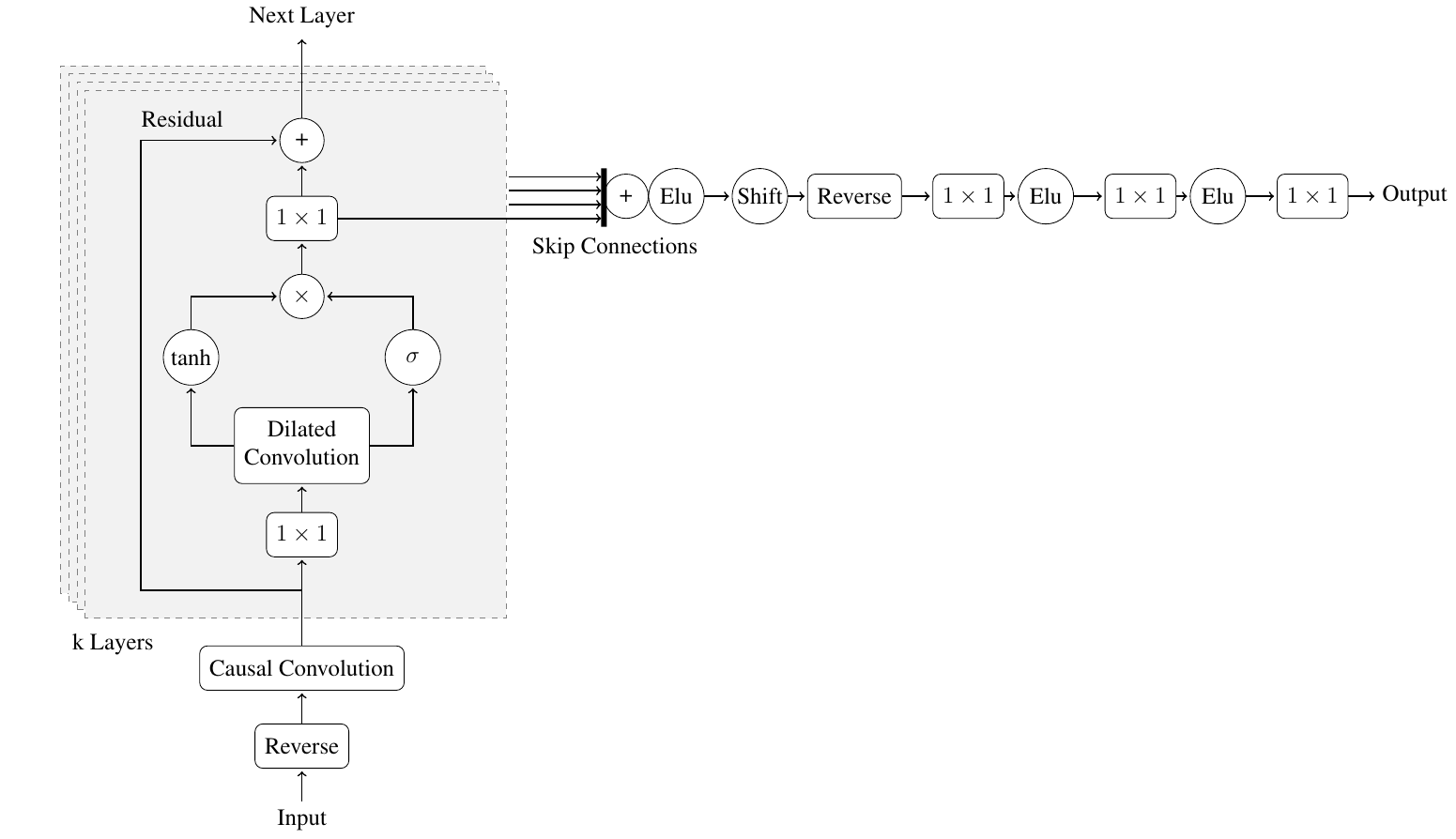}
    \caption{Overview of the blind-spot WaveNet model used in this work.}
    \label{fig:blind_spot_wavenet}
\end{figure*}

We use the blind-spot neural network, whose receptive field does not include the center pixel, to prevent the model from learning the identity. This is because we use the whitened strain data as both input and target. 
Such a network is self-supervised, trained by predicting the output value based on the surrounding context without requiring an additional reference target. Laine et al.~\cite{laine2019high} proposed a method to implement the blind-spot neural network under the current deep learning framework by using a neural network with the receptive field restricted in different directions and augmenting the input data to have different rotation angles. The restriction on the receptive field can be easily done by shifting the output of a normal convolution~\cite{2016arXiv160605328V, 2016arXiv160106759V, 2017arXiv170105517S}. An additional one-pixel offset is applied to create the blind spot. Therefore, the neural network can acquire the context information from the surrounding pixels except for the center pixel.

In this work, we apply the blind-spot technique in Reference~\cite{laine2019high} on the WaveNet-based model~\cite{2016arXiv160903499V}, and the structure of the model we use is shown in Figure~\ref{fig:blind_spot_wavenet}. We first create a time-reversed copy of the input data and stack them to the minibatch axis. Thus, the batch size of the data is twice as large as the original input data. Then, we perform the causal convolution~\cite{2016arXiv160903499V} to obtain $1440$ feature maps, in which each pixel only correlates to one side of its surroundings. The data then pass through $12$ convolution stacks; each stack has dilated causal convolution layers with residual connections~\cite{2015arXiv151203385H}, each with kernel size $3$ and the dilation rate cyclic switched in $1, 2, 4$. 
In each dilated causal convolution layer, we first perform the $1\times1 $ convolution to downsize the number of feature maps to $360$ before the causal convolution to save computational resources. We then apply a $\tanh$ and a sigmoid activation separately on the data and multiply them together. After the multiplication, we perform another $1\times1 $ convolution on the data to increase the number of feature maps back to $1440$. Next, we add the input data to the output of the convolution and pass the data to the next layer. We finally aggregate the last $1\times1 $ convolution output in each layer to obtain the skip connections.

After applying an Elu activation function~\cite{clevert2015fast} on the skip connections, we shift the data by one pixel to the right to create the blind spot. We then split the minibatch axis into two pieces, undo the reverse operation on the second piece, and stack them to the feature axis. Therefore, the data now has the same batch size as the original input data, but the number of feature maps is twice as large as the original skip connections. Finally, we pass the data to three $1\times1 $ convolution layers with the number of feature maps $2880$, $1440$, and $1$, each followed by an Elu activation except for the last layer. The output of the last 1x1 convolution is the cleaned data. In this work, we do not rigorously tune the model's hyperparameters, leaving space for optimization in the future. 

\subsection{\label{subsec:data_preparation}Data Preparation}
\begin{table}[tbp]
\centering
\begin{tabular}{c c}
    \hline\hline
    parameter name & distribution \\[0.5ex]
    \hline
    BBH approximant & SEOBNRv4T, IMRPhenomD \\
    NSBH approximant & TaylorT4, IMRPhenomBHNS \\
    BNS approximant & TaylorT4 \\
    $m_\text{BH}$ & uniform on $[3, 75]\; M_\odot$ \\
    $m_\text{NS}$ & uniform on $[1.4, 3]\; M_\odot$ \\
    mass ratio q & uniform on $[1, 10]$ \\
    right ascension & uniform on $[0, 2\pi]$ \\
    declination & uniform on $[-\pi/2, \pi/2]$ \\
    polarization angle $\psi$ & uniform on $[0, 2\pi]$ \\
    $\rho_\text{opt}$ for BBH & $[5, 25]$ \\
    $\rho_\text{opt}$ for NSBH and BNS & $[10, 35]$ \\
    \hline\hline
\end{tabular}
\caption{\label{table:gwparas}The parameters of GW signals used in the training dataset.}
\end{table}

\begin{figure}[tb]
    \centering
    \includegraphics[width=\linewidth]{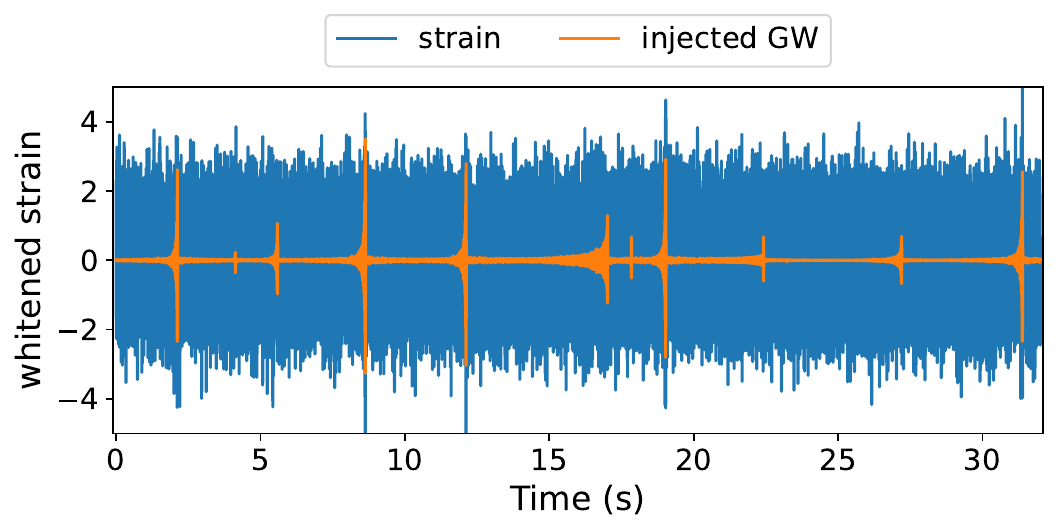}
    \caption{\label{fig:sample}An example of a $32$-second segment in the training dataset. The blue curve is the whitened strain data, and the orange curve indicates the injected GW signals. 
    }
\end{figure}

We use the public LIGO Hanford and Livingston data during O2, and O3 runs available on the Gravitational Wave Open Science Center (GWOSC)~\cite{RICHABBOTT2021100658} as the background noise for the training dataset. We use the strain data having the quality flag $\texttt{CBC\_CAT3}=100$ with a sampling rate of $4096$ Hz. We randomly chose $120$k data segments of $32$ seconds from these strain data as our training dataset.

Furthermore, to help the model recognize the feature of GW signals, we randomly inject $8$ to $10$ simulated GW signals from compact binary coalescence into each segment, using the PyCBC~\cite{alex_nitz_2023_8190155} package with the parameters listed in Table~\ref{table:gwparas}. The time interval between each merger event in a segment is randomly chosen from $3$ to $5$ seconds. We use the \texttt{SEOBNRv4T}~\cite{Bohe:2016gbl} and the \texttt{IMRPhenomD}~\cite{PhysRevD.93.044006, PhysRevD.93.044007} approximant for BBH signal generation, the \texttt{TaylorT4}~\cite{PhysRevD.80.084043} and the \texttt{IMRPhenomBHNS}~\cite{PhysRevD.92.084050, PhysRevD.100.044003} approximant for NSBH, and the \texttt{TaylorT4} approximant for BNS, respectively. We believe that using different approximants for binary systems can help increase the variety of the features of GW signals, enhancing the model's ability to extract GW features in real data. The binary is assumed to be spinless and arbitrarily located in the sky with an arbitrary polarization angle $\psi$. We choose the component masses range in $3 M_\odot$ to $75 M_\odot$ for black hole masses $m_\text{BH}$, and $1.4 M_\odot$ to $3 M_\odot$ for neutron star masses $m_\text{NS}$. The mass ratio $q=m_1/m_2$ of the component masses in the binary is constrained to be less than $10$. We set the starting frequency of the simulated signal to be $15$ Hz. When injecting the GW signal $h(t)$ into the strain data, we scale its amplitude so that the optimal SNR, $\rho_\text{opt}$, defined by \cite{Sathyaprakash:2009xs}
\begin{equation}
    \rho^2_\text{opt} = 4\int_0^\infty\frac{\left|\tilde{h}(f)\right|^2}{S_n(f)}df,
\end{equation}
lies within the range of $5$ to $25$ for BBH signals and $10$ to $35$ for NSBH and BNS signals.
Figure~\ref{fig:sample} shows an example of a $32$-second segment in the training dataset. The blue curve is the whitened strain data, and the orange curve indicates the injected GW signals.
After injection, we whiten the strain data using the PSD estimated from the $128$ seconds of data chosen around the segment. Finally, we save each segment in individual files. Note that we do not preserve individual simulated GW signals since we do not need them as training labels. 

For the test dataset used in Section~\ref{subsec:mock_test}, we randomly pick $512$ segments of $2$ seconds from the strain data described above and inject GW signals into them with $\rho_\text{opt}$ ranging from $5$ to $25$ in steps of $0.5$ for BBH, and $8$ to $35$ in steps of $0.5$ for NSBH and BNS respectively. Unlike the training dataset, we only inject one GW signal for each sample, which peaks at around a third quarter of the segment and saves the GW signal alone for the performance evaluation.

To maximize the model's ability for feature extraction, besides GW signals from CBCs, one may also inject other types of GW signals, such as GWs from CCSNe or continuous waves from rotating neutron stars, into the training data. In this work, we only inject GW signals from CBCs due to the difficulty of generating the simulated signals from other sources.

\subsection{Training}\label{subsec:training}
We implement our model using the Tensorflow~\cite{tensorflow2015-whitepaper} package, the ADAMax optimizer \cite{2014arXiv1412.6980K} with default parameters, an initial learning rate of $0.0003$, and the size of minibatch 8. Each minibatch consists of $2$-second segments randomly cropped from the training dataset described in Section~\ref{subsec:training}. We train our model using $0.5$M mini-batches and ramp down the learning rate to $10^{-6}$ using the cosine schedule during the last $20$\% of training.
Before feeding into the model, we standardize the data to have zero mean and unit variance. Although standardization may alter the scale of the Gaussian noise spectrum $S(f)$, the noise spectrum is constant for whitened data and does not contribute to the gradient update. Moreover, standardization can stabilize the training process and make the model more easily converge. When calculating the loss function, we first apply a Tukey window with $\alpha=0.25$ to the residual $r(t)$ before the Fourier transforms to avoid spectral leakage. In this work, we sum up all the power ranging from $0$ Hz to the Nyquist frequency of the data, which is $2048$ Hz.

During the training, we monitored the loss function to check the convergence without using a validation dataset since there are no proper metrics to monitor whether the model is overfitting on extracting non-Gaussian features. Instead, we directly evaluate the model performance by running the test described in Section~\ref{sec:results} after training.

We trained our model on the cluster with eight NVIDIA\texttrademark\ RTX 3080 GPUs in the Center of Informatics and Computation in Astronomy (CICA). The training process takes about $120$ hours.

\section{\label{sec:results}Results}
\subsection{\label{subsec:mock_test}Quantitative Performance on Mock Data}
 Following the common metric used in GW denoising literature, we use the overlap $\mathcal{O}(h,s)$ to quantify the performance of the model of reconstructing the clean signal, defined by~\cite{LIGOScientific:2020ibl}
\begin{equation}
    \mathcal{O}(h,s) = \frac{(h|s)}{\sqrt{(h|h)(s|s)}},
\end{equation}
where $s$ is the cleaned signal and $h$ is the injected GW template.
We calculate the overlap in two ways: the first is to calculate the overlap using the full $2$-second data segment to see the overall match between the cleaned signal and the injected GW signal. The second is to calculate the overlap using the data segment of $0.25$s around the peak of the injected GW signal to see how well the model can recover the waveform around the merger event for detection and parameter estimation.
The evaluation is done on a personal computer with an NVIDIA\texttrademark\ RTX $3060$ GPU. 

\begin{figure*}[tbp]
    \centering
    \subfloat[\label{fig:overlap_BBH}BBH]{\includegraphics[width=0.48\linewidth]{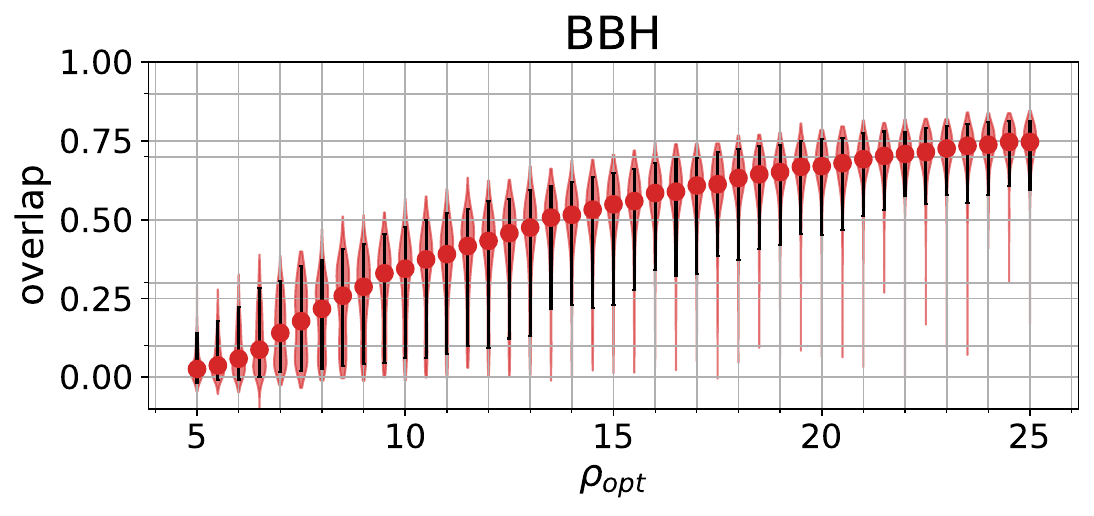}
    \hfill
    \includegraphics[width=0.48\linewidth]{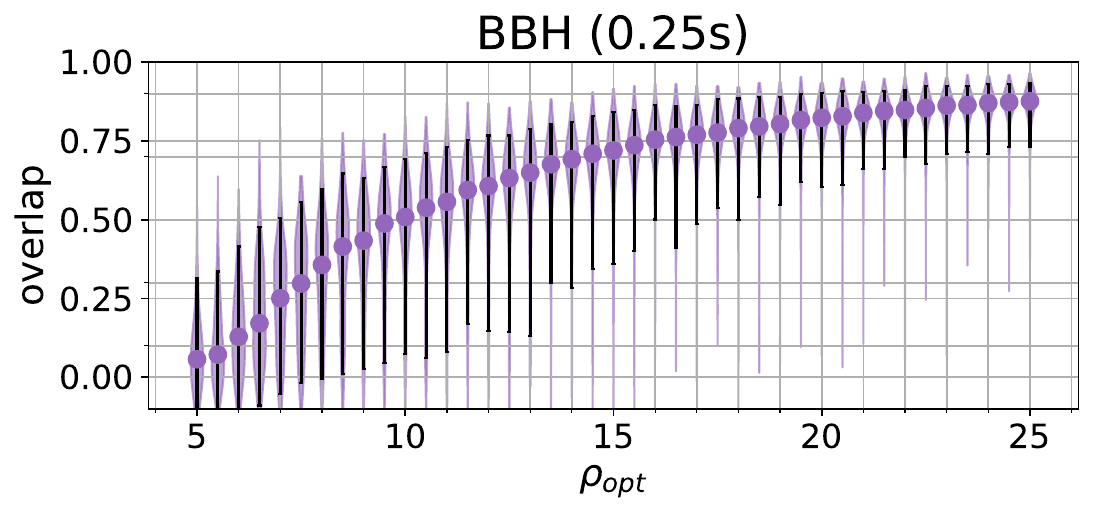}}
    \\
    \subfloat[\label{fig:overlap_BHNS}BH-NS]{\includegraphics[width=0.48\linewidth]{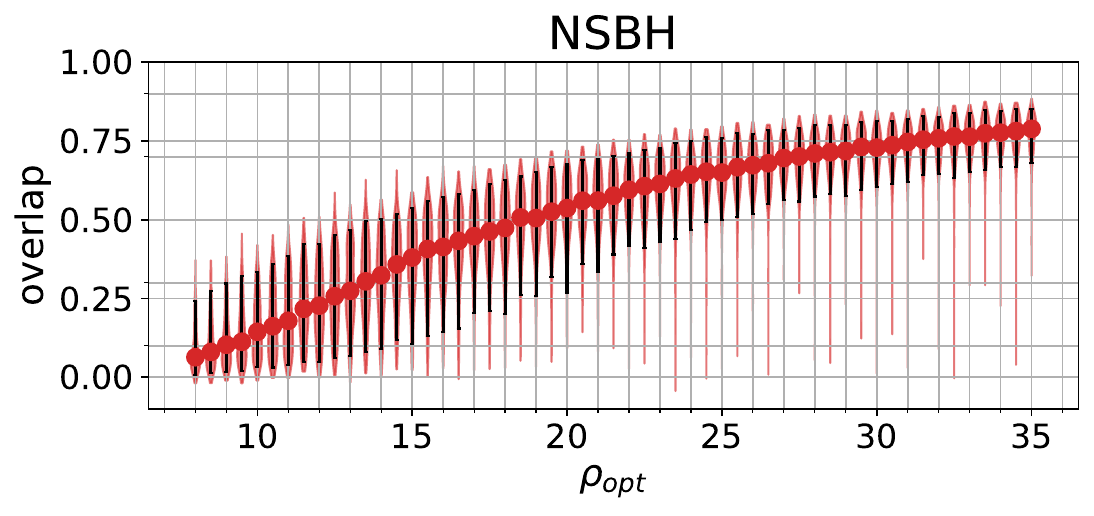}
    \hfill
    \includegraphics[width=0.48\linewidth]{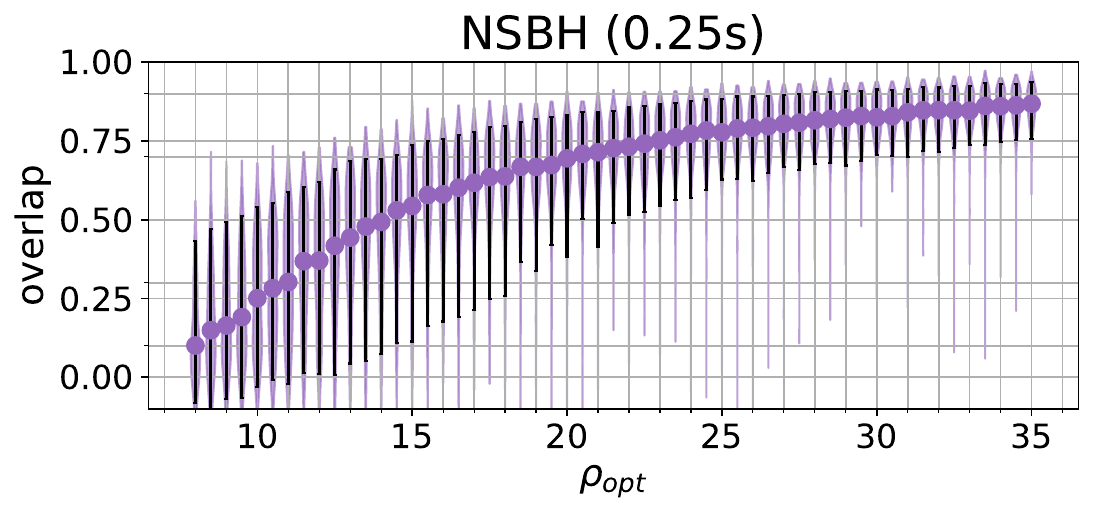}}\\
    \subfloat[\label{fig:overlap_BNS}BNS]{\includegraphics[width=0.48\linewidth]{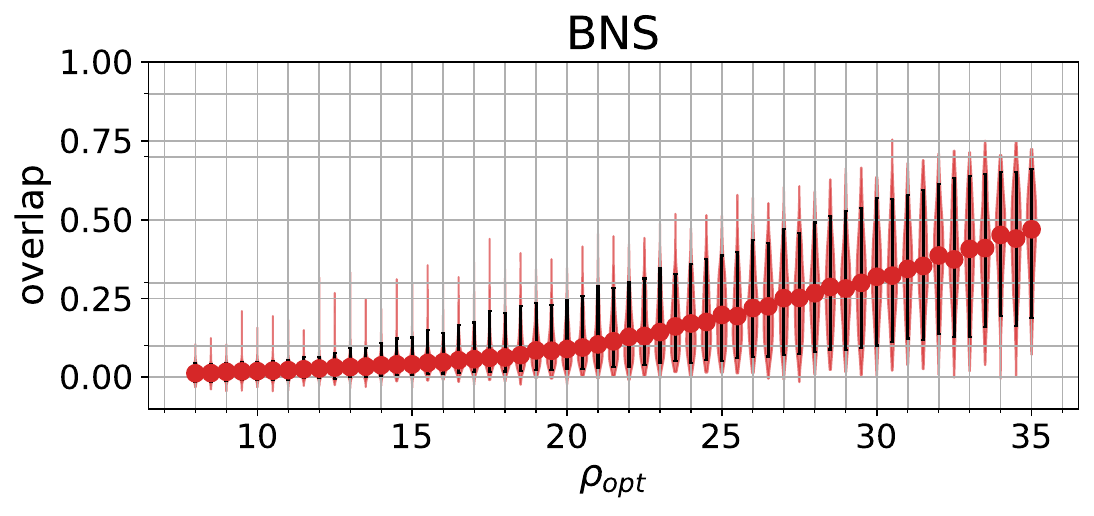}
    \hfill
    \includegraphics[width=0.48\linewidth]{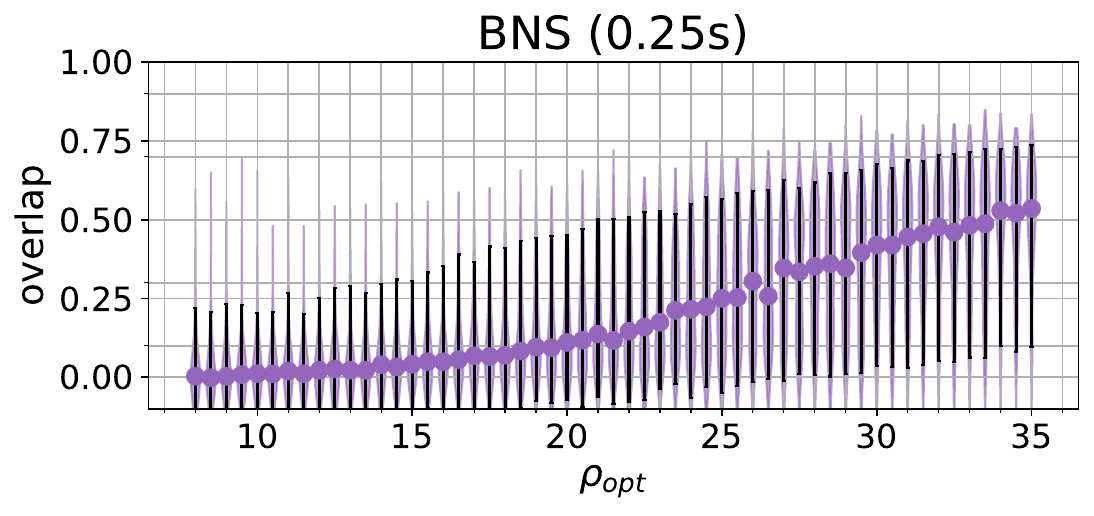}}
    \caption{\label{fig:overlap_result}The overlap result of (a) the BBH test dataset, (b) the NSBH test dataset, and (c) the BNS test dataset. The violin plots are the overlap distribution of all $512$ samples in different $\rho_\text{opt}$ bins. The dot in each bin is the mean of the overlap distribution. The black error bars represent the $90\%$ confidence interval of the distribution. The left panel is the overlap calculated using the full $2$-second data segment, and the right panel is the overlap calculated using the data segment of $0.25$s around the peak of the injected GW signal.}
\end{figure*}

We present the overlap distribution on the BBH, NSBH, and BNS datasets in Figure~\ref{fig:overlap_BBH}, \ref{fig:overlap_BHNS}, and \ref{fig:overlap_BNS}, respectively. The area in each SNR bin indicates the overlap distribution of $512$ samples in the SNR dataset, and the dot presents the mean of these values. The black error bars represent the $90\%$ confidence interval of the distribution. The left panel in Figure~\ref{fig:overlap_result} is the overlap calculated using the full $2$-second data segment, and the right panel is the overlap calculated using the data segment of $0.25$s around the peak of the injected GW signal.

As we expected, the mean overlap improves when SNR increases. However, there is a large scattering of overlaps in all datasets, including signals with high optimal SNR, and the maximum values of the mean overlap for BBH and NSBH datasets are only around $0.8$ for the $2$-second data. The maximum mean overlap values for BBH and NSBH datasets improve to $0.9$ when using the $0.25$s segment around the GW peak. We suspect that the large scattering in the overlap distribution is due to the noise fluctuation in the background data since our model can also probe non-Gaussian noise features in the data. On the other hand, the low mean overlap for $2$-second data may be due to the inability of the model to recover the low-amplitude waveform at the early inspiral stage or the remaining high-frequency noise in the denoised signal, which downgrades the overlap, as shown in Section~\ref{subsec:real_test}.

The mean overlaps of the BNS dataset are lower than the BBH and NSBH datasets and do not improve significantly when using $0.25$ second data segments. We believe this is because the BNS waveform has a much longer duration and wider frequency range and, therefore, can have a larger optimal SNR with a smaller amplitude around the merger event, which the model cannot cover well. In addition, the BNS waveform has higher frequencies around the merger and, therefore, can be more easily affected by the high-frequency noise remaining in the denoised signal.

\begin{figure}[tbp]
    \centering
    \includegraphics[width=0.95\linewidth]{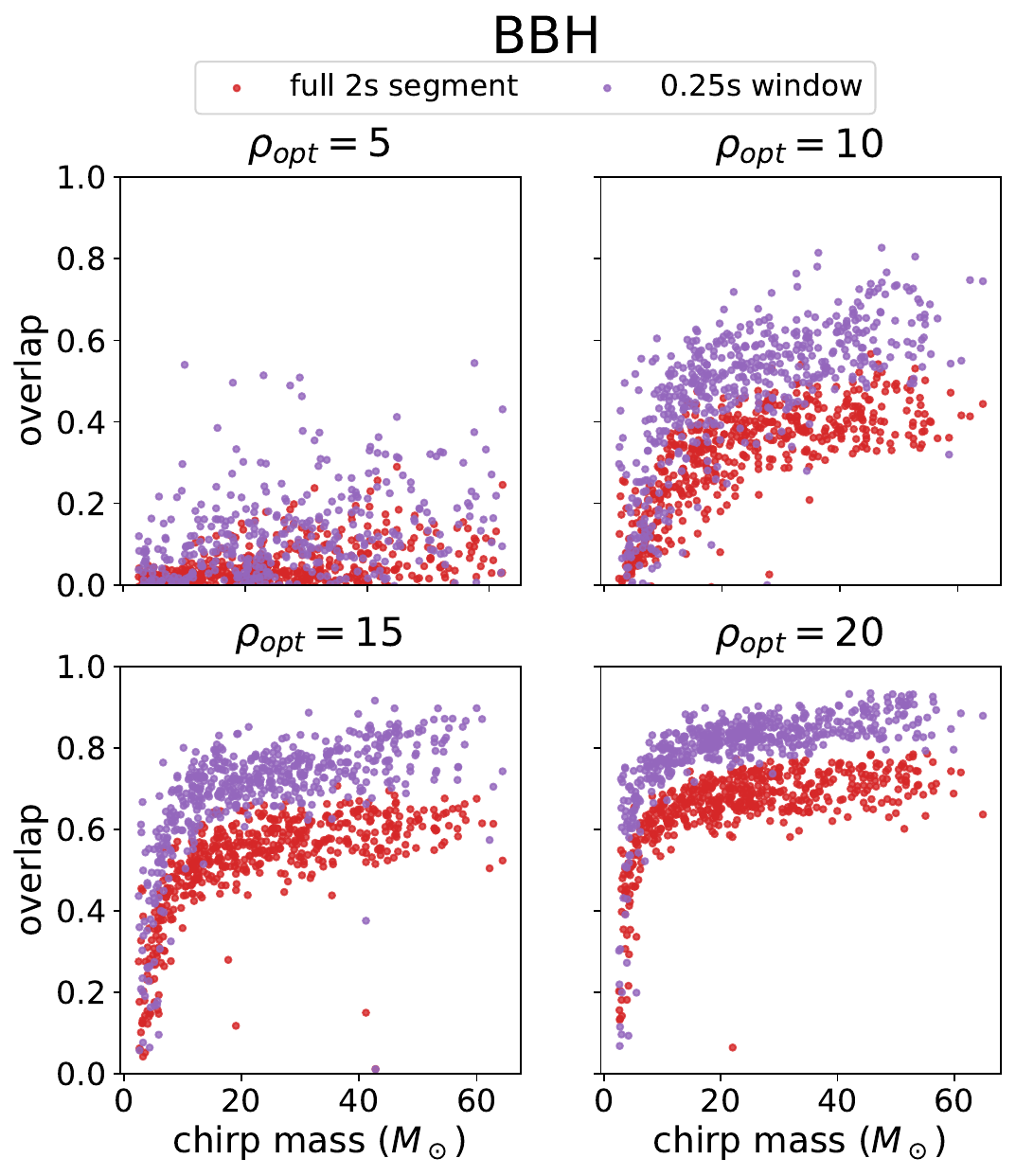}
    \caption{\label{fig:overlap_chirp_mass_BBH}The overlap versus chirp mass of the BBH test dataset with $\rho_\text{opt}=5, 10, 15, 20$, respectively. The red dots are the overlap results using the full $2$-second data segment, and the purple dots are the overlap results using the data segment of $0.25$s around the peak of the injected GW signal. Our model performs better in recovering GW signals with large chirp masses and around the merger event.
    }
\end{figure}

\begin{figure}[tbp]
    \centering
    \includegraphics[width=0.95\linewidth]{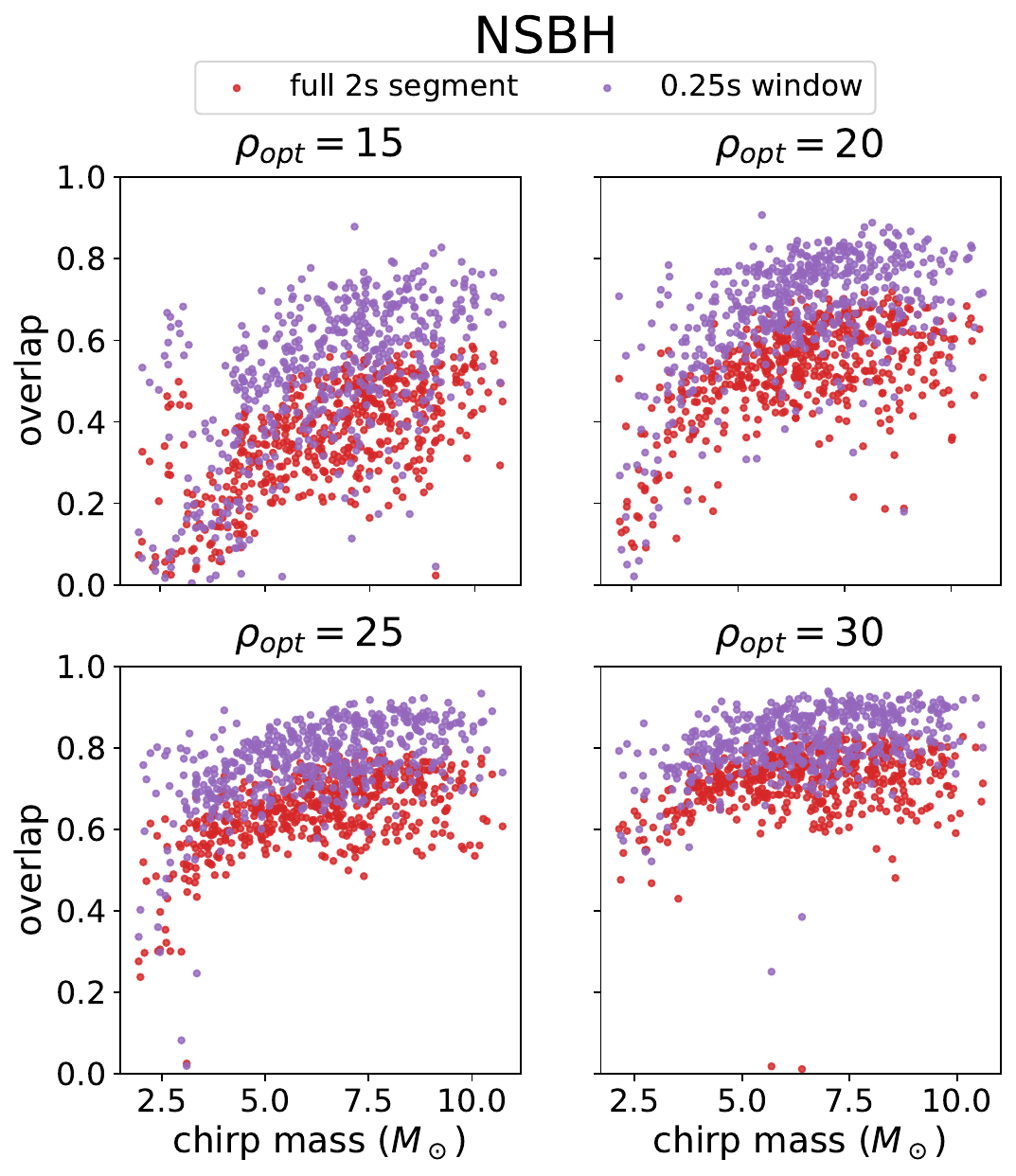}
    \caption{\label{fig:overlap_chirp_mass_BHNS}The overlap versus chirp mass of the NSBH test dataset with $\rho_\text{opt}=15, 20, 25, 30$, respectively. Our model still performs better in recovering GW signals with large chirp masses and around the merger event.}
\end{figure}

\begin{figure}[tbp]
    \centering
    \includegraphics[width=0.95\linewidth]{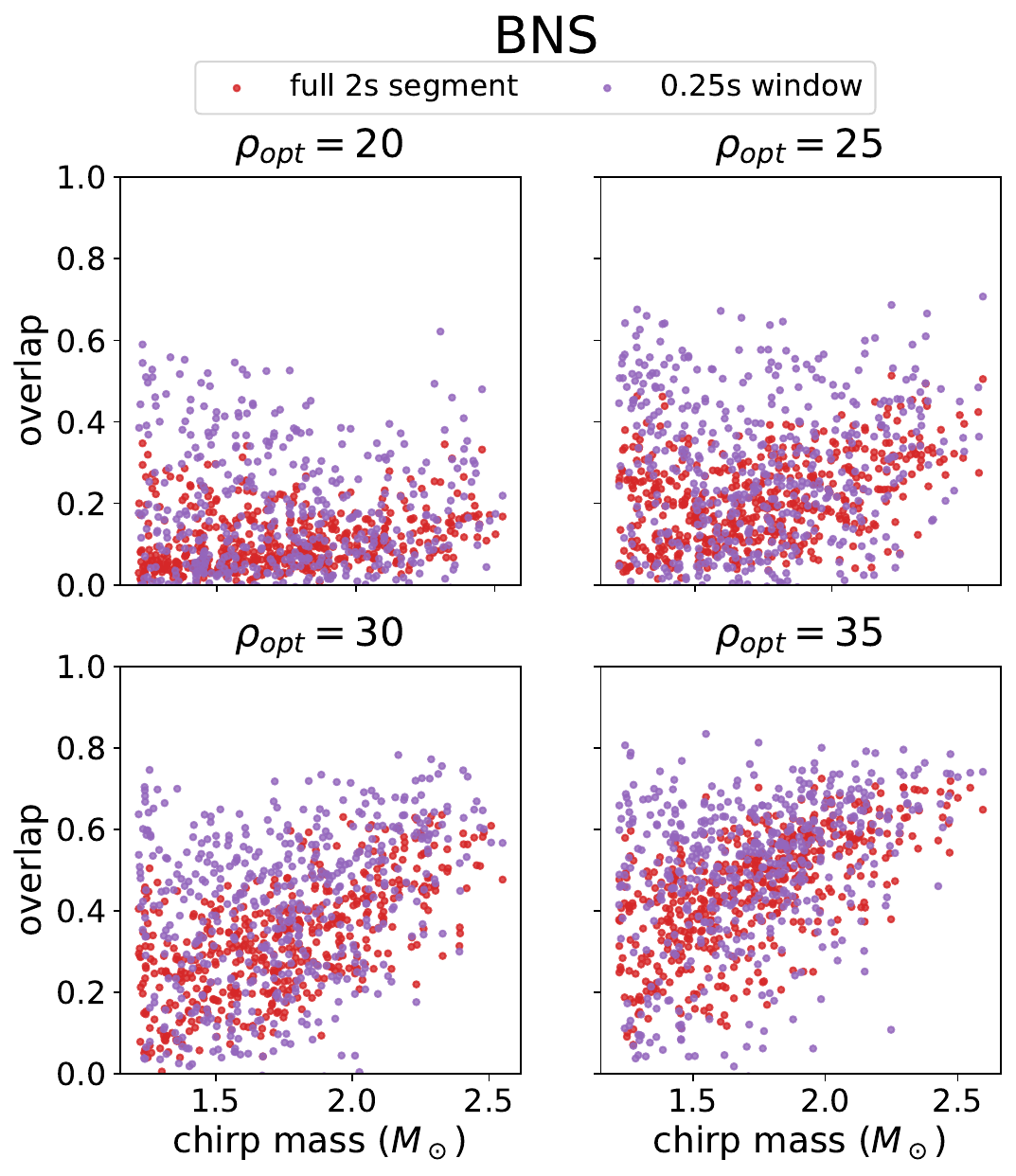}
    \caption{\label{fig:overlap_chirp_mass_BNS}The overlap versus chirp mass of the BNS test dataset with $\rho_\text{opt}=20, 25, 30, 35$, respectively. Again, our model performs better in recovering GW signals with large chirp masses, but there is no significant improvement in overlap when using the data segment around the merger event. 
    }
\end{figure}

To further investigate the large scattering in the overlap distributions, we consider the model performance against the signal frequency. The frequency evolution of the inspiral waveform has the form~\cite{andp.201600209}
\begin{align}\label{eq:frequency_evolution}
    f_\text{GW}^{-8/3}(t) = & \frac{(8\pi)^{8/3}}{5}\left(\frac{G\mathcal{M}_c}{c^3}\right)^{5/3}(t_c-t)\nonumber\\
    &+ \text{higher order corrections},
\end{align}
where $t_c$ is the coalescence time, $G$ is the gravitational constant, $c$ is the speed of light, and $\mathcal{M}_c$ is the chirp mass
\begin{equation}
    \mathcal{M}_c = \frac{(m_1m_2)^{3/5}}{(m_1+m_2)^{1/5}}.
\end{equation} 
From Equation~\eqref{eq:frequency_evolution}, we can see that a binary system with a larger chirp mass will have a lower frequency at a given time $t$. Therefore, we compare the overlap value with the chip mass in Figures~\ref{fig:overlap_chirp_mass_BBH}, \ref{fig:overlap_chirp_mass_BHNS} and \ref{fig:overlap_chirp_mass_BNS} respectively. We choose the samples with $\rho_\text{opt}=5, 10, 15, 20$ for the BBH dataset,  $\rho_\text{opt}=15, 20, 25, 30$ for the NSBH dataset and  $\rho_\text{opt}=20, 25, 30, 35$ for the BNS dataset. The red dots are the overlap results using the full $2$-second data segment, and the purple dots are the overlap results using the data segment of $0.25$s around the peak of the injected GW signal.
At high $\rho_\text{opt}$, our model performs well to recover GW signals with large chirp mass, which has lower frequencies around the merger event. On the other hand, our model performs poorly when recovering GW signals with small chirp mass, which has higher frequencies around the merger event. In addition, there are a few outliers in these figures, which may be due to the presence of glitches in the data.

Figures~\ref{fig:overlap_result}, \ref{fig:overlap_chirp_mass_BBH}, \ref{fig:overlap_chirp_mass_BHNS} and \ref{fig:overlap_chirp_mass_BNS} indicate that our model performs not well on low-amplitude signals, and GW signals with small chirp mass tend to have low amplitude while the optimal SNR is unchanged. This can be due to the difficulty of distinguishing the low-amplitude feature from the noise background using Gaussian likelihood in the frequency domain, as our model does not know the shape of the GW waveform at the training stage. To improve the performance on low-amplitude signals, we may have to use a different noise model or include more kinds of input data, such as phase information. We leave this as future work.

The left panel in Figure~\ref{fig:overlap_result} shows that our model performs better in extracting GW features around the merger event. In the future, we will investigate which overlap values can be used as the detection and parameter estimation threshold. However, although our model cannot recover the full part of GW waveforms, it can still be used as the feature extractor for detection and parameter estimation. 

\subsection{\label{subsec:real_test}Performance on Real Events}
In this section, we test our model with the confidently detected events in the Gravitational Wave Transient Catalog (GWTC) reported by the LVK Collaboration during O1, O2, and O3 observation runs~\cite{LIGOScientific:2018mvr, LIGOScientific:2020ibl, 2021arXiv210801045T, KAGRA:2021vkt}. 
We use the $4096$ seconds $4$KHz strain files containing the merger events available in GWOSC and choose the $32$-second segments, each with the event time at the center. Each segment is whitened using the PSD estimated from the $64$ seconds of data $16$ seconds before the event time. For BNS events GW170817 and GW190425, we use the $64$-second data, which is $128$ seconds before the event time, to estimate the PSD. 
We slice the center $2$-second data of the segment as the model input. There is a loud glitch in the L1 data before the event GW170817, and we use the method described in Reference~ \cite{Pankow:2018qpo} to mitigate the glitch.
To quantify the model performance, we generate the waveform templates using the approximant \texttt{SEOBNRv4T} for BBHs, \texttt{IMRPhenomBHNS} for BH-NSs, and \texttt{TaylorT4} for BNSs, respectively. We use the best-fit parameters of the waveforms provided on the GWOSC. To reconstruct the event waveforms, we perform the vanilla matched-filtering and then shift the template's time and phase to align with the strain data.

\begin{figure}[tbp]
    \centering
    \includegraphics[width=\linewidth]{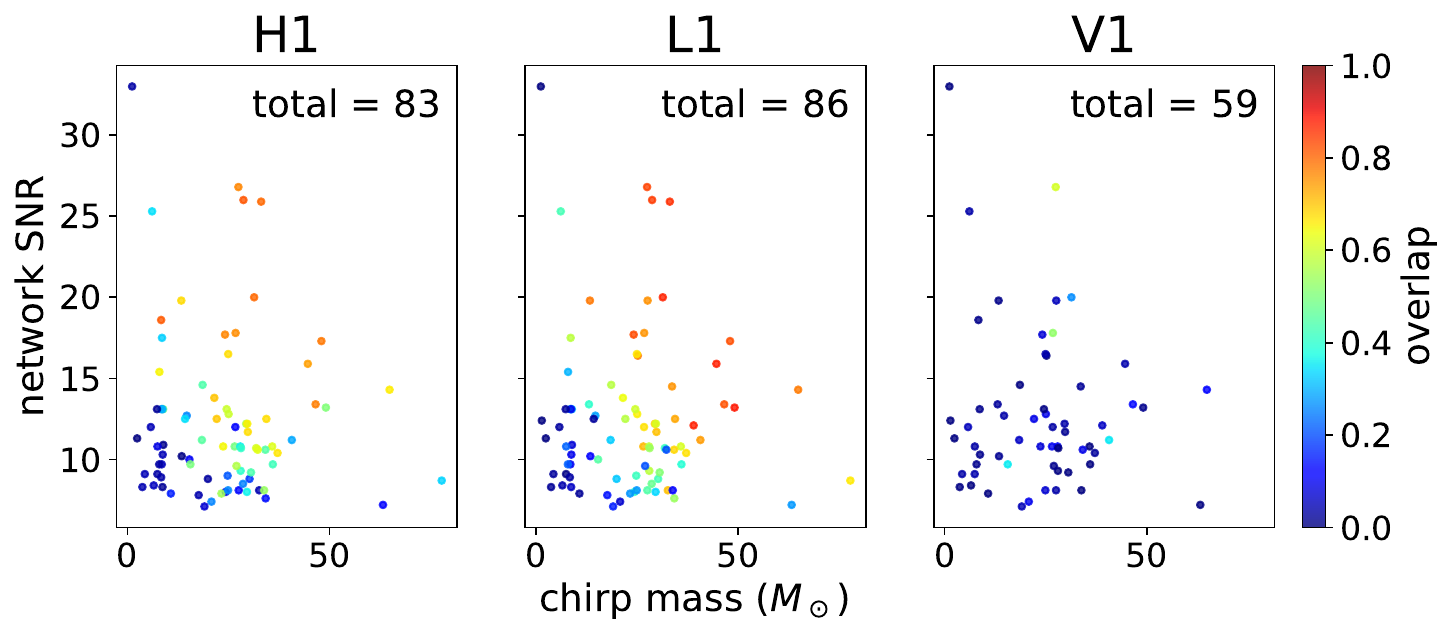}
    \caption{\label{fig:overlap_chirp_snr_real}The scatter plot of the overlap versus the network SNR and the chirp mass of the events. The color of the dots represents the value of overlaps. The overlaps are calculated using the $0.25$s data segment around the peak of the reconstructed waveform.}
\end{figure}

We plot the result as the scatter plots of overlaps against the network SNR and the chirp mass of the source in Figure~\ref{fig:overlap_chirp_snr_real}. The color of the dots represents the overlap value between $0$ and $1$. The overlaps are calculated using the $0.25$s segment around the peak of the reconstructed waveform. For LIGO H1 and L1 detector data, our model can denoise about $37.8$\% and $48.81$\% of the events to have overlaps exceeding $0.5$, respectively. However, for the Virgo data, our model can only denoise $4.35$\% of the events to have overlap exceeding $0.5$. This may be because the Virgo detector is less sensitive than the LIGO detectors, and the GW features in the Virgo time series data are not significant enough to be captured by our model.
In addition, from Figure~\ref{fig:overlap_chirp_snr_real}, we can see that our model predictions mostly have high overlaps on the LIGO H1 and L1 event data with high chirp masses or network SNRs and have low overlaps on the events with low chirp masses or network SNRs, which is in agreement with the mock data results in Section~\ref{subsec:mock_test}. The blue dot in the top left of each plot is the GW170817 BNS event, which our model fails to denoise in all detector data, although having the highest network SNR. This also agrees with the mock data results that our model performs poorly on BNS signals.

\begin{figure}[tbp]
    \centering
    \subfloat[\label{fig:gw150914_full}]{\includegraphics[width=\linewidth]{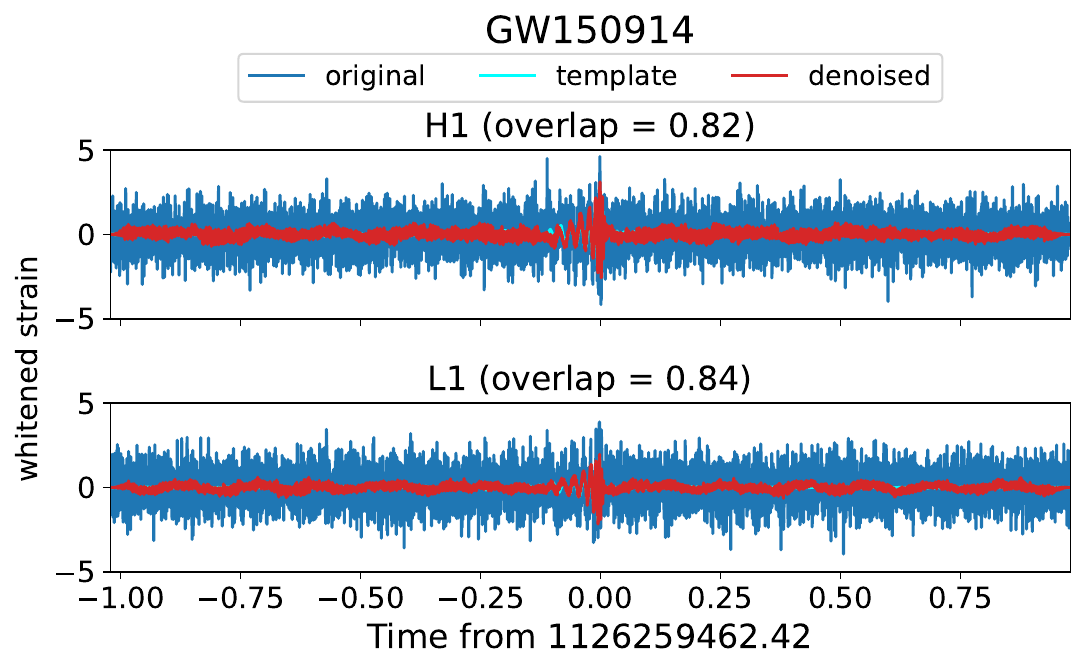}}
    \\
    \subfloat[\label{fig:gw150914_zoom}]{\includegraphics[width=\linewidth]{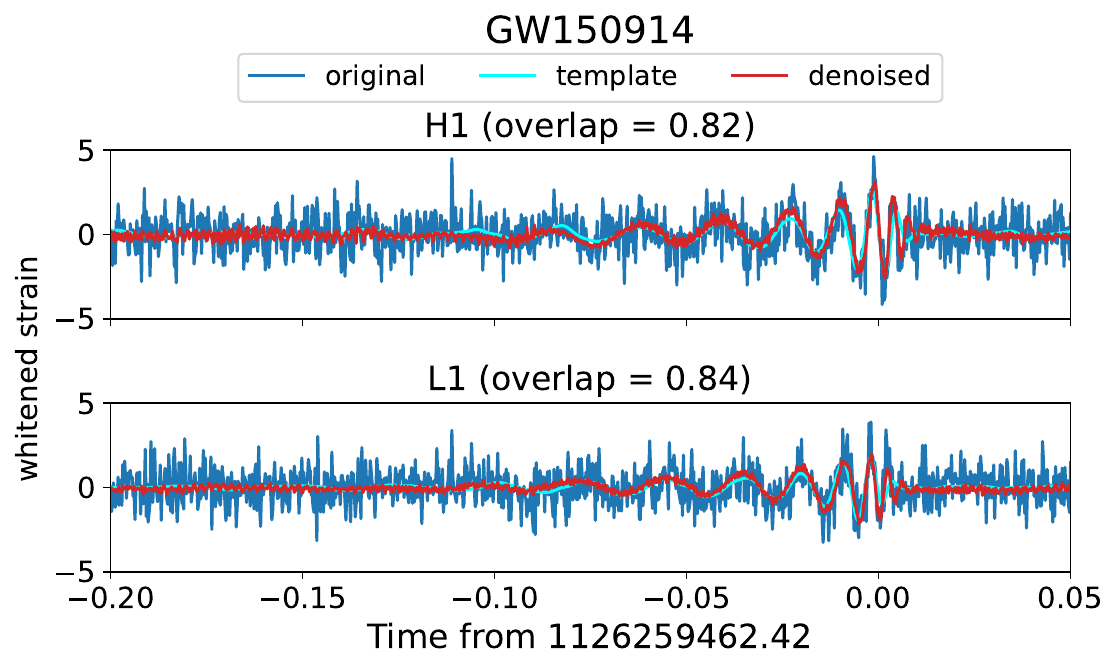}}
    \caption{\label{fig:gw150914}(a) The denoised results of $2$-second H1 and L1 data around the event GW150914. The blue curves are the whitened strain data, the red curves are the denoised results, and the cyan curves are the reconstructed waveforms. (b) The $0.25$s windows around the peak of the reconstructed waveform, which we use to calculate overlaps.}
\end{figure}

\begin{figure}[tbp]
    \centering
    \includegraphics[width=\linewidth]{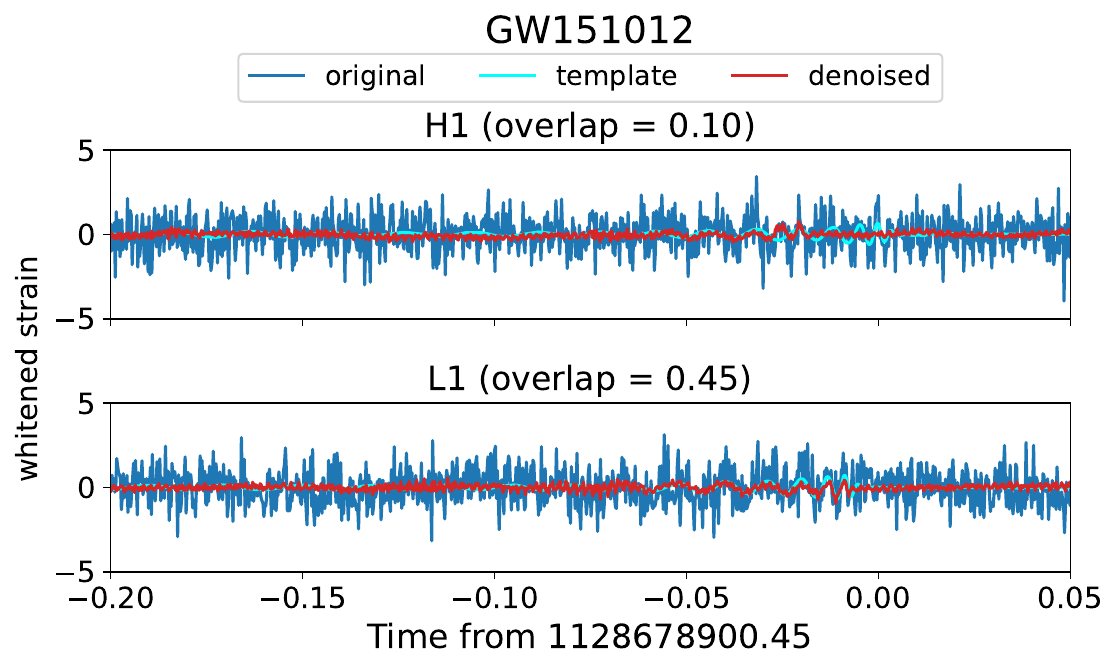}
    \caption{\label{fig:gw151012}The $0.25$s windows of the denoised results of H1 and L1 data around GW151012.}
\end{figure}

\begin{figure}[tbp]
    \centering
    \includegraphics[width=\linewidth]{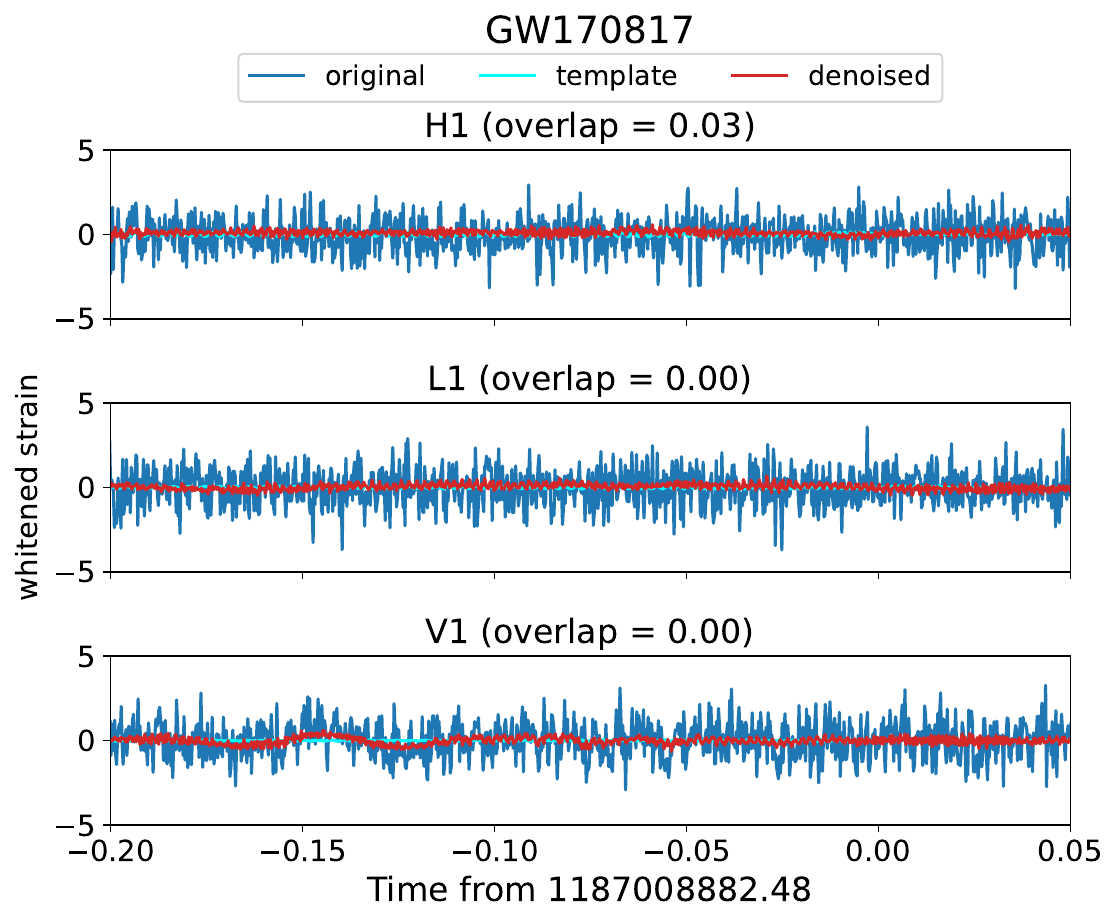}
    \caption{\label{fig:gw170817}The $0.25$s windows of the denoised results of H1, L1 and V1 data around GW170817.}
\end{figure}

To see how the denoised results behave, we plot the denoised result of the $2$-second H1 and L1 data around the event GW150914 in Figure~\ref{fig:gw150914_full}, and in Figure~\ref{fig:gw150914_zoom} we zoom the result of each detector data to the $0.25$s window around the peak of the reconstructed waveform which we use to calculate overlap.
The blue curves are the whitened strain data, the red curves are the denoised results, and the cyan curves are the reconstructed waveforms. The overlap is 0.82 for the H1 data and 0.84 for the L1 data. Figure~\ref{fig:gw150914} shows that our model can effectively remove the Gaussian noise and recover the GW features around the loud merger event, which can be used as a feature extractor for detection and parameter estimation tasks.
However, we can see that the denoised results still contain some high-frequency noises, which decrease the overlap. In addition, the mismatches between the denoised results and the reconstructed waveforms at the early inspiral stage and the ring-down stage also downgrade the overlap. Our denoised results tend to follow the trend of low-frequency data fluctuation since our model does not know GW waveforms. The same phenomenon can be observed in the denoised results of event GW151012, as shown in Figure~\ref{fig:gw151012}, where the denoised result of H1 data peaks earlier than the reconstructed waveform, which may bias the parameter estimation results.

On the other hand, our model cannot preserve the GW features of the BNS event GW170817, as shown in Figure~\ref{fig:gw170817}. The overlap is 0.03 for the H1 data and 0.0 for the L1 and V1 data. We can see that the amplitudes of the reconstructed waveforms are insignificant in all detector time series data and may be recognized by the model as parts of Gaussian noise. Therefore, we may need to seek other information rather than time series data alone to extract the GW features for BNS waveforms.
We present the denoised results of the remaining confidently-detected GW events during O1, O2, and O3 runs in Appendix~\ref{sec:appendix_a}.

\subsection{\label{subsec:glitch_test}Testing on Transient Noises}

\begin{figure*}[tbp]
    \centering
    {\includegraphics[width=0.99\linewidth]{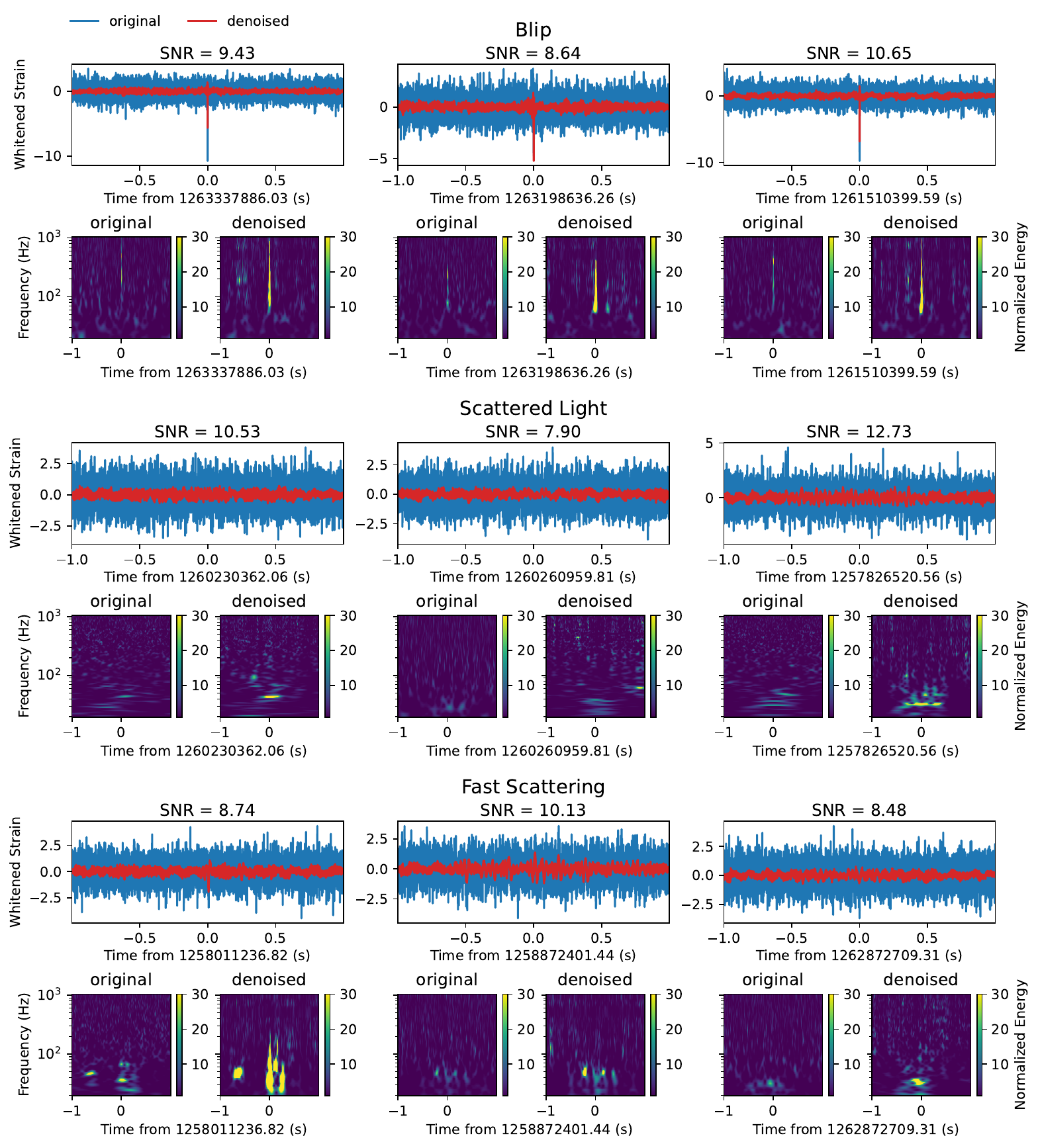}}
    \caption{\label{fig:glitches_H1}The denoised results of the 2-second H1 O3b data containing three types of low-frequency glitches: Blip (top row), Scattered Light (middle row), and Fast Scattering (bottom row). The strain segments are randomly chosen according to the glitch timestamps in the Gravity Spy O3b ML classification results with ML confidence larger than $0.95$ and glitch SNR between $7.5$ and $15$. The glitch SNRs are calculated by the Omicron pipeline. Each panel contains denoised results of three strain segments. The upper part of the denoised result of each segment is the time series data, the blue curve is the whitened strain data, and the red curve is the denoised results. The two plots below each time series plot are the spectrogram of the original strain data (left) and the denoised data (right). Our model can extract low-frequency glitches and emphasize their morphology in the spectrogram.
    }
\end{figure*}

\begin{figure*}[tbp]
    \centering
    {\includegraphics[width=0.99\linewidth]{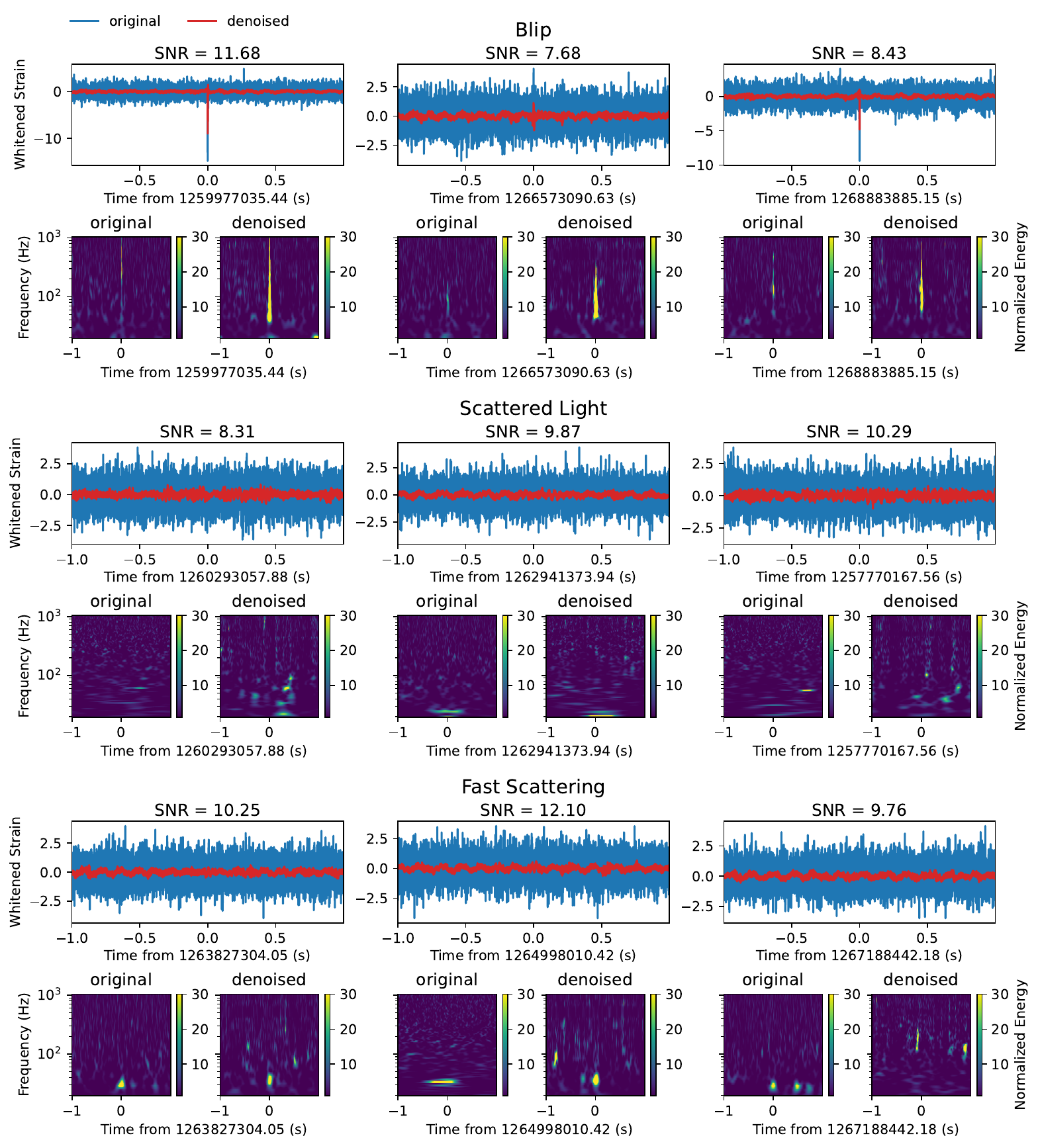}}
    \caption{\label{fig:glitches_L1}The denoised results of the 2-second L1 O3b data containing three types of low-frequency glitches: Blip (top row), Scattered Light (middle row), and Fast Scattering (bottom row). Again, the strain segments are randomly chosen according to the glitch timestamps in the Gravity Spy O3b ML classification results with ML confidence larger than $0.95$ and glitch SNR between $7.5$ and $15$. The conventions in plots are the same as those in Figure~\ref{fig:glitches_H1}.}
\end{figure*}

We also tested our model with the strain data containing glitches, as we expect there to be some glitches in the background data we used. These non-Gaussian features can be probed when training the model with the Gaussian noise likelihood.
We tested our model with three types of low-frequency glitches in LIGO O3b data, each with a different morphology in the spectrogram: Blip, Scattered Light, and Fast Scattering. To find the timestamp of these glitches, we refer to the public classification results of H1 and L1 O3b data from the Gravity Spy project~\cite{Glanzer_2023, Zevin_2017, BAHAADINI2018172, PhysRevD.99.082002, Soni_2021}, which is the machine learning project to classify the glitches found by the Omicron pipeline~\cite{ROBINET2020100620}. For each type of glitch, we randomly chose three glitch events with their SNR lying between $7.5$ and $15$ and their ML confidence larger than $0.95$.

We present the denoised results of $2$-second H1 and L1 data segments containing Blip, Scattered Light, and Fast Scattering glitches in Figure~\ref{fig:glitches_H1} and \ref{fig:glitches_L1}, respectively. Each figure contains three rows of plots corresponding to Blip, Scattered Light, and Fast Scattering glitches. In each row, there are denoised results of three strain segments. For each segment, we present the original strain time series (blue curve) and the denoised time series (red curve) in the top panel, and in the bottom panel, we plot the spectrograms of the original strain data on the left side, and the denoised data on the right side, respectively.

Our model can extract the structure of Blip glitches from time series data and emphasize their morphologies in the spectrogram for both H1 and L1 data. However, for Scattered Light and Fast Scattering glitches with low SNR, the extracted structures are insignificant in time series data, and some of their morphologies are also slightly altered in the spectrogram, especially in L1 data, which may affect the accuracy and confidence of glitch classification tasks. The removal of low SNR glitches may be due to their low amplitude in strain data; therefore, our model does not pick up their non-Gaussianities. In addition, Fast Scattering glitches are more frequent during the O3 run~\cite{smith44803, fastScatter}, whereas our training dataset contains mostly O2 data.

From the perspective of GW detection, although our model also extracts glitch features, most of the glitches have very different structural characteristics from CBC signals, which we believe can be easily distinguished by a simple classifier model after rescaling the result. However, our model will not extract the GW signal alone when the signal overlaps with a glitch. The method to separate GW signals from glitches is left for future work.

\subsection{Early Detection}\label{subsec:early_detection}
\begin{figure}[tbp]
    \centering
    \includegraphics[width=\linewidth]{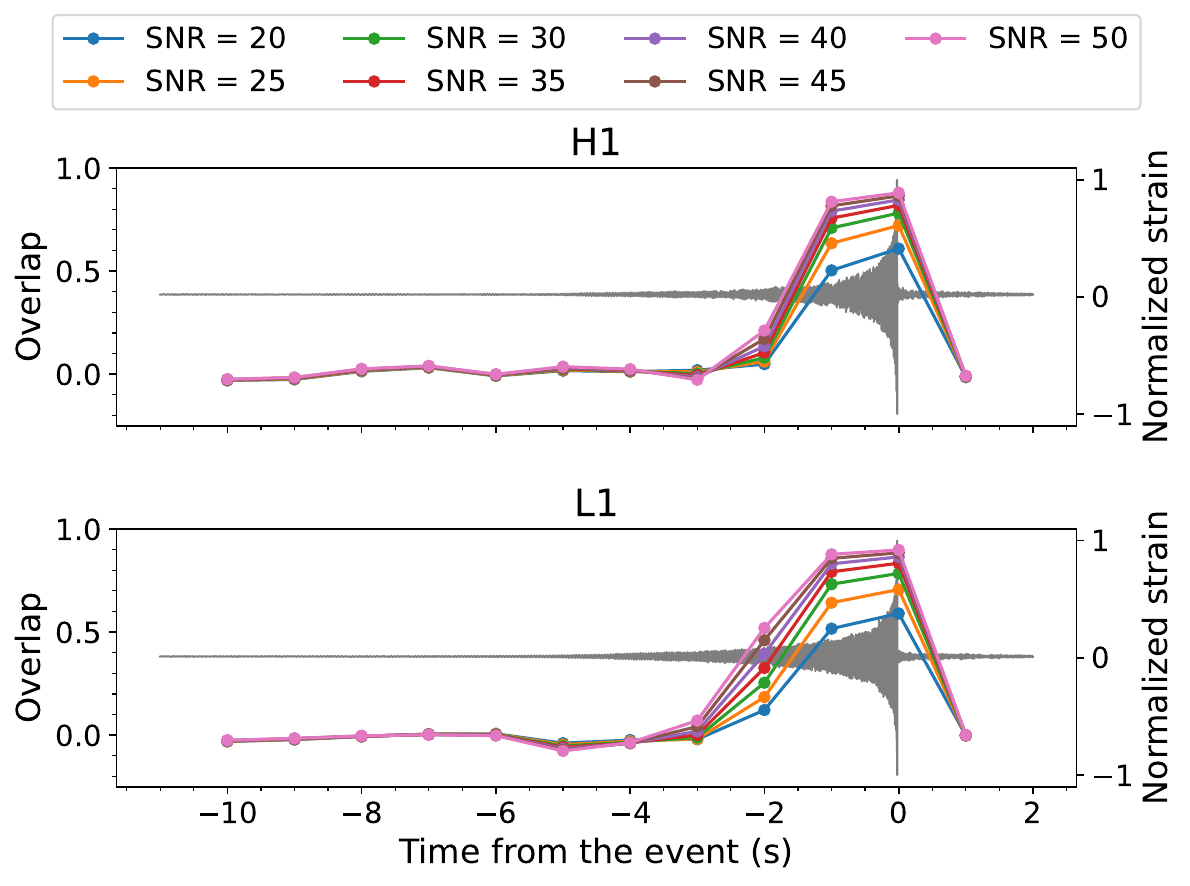}
    \caption{\label{fig:early_det_bbh}The overlap result of the time windows containing a BBH signal with component masses $m_1 = 11.0\, M_\odot$ and $m_2 = 7.6\, M_\odot$. The gray curve is a normalized waveform to indicate the signal's shape and position in the strain data. 
    }
\end{figure}

\begin{figure}[tbp]
    \centering
    \includegraphics[width=\linewidth]{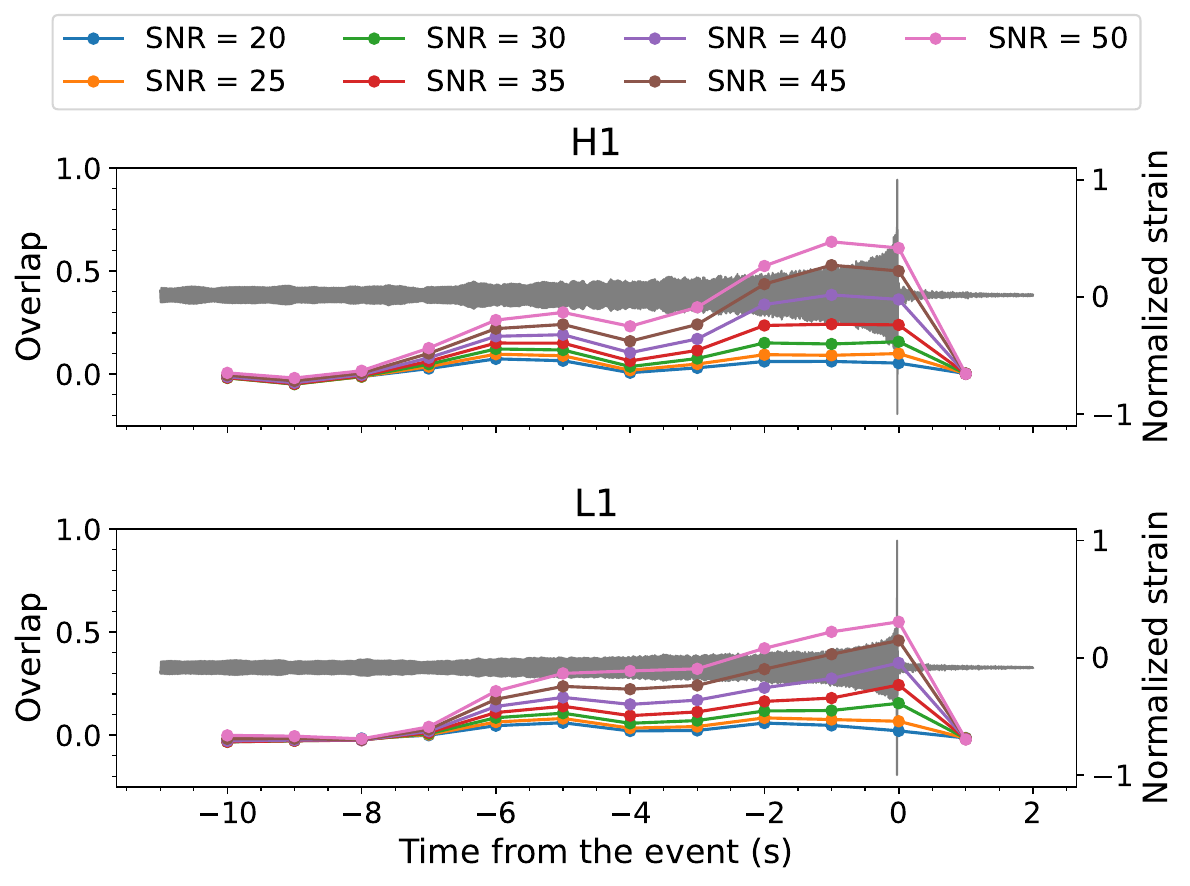}
    \caption{\label{fig:early_det_bns}The overlap result of the time windows containing a BNS signal with component masses $m_1 = 1.7\, M_\odot$ and $m_2 = 1.5\, M_\odot$. The gray curve is a normalized waveform to indicate the signal's shape and position in the strain data.}
\end{figure}

\begin{figure}[tbp]
    \centering
    \includegraphics[width=\linewidth]{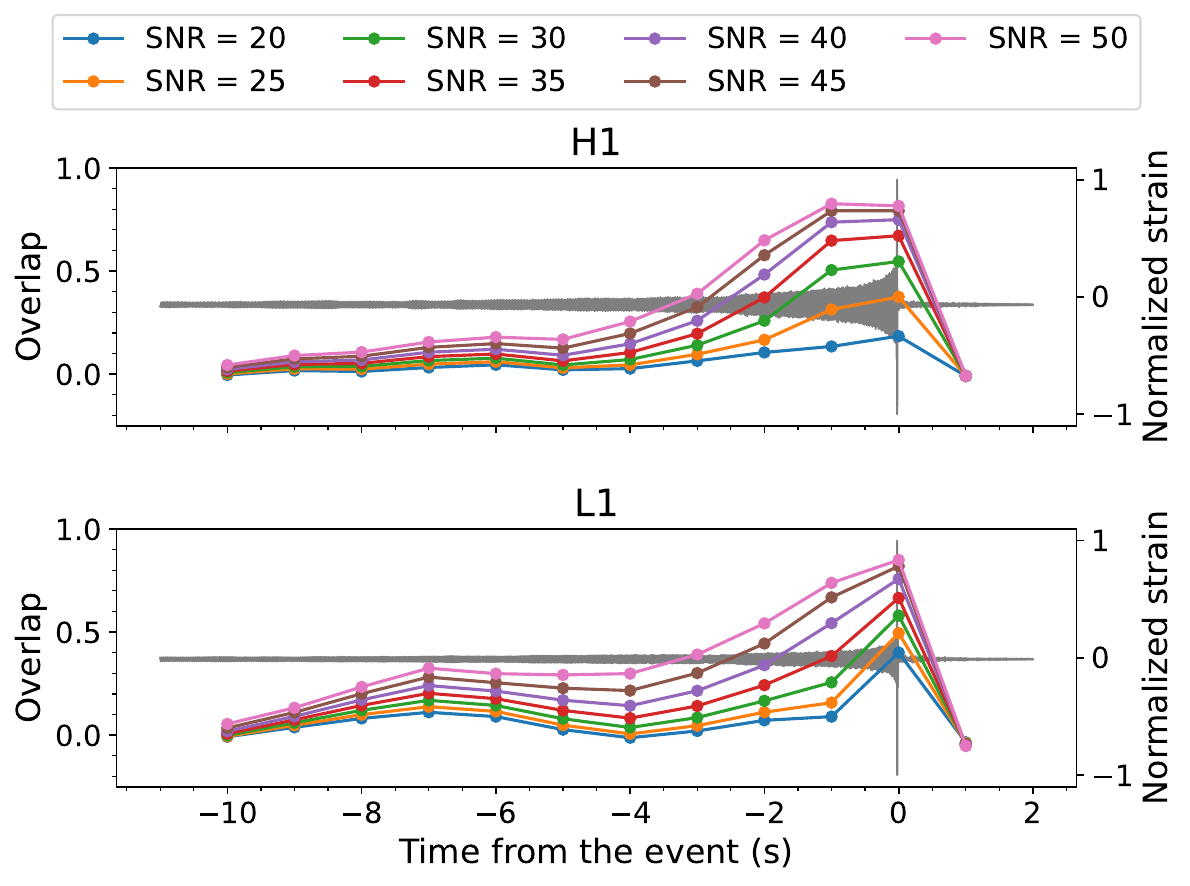}
    \caption{\label{fig:early_det_nsbh}The overlap result of the time windows containing an NSBH signal with component masses $m_1 = 5.9\, M_\odot$ and $m_2 = 1.44\, M_\odot$. The gray curve is a normalized waveform to indicate the signal's shape and position in the strain data.}
\end{figure}

Since the input data during training may contain only inspiral signals, it is interesting to see whether our model can extract GW inspiral features when the signal is loud. 
Therefore, we performed three tests. We randomly chose a 16-second data segment from H1 and L1 O3b strain data in each test. For the first test, we injected a BBH signal similar to the event GW170608 with component masses $m_1 = 11.0\, M_\odot$ and $m_2 = 7.6\, M_\odot$, and for the second test, we injected a BNS signal with component masses $m_1 = 1.7\, M_\odot$ and $m_2 = 1.5\, M_\odot$. Finally, for the third test, we injected an NSBH signal similar to the event GW200115\_042309 with component masses $m_1 = 5.9\, M_\odot$ and $m_2 = 1.44\, M_\odot$. The signals were generated with the optimized orientation of the H1 detector. During the test, we adjust their optimal SNR from $20$ to $50$ in step $5$. We then slice and whiten the data into several $2$-second time windows with $1$-second overlap and feed them into our model. We calculate the overlap between the denoised result and the injected GW signal in each window.

We plot the test results in Figure~\ref{fig:early_det_bbh}, Figure~\ref{fig:early_det_bns}, and Figure~\ref{fig:early_det_nsbh}. In each plot, the blue, orange, green, red, purple, brown, and pink dots correspond to the overlap results of each time window with injections of SNR $20, 25, 30, 35, 40, 45$, and $50$, respectively. The gray curve is a normalized waveform to indicate the shape and position of the injected waveform in the strain data. The x-axis is the time distance of the center of time windows from the merger time. 
When the optimal SNR of the signal is larger than $50$, our model can extract GW features with overlap larger than $0.2$ in both detector data $2$ seconds before the merger for BBH signals, $6$ seconds for BNS signals, and $4$ seconds before the merger for NSBH signals, respectively. 
Although the result of the BBH signal is inadequate for low-latency early-warning systems since we are using a $2$-second window and need additional $1$-second data at both sides for whitening, our model can extract inspiral features of NSBH and BNS events for other early-warning detection models when the SNR is high.

\begin{figure*}[tbp]
    \centering
    \includegraphics[width=0.99\linewidth]{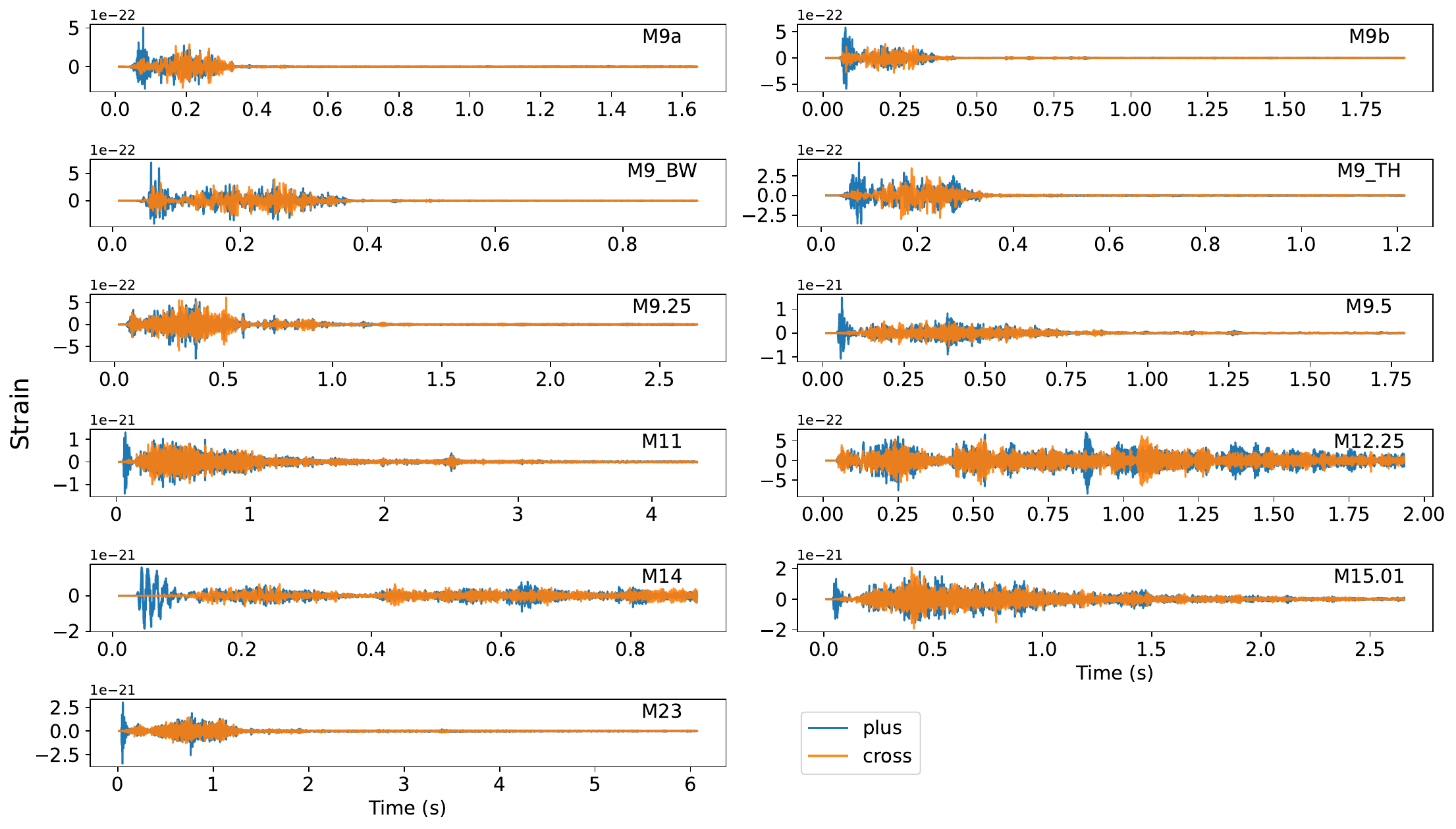}
    \caption{\label{fig:ccsn_waveform}The plus (blue curves) and cross (yellow curves) polarizations of CCSN waveforms for testing our model in this work. The waveforms are generated from 3D simulations in Ref.~\cite{PhysRevD.107.103015} with various progenitor mass models. We applied an anti-aliasing filter before downsampling to $4096$ Hz and a low-pass filter at $15$ Hz to remove low-frequency components making large deviations from $0$. The waveforms are observed along the positive $x$ axis of the simulation frame, and we scale the waveforms so that the sources are placed at a distance of $1$kpc.}
\end{figure*}

\subsection{\label{subsec:ccsn}Test on CCSN waveforms}
Our previous results show that our model performs better on short-duration signals. Therefore, our model on CCSN waveforms is worth testing, although we did not put them in the training dataset. 
We test our model using the publicly available GW waveforms from Vartanyan's simulations~\cite{PhysRevD.107.103015}. The simulations contain $11$ neutrino-driven, non-rotating progenitor models, including four $9M_\odot$ models simulated on different computing clusters and models with progenitor masses $9.25$, $9.5$, $11$, $12.25$, $14$, $15.01$, and $23 M_\odot$. All progenitor models are employed with $12$ neutrino energy groups and the SFHo equation of state~\cite{Steiner_2013, PhysRevD.107.103015}. Among these models, the $12.25 M_\odot$ and $14 M_\odot$ progenitors failed to explode. The waveforms are observed along the positive $x$ axis in the simulation frame. Also, note that the waveforms we used only come from the matter terms of the simulation and do not contain signals from neutrino memory~\cite{ccsn_web}.

The original waveform data are sampled at various sampling rates. To match the sampling rate we used in this work, we first applied a Butterworth low-pass filter at $2048$ Hz to prevent aliasing, then downsampled the data to the sampling rate of $4096$ Hz. We also used a high-pass filter at $15$ Hz to remove the low-frequency components in the data; otherwise, the data can have a large offset from $0$, impacting the model performance. 
We randomly chose 4-second data segments from H1 and L1 O3b strain data to perform the test. We injected a CCSN signal located at the optimal direction of each detector, with its starting time at the $1.25$ second of the data. We then whitened the data and kept the center 2-second segment as the model input. During the test, we use the waveforms from CCSN models described above and adjust their $\rho_\text{opt}$ from $10$ to $40$. We calculate the overlap between model predictions and injected GW signals. Although, in reality, it is unlikely to search for CCSN signals using matched filtering, the matched filter SNR can help us scale the waveform better so that the statistics of the strain data will not deviate too much from the training dataset.

\begin{figure*}[tbp]
    \centering
    \includegraphics[width=0.99\linewidth]{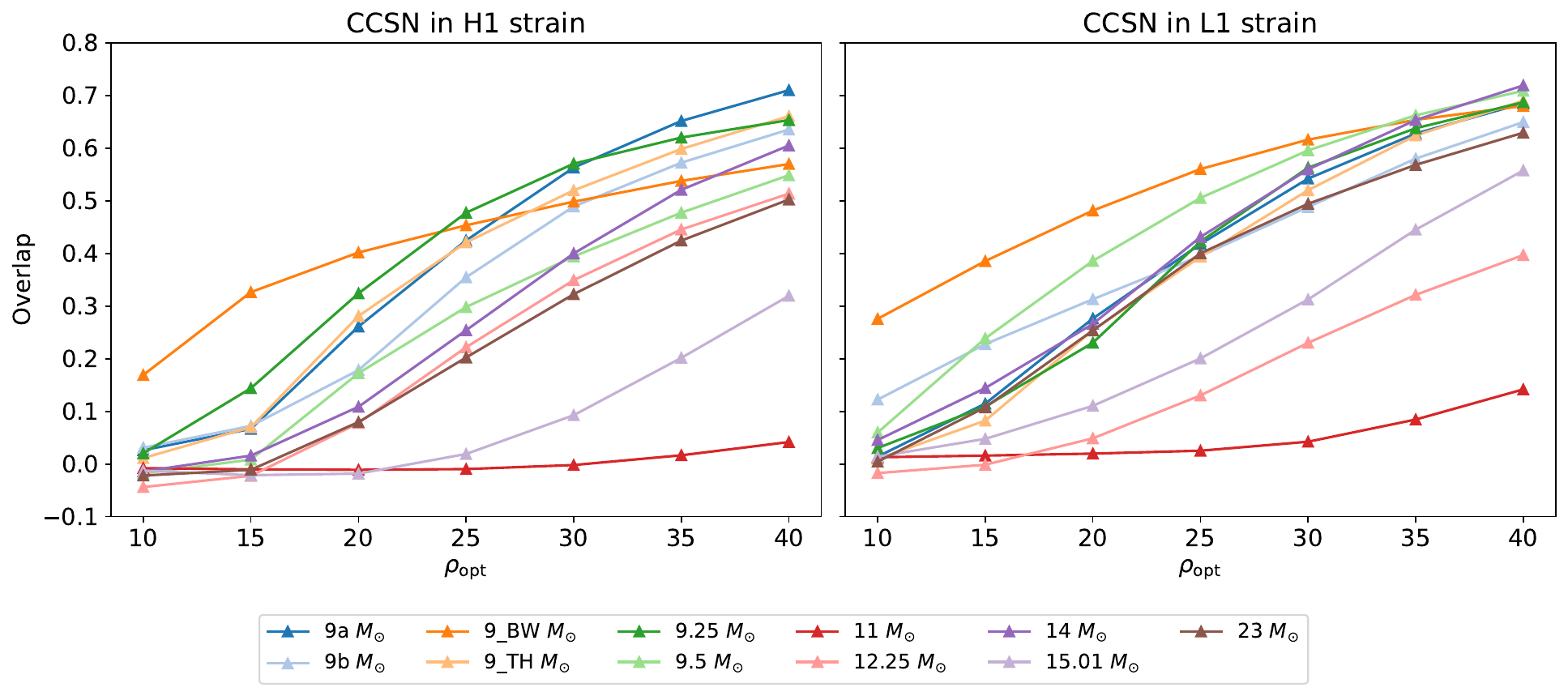}
    \caption{\label{fig:ccsn_snr_test}The performance of our model on different CCSN models at different optimal SNR.}
\end{figure*}

We present the test result in Figure~\ref{fig:ccsn_snr_test}. In each plot, the blue, light blue, orange, light orange, green, light green, red, pink, violet, light purple, and purple triangle dots correspond to the overlap results of progenitor models $9a\; M_\odot$, $9b\; M_\odot$, $9_\text{BW}\; M_\odot$, $9_\text{TH}\; M_\odot$, $9.25\; M_\odot$, $9.5\; M_\odot$, $11\; M_\odot$, $12.25\; M_\odot$, $14\; M_\odot$, $15.01\; M_\odot$, and $23\; M_\odot$, respectively. Our model can partially extract features of all waveform models when $\rho_\text{opt}\geq 25$ except for the $15.01\; M_\odot$ model in the H1 detector and $11\; M_\odot$ model in both detectors. The poor performance of these progenitor models is due to their longer waveform duration and, therefore, smaller amplitudes when scaling to the same SNR.

The test result shows that our model can extract short-duration CCSN signals even if it does not see them during the training stage. In the future, we will inject more CCSN waveforms into the training dataset to improve the model performance and robustness against CCSN signals.

\section{\label{sec:conclusion}Conclusion}
We developed a self-supervised learning denoiser model to extract GW features from the time series data without explicitly providing cleaned GW waveforms during training. We trained our model using the Gaussian noise likelihood assumption, and our model can extract loud BBH and BHNS features with average overlap greater than 0.5 in the 0.25s window when the signal SNR is larger than 10 and 14, respectively. However, we may need additional types of input data to denoise BNS signals in the future. 

We also tested our model on the actual events detected by LIGO and Virgo during O1, O2, and O3 observation runs. Our model can denoise about $38$\% and $49$\% of the events to have overlaps exceeding $0.5$ in the H1 and L1 data, respectively. However, for the Virgo data, our model can only denoise $4.35$\% of the events to have overlap exceeding $0.5$. We also tested our model on the strain data containing glitches and found that our model can effectively reveal the glitch structures that can be easily distinguished from CBC signals. In addition, our model shows the potential to extract features and allow detection models to send early warning alerts for BNS and NSBH events when the SNR is high. Furthermore, our model can partially extract loud CCSN features even if not added to the training dataset.

Our work shows the potential of using self-supervised learning to extract GW features from time series data by training with injected strain data alone. In the future, we will investigate the improvement in detection and parameter estimation tasks with the integration of our feature extractor and seek a way to improve the model's ability to extract BNS and low-amplitude features. 
We will also investigate the possibility of using our model to extract GW features from other types of GW signals, such as GWs from other CCSN models or continuous waves from spinning neutron stars. 

\begin{acknowledgments}
    This research has made use of data or software obtained from the Gravitational Wave Open Science Center (gw-openscience.org), a service of LIGO Laboratory, the LIGO Scientific Collaboration, the Virgo Collaboration, and KAGRA. LIGO Laboratory and Advanced LIGO are funded by the United States National Science Foundation (NSF) as well as the Science and Technology Facilities Council (STFC) of the United Kingdom, the Max-Planck-Society (MPS), and the State of Niedersachsen/Germany for support of the construction of Advanced LIGO and construction and operation of the GEO600 detector. Additional support for Advanced LIGO was provided by the Australian Research Council. Virgo is funded, through the European Gravitational Observatory (EGO), by the French Centre National de Recherche Scientifique (CNRS), the Italian Istituto Nazionale di Fisica Nucleare (INFN) and the Dutch Nikhef, with contributions by institutions from Belgium, Germany, Greece, Hungary, Ireland, Japan, Monaco, Poland, Portugal, Spain. The construction and operation of KAGRA are funded by Ministry of Education, Culture, Sports, Science and Technology (MEXT), and Japan Society for the Promotion of Science (JSPS), National Research Foundation (NRF) and Ministry of Science and ICT (MSIT) in Korea, Academia Sinica (AS) and the National Science and Technology Council (NSTC) in Taiwan.

    We acknowledge the support from NSTC through grants 112-2811-M-007-041 and 112-2112-M-007-042. This work used high-performance computing facilities operated by the Center for Informatics and Computation in Astronomy (CICA) at National Tsing Hua University. This equipment was funded by the Ministry of Education of Taiwan, NSTC, and National Tsing Hua University.
\end{acknowledgments}

\bibliography{Draft}

\newpage
\appendix
\renewcommand{\thefigure}{A\arabic{figure}}
\setcounter{figure}{0}
\section{\label{sec:appendix_a}Denoised Results of other Events}
In this appendix, we present the rest of the denoised results of the confidently detected events on the Gravitational Wave Transient Catalog (GWTC) reported by the LVK Collaboration during O1, O2, and O3 observation runs, in addition to the events GW150914, GW151012, and GW170817 described in Section~\ref{subsec:real_test}. Although the input data are the 2-second segments, for each event, we focus on the 0.25s data segment around the peak of the reconstructed waveform, which is used for calculating the overlap. For O1 and O2 events, we use the name listed on GWTC. Since starting from O3, GW candidates have been uploaded to the Gravitational-wave candidate event database (GraceDB). In addition to the GWTC name, we attached the GraceDB name as a reference for O3 events. 

\begin{figure*}[tbp]
    \centering
    \subfloat[\label{fig:appendix_1_1}]{\includegraphics[width=0.48\linewidth]{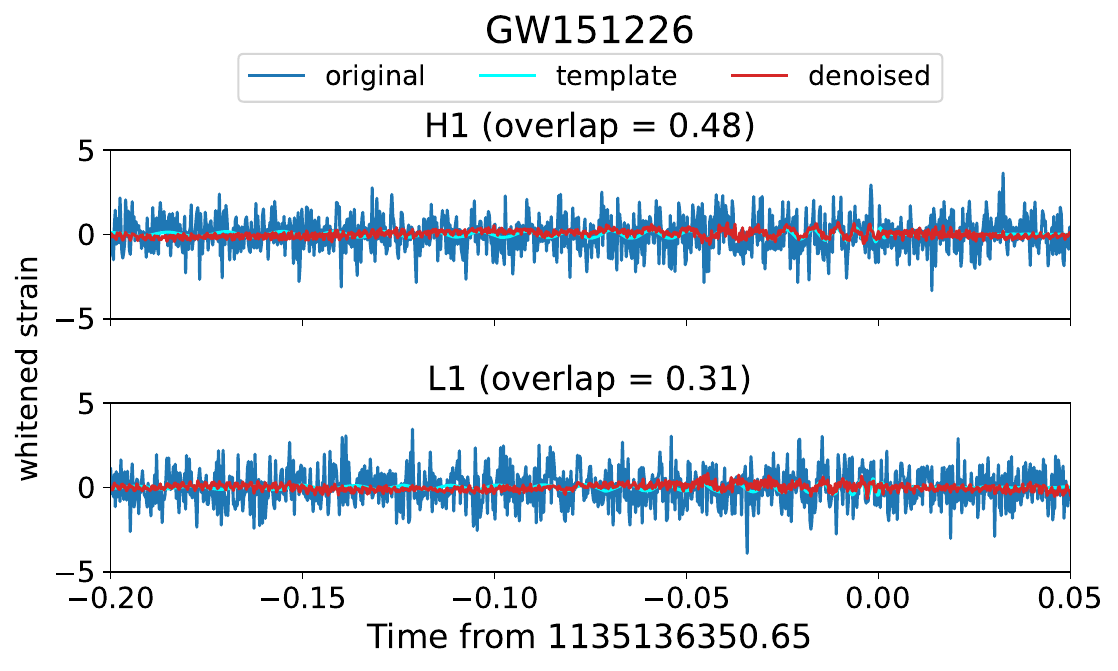}}
    \hfill
    \subfloat[\label{fig:appendix_1_2}]{\includegraphics[width=0.48\linewidth]{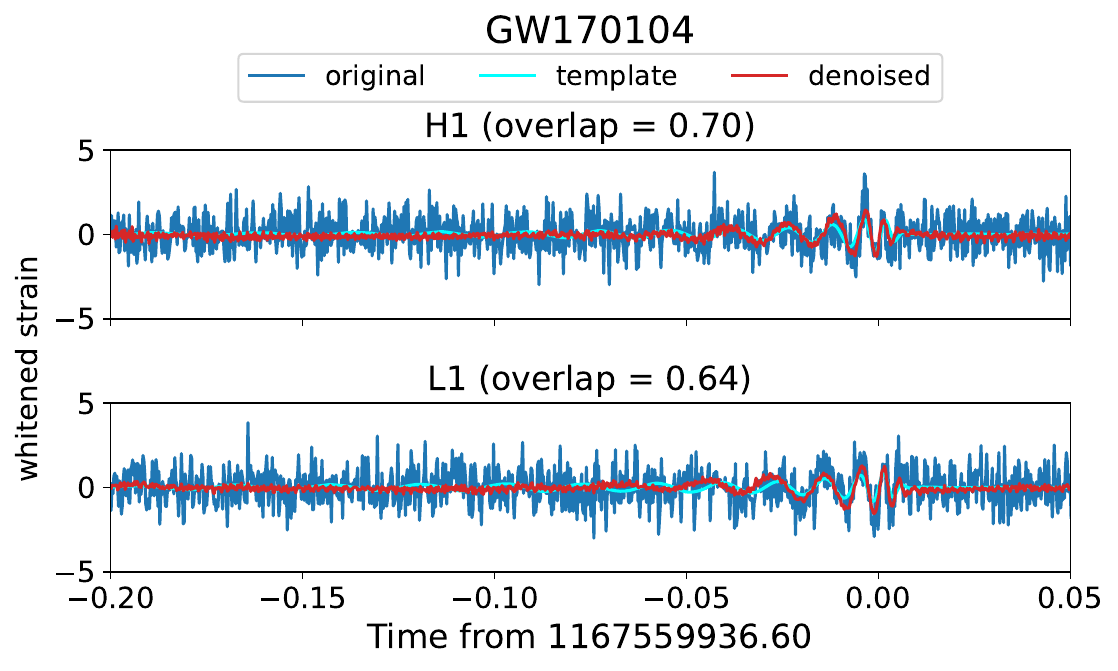}} 
    \caption{\label{fig:appendix_1}The $0.25$s windows of the denoised results of the H1 and L1 data around GW151226 and GW170104.}
\end{figure*}

\begin{figure*}[tbp]
    \centering
    \subfloat[\label{fig:appendix_2_1}]{\includegraphics[width=0.48\linewidth]{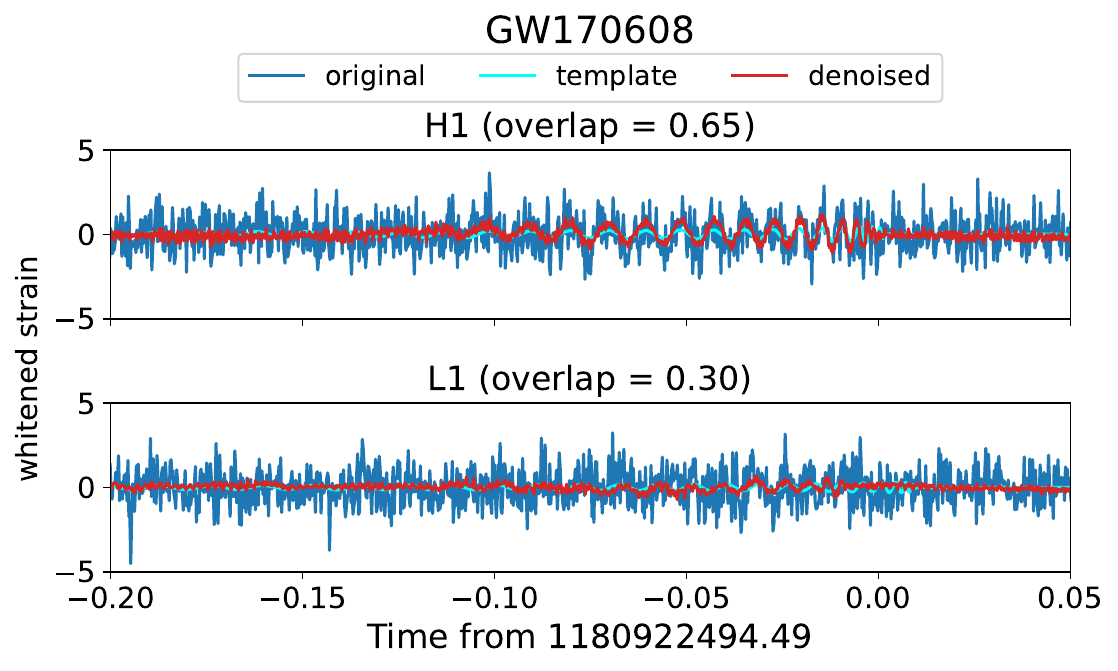}}
    \hfill
    \subfloat[\label{fig:appendix_2_2}]{\includegraphics[width=0.48\linewidth]{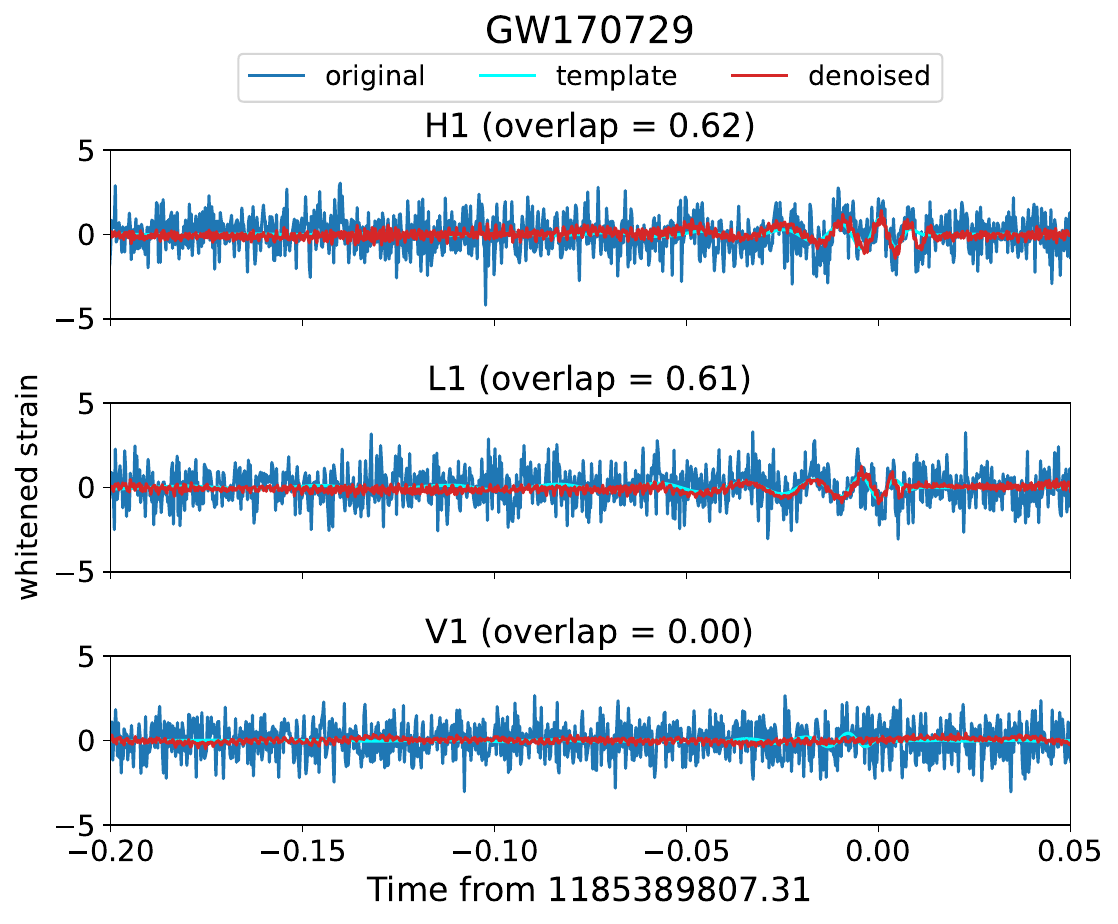}}
    \caption{\label{fig:appendix_2}The $0.25$s windows of the denoised results of the H1, L1 and V1 data around GW170608 and GW170729.}
\end{figure*}

\begin{figure*}[tbp]
    \centering
    \subfloat[\label{fig:appendix_3_1}]{\includegraphics[width=0.48\linewidth]{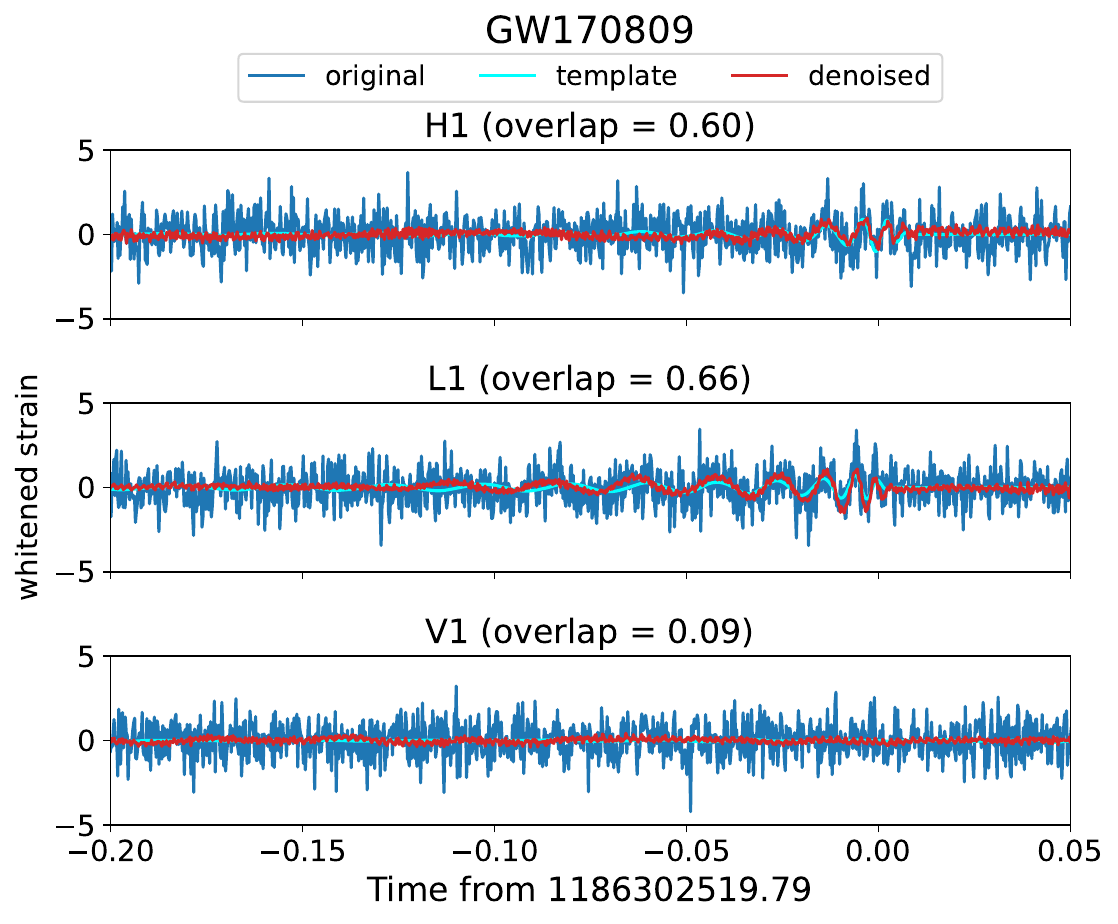}}
    \hfill
    \subfloat[\label{fig:appendix_3_2}]{\includegraphics[width=0.48\linewidth]{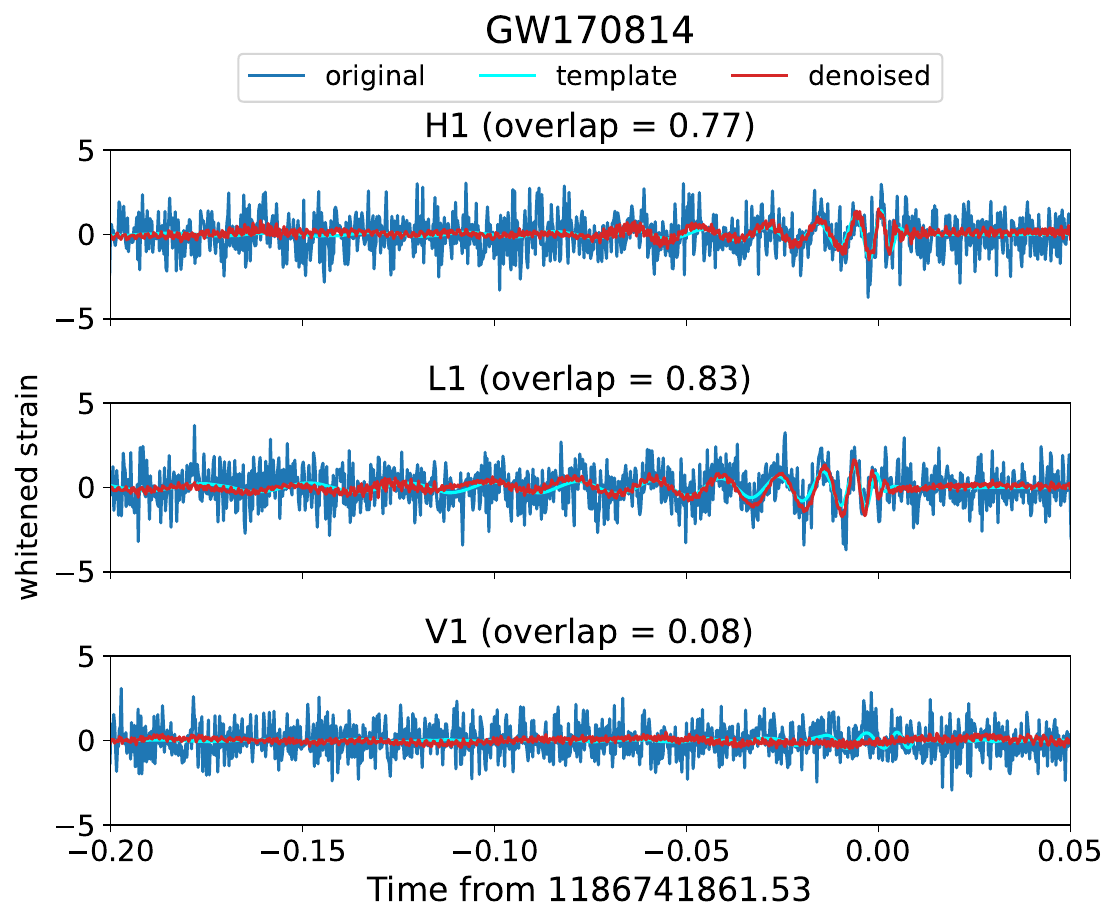}}
    \caption{\label{fig:appendix_3}The $0.25$s windows of the denoised results of the H1, L1 and V1 data around GW170809 and GW170814.}
\end{figure*}

\begin{figure*}[tbp]
    \centering
    \subfloat[\label{fig:appendix_4_1}]{\includegraphics[width=0.48\linewidth]{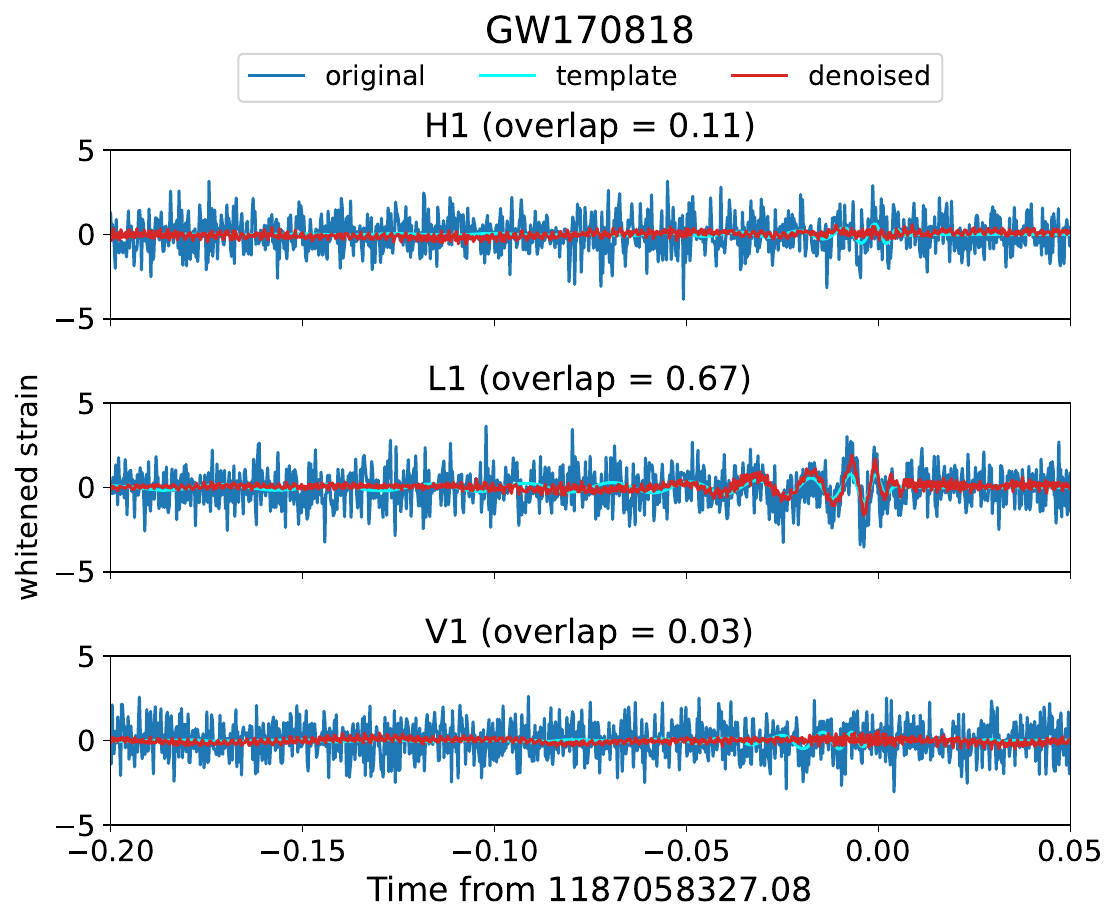}}
    \hfill
    \subfloat[\label{fig:appendix_4_2}]{\includegraphics[width=0.48\linewidth]{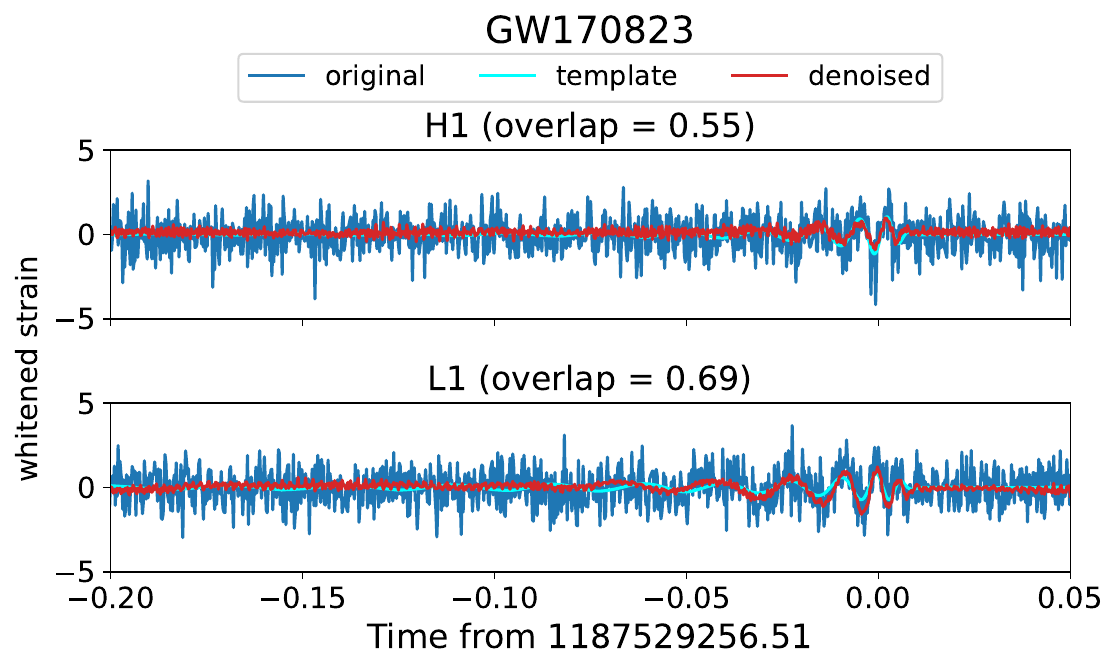}}
    \caption{\label{fig:appendix_4}The $0.25$s windows of the denoised results of the H1, L1 and V1 data around GW170818 and GW170823.}
\end{figure*}

\begin{figure*}[tbp]
    \centering
    \subfloat[\label{fig:appendix_5_1}]{\includegraphics[width=0.48\linewidth]{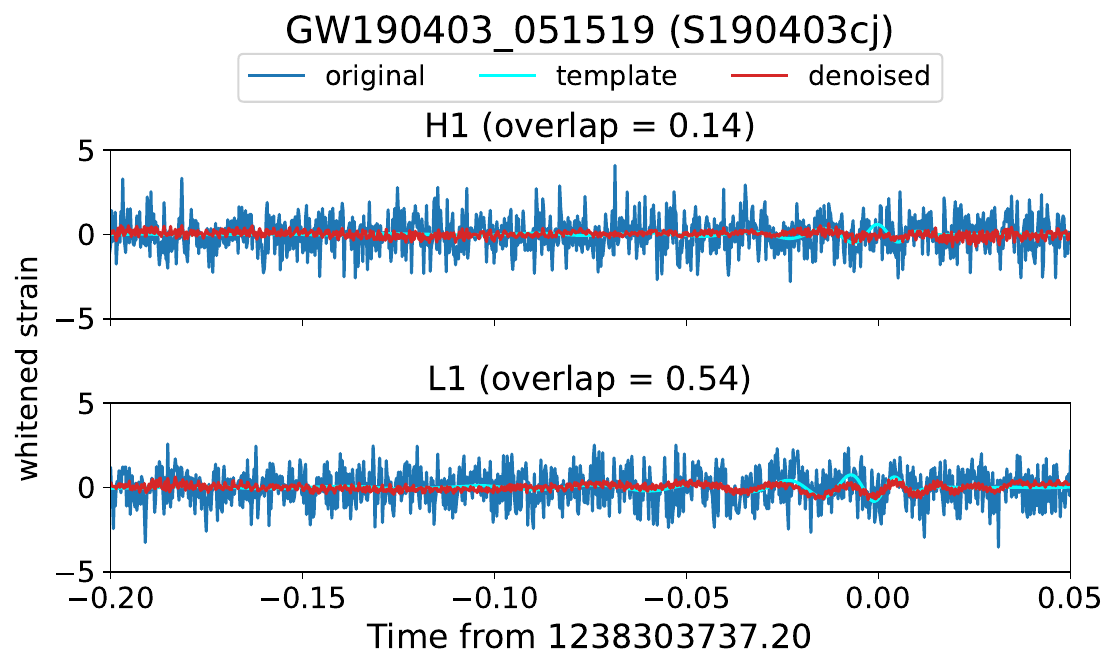}}
    \hfill
    \subfloat[\label{fig:appendix_5_2}]{\includegraphics[width=0.48\linewidth]{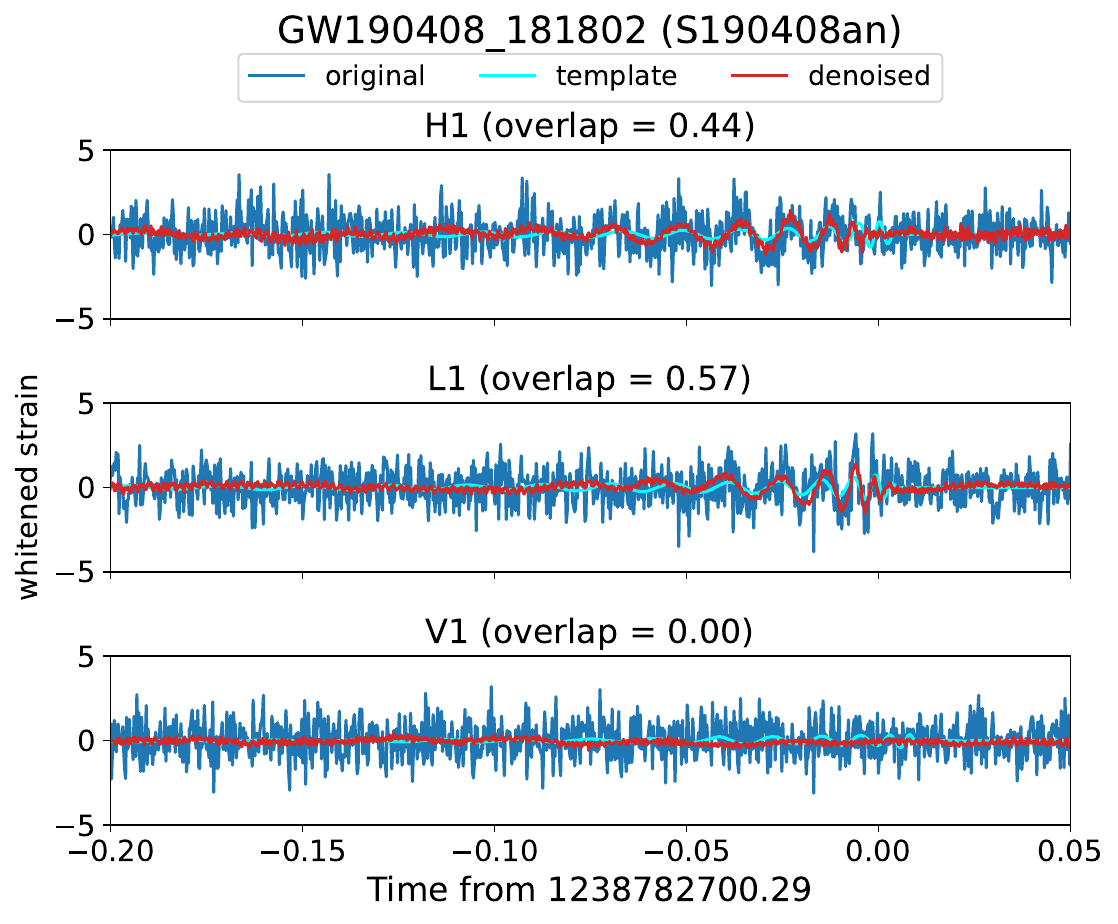}}
    \caption{\label{fig:appendix_5}The $0.25$s windows of the denoised results of the H1, L1 and V1 data around S190403cj and S190408an.}
\end{figure*}

\begin{figure*}[tbp]
    \centering
    \subfloat[\label{fig:appendix_6_1}]{\includegraphics[width=0.48\linewidth]{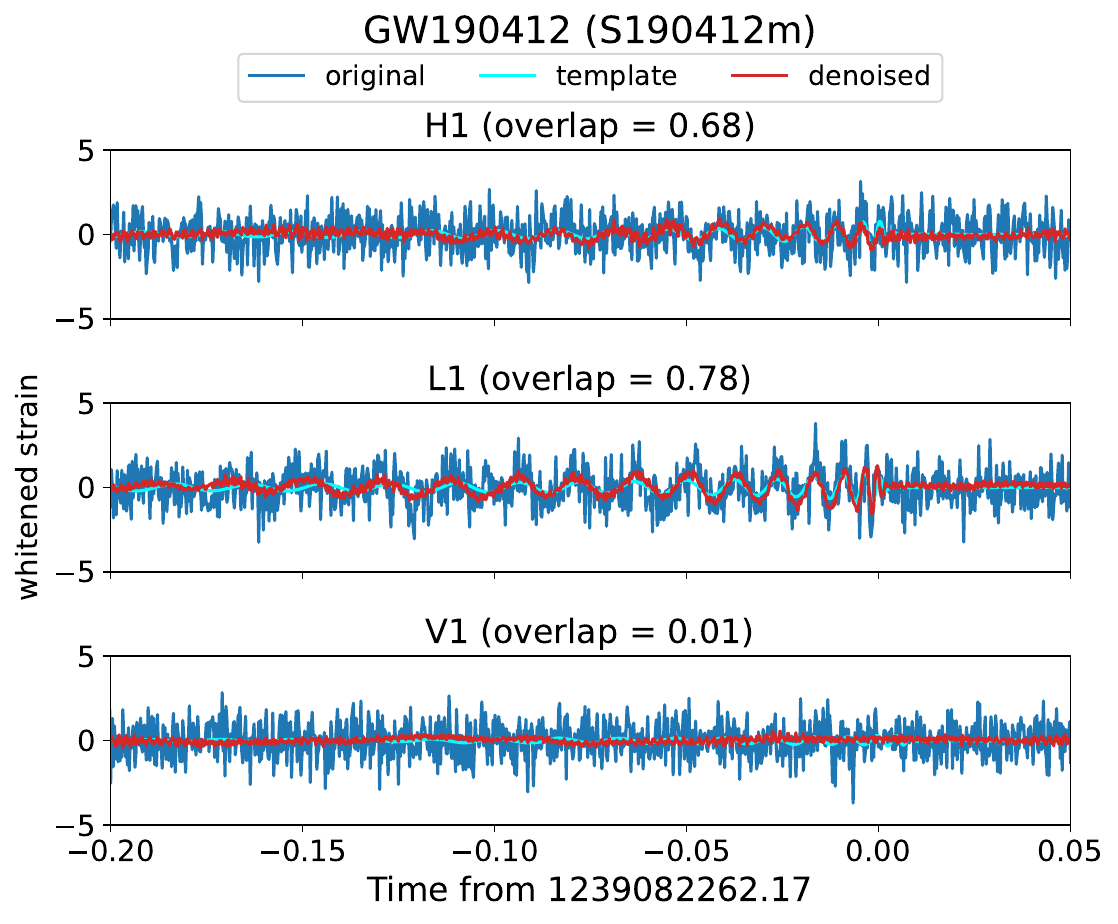}}
    \hfill
    \subfloat[\label{fig:appendix_6_2}]{\includegraphics[width=0.48\linewidth]{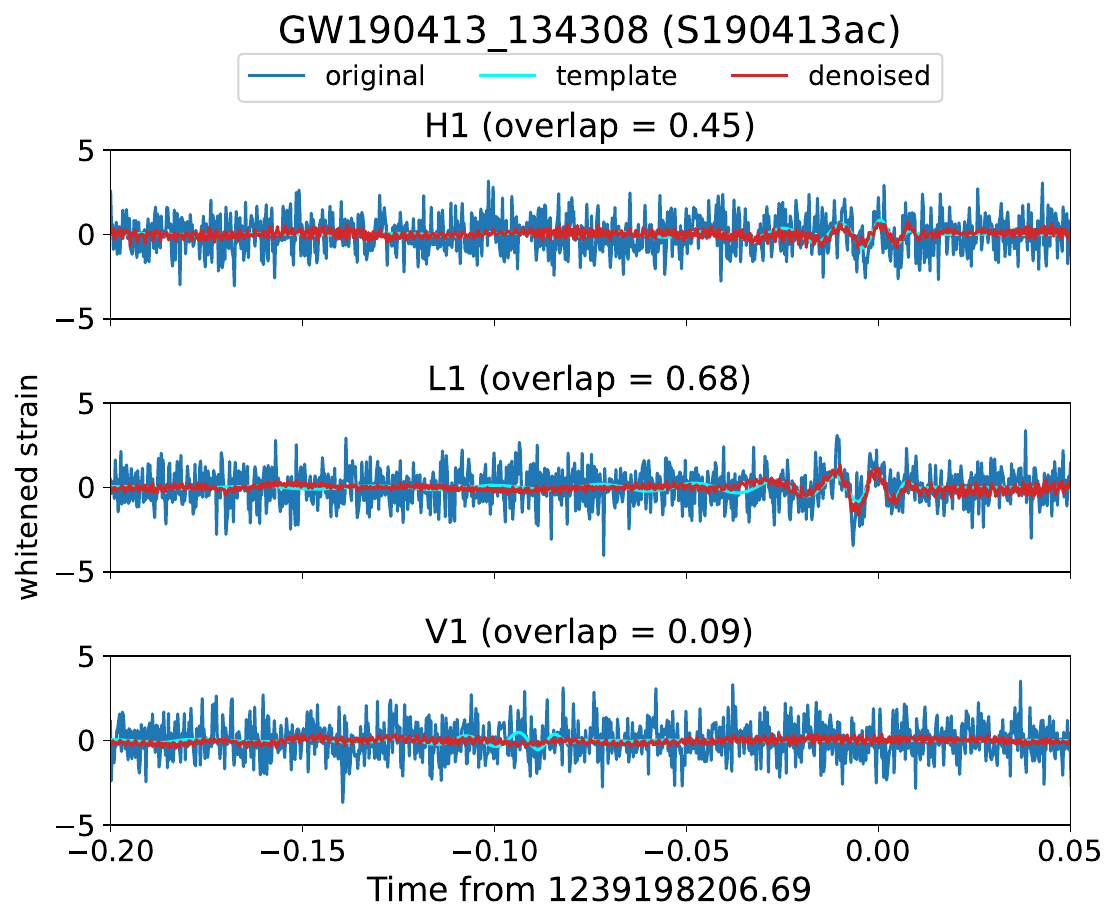}}
    \caption{\label{fig:appendix_6}The $0.25$s windows of the denoised results of the H1, L1 and V1 data around S190412m and S190413ac.}
\end{figure*}

\begin{figure*}[tbp]
    \centering
    \subfloat[\label{fig:appendix_7_1}]{\includegraphics[width=0.48\linewidth]{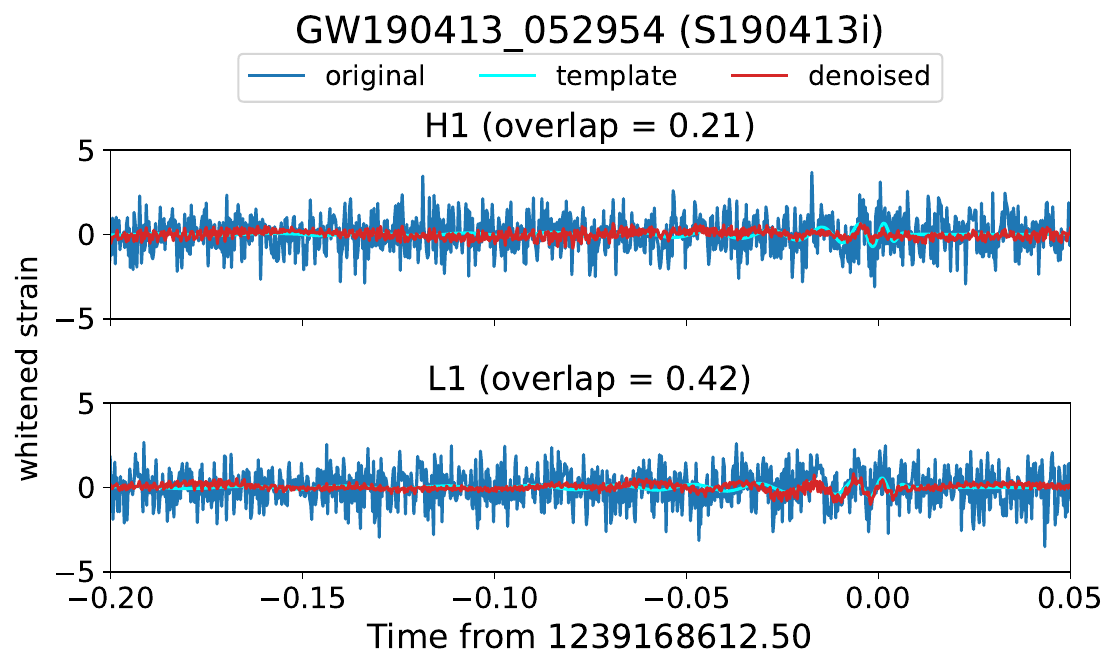}}
    \hfill
    \subfloat[\label{fig:appendix_7_2}]{\includegraphics[width=0.48\linewidth]{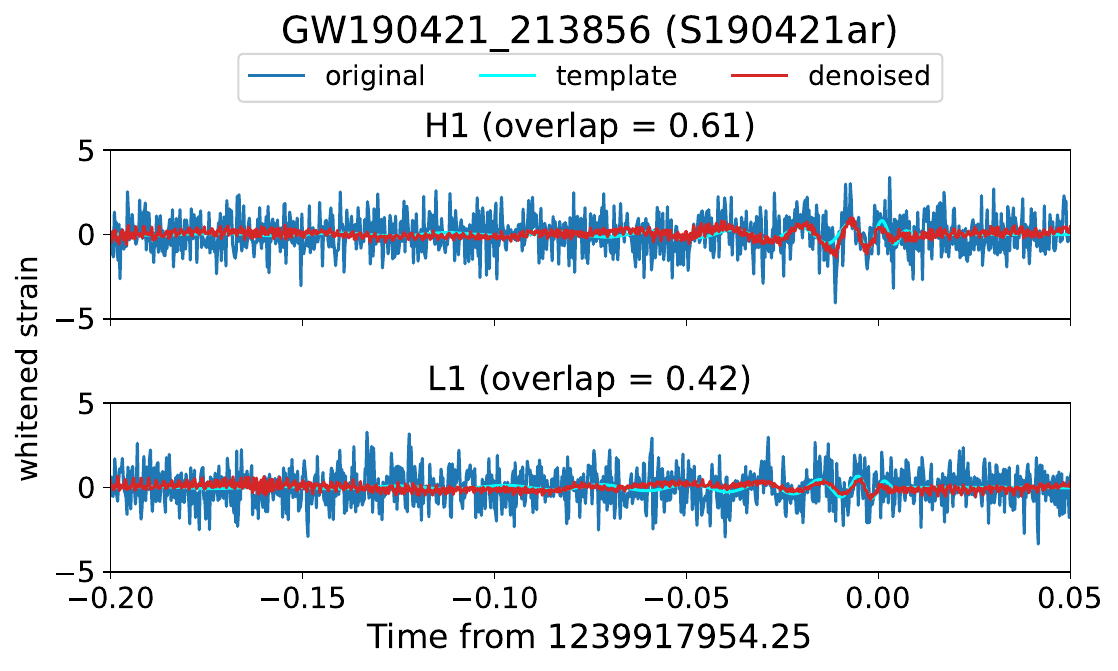}}
    \caption{\label{fig:appendix_7}The $0.25$s windows of the denoised results of the H1 and L1 data around S190413i and S190421ar.}
\end{figure*}

\begin{figure*}[tbp]
    \centering
    \subfloat[\label{fig:appendix_8_1}]{\includegraphics[width=0.48\linewidth]{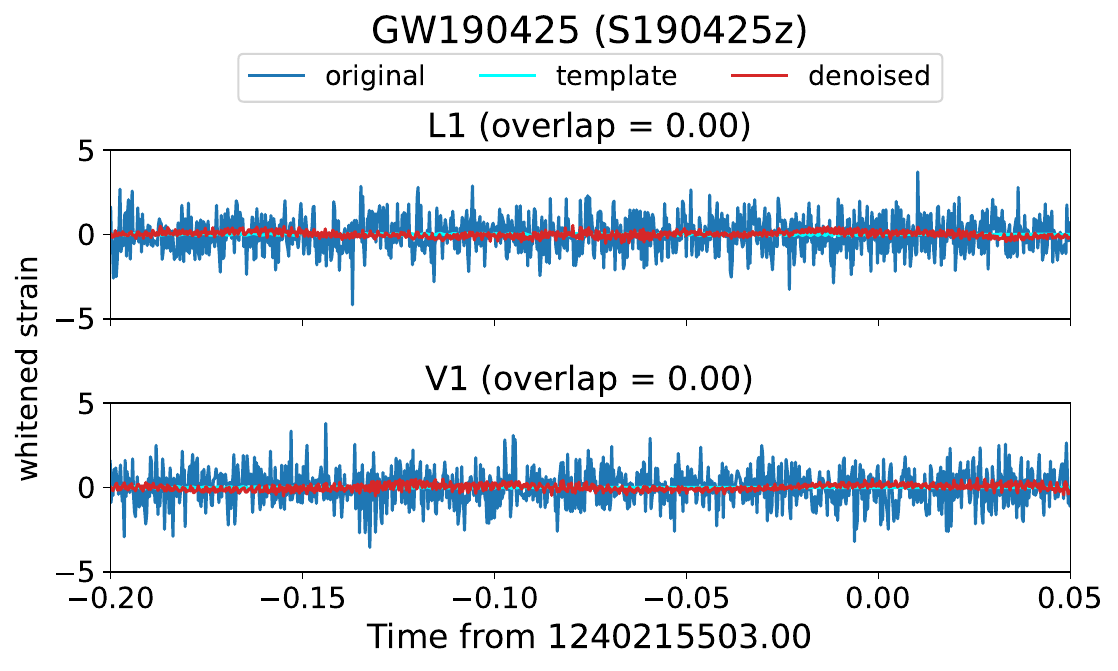}}
    \hfill
    \subfloat[\label{fig:appendix_8_2}]{\includegraphics[width=0.48\linewidth]{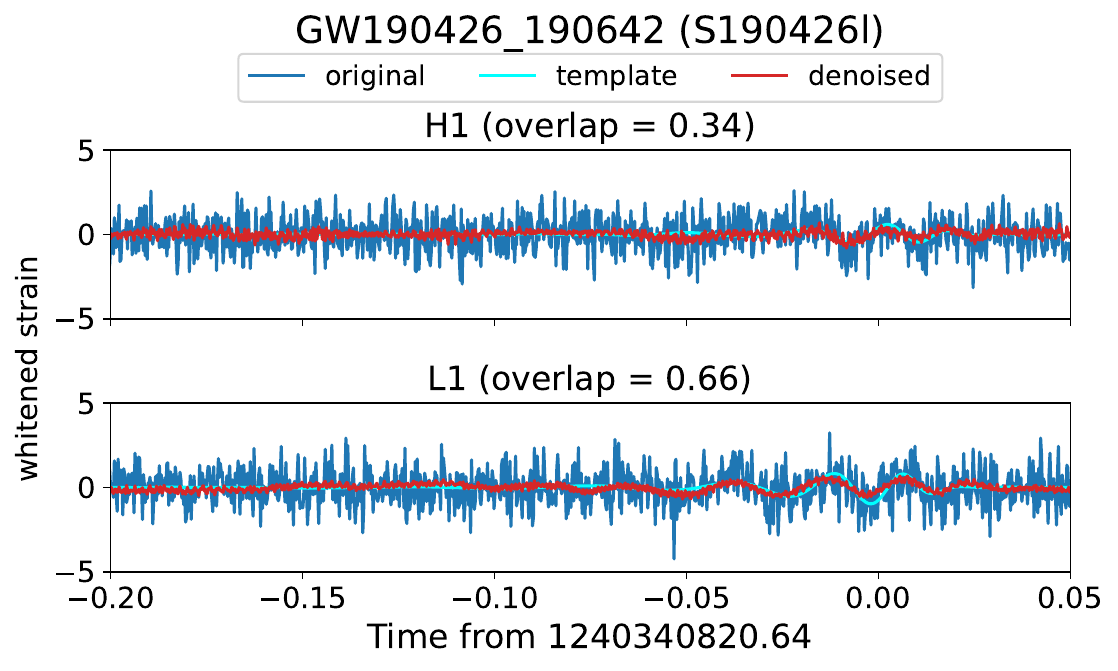}}
    \caption{\label{fig:appendix_8}The $0.25$s windows of the denoised results of the H1 and L1 data around S190425z and S190426l.}
\end{figure*}

\begin{figure*}[tbp]
    \centering
    \subfloat[\label{fig:appendix_9_1}]{\includegraphics[width=0.48\linewidth]{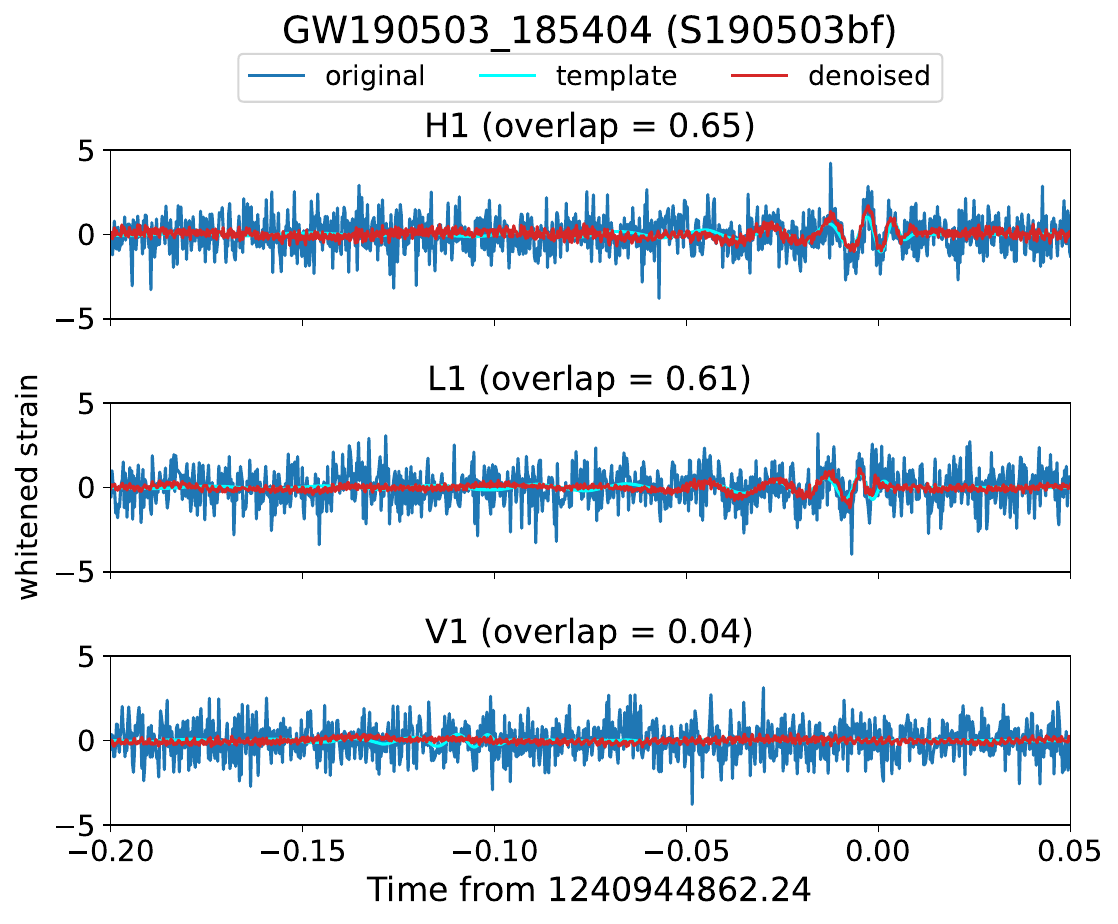}}
    \hfill
    \subfloat[\label{fig:appendix_9_2}]{\includegraphics[width=0.48\linewidth]{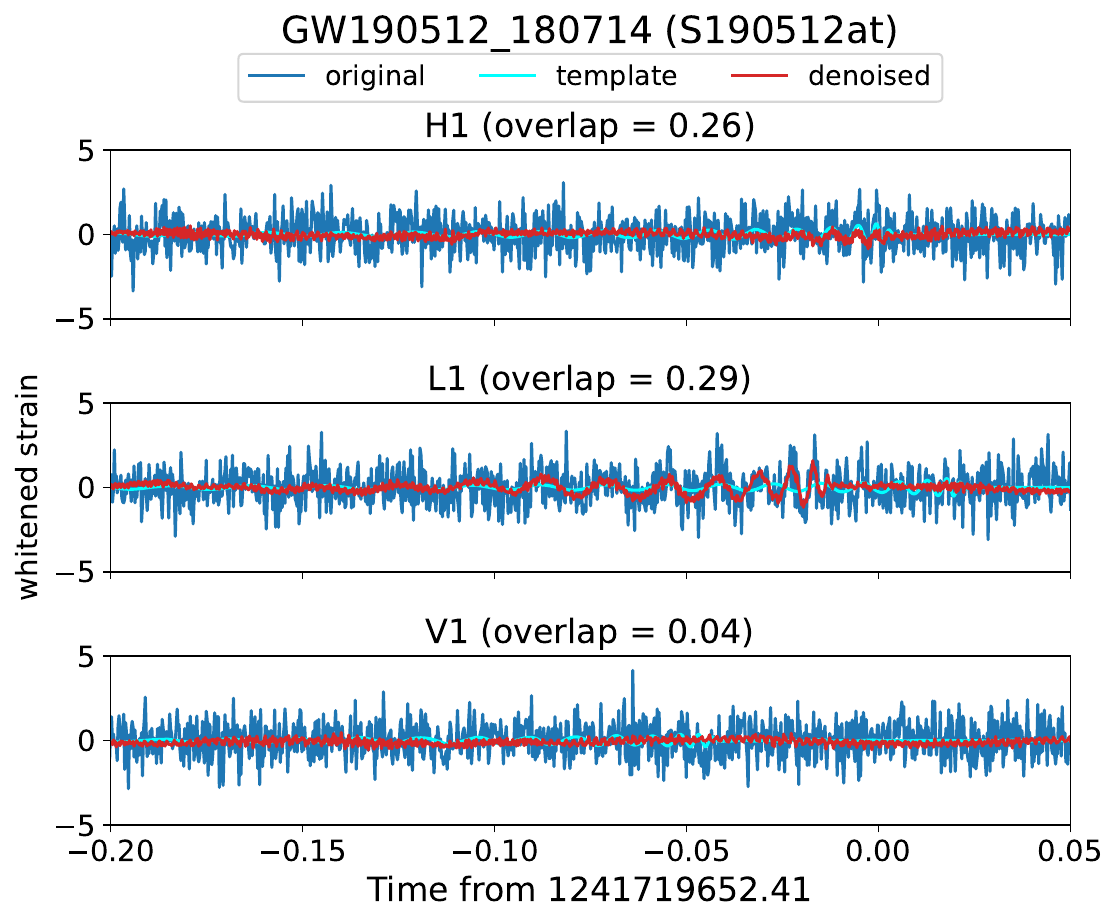}}
    \caption{\label{fig:appendix_9}The $0.25$s windows of the denoised results of the H1, L1 and V1 data around S190503bf and S190512at.}
\end{figure*}

\begin{figure*}[tbp]
    \centering
    \subfloat[\label{fig:appendix_10_1}]{\includegraphics[width=0.48\linewidth]{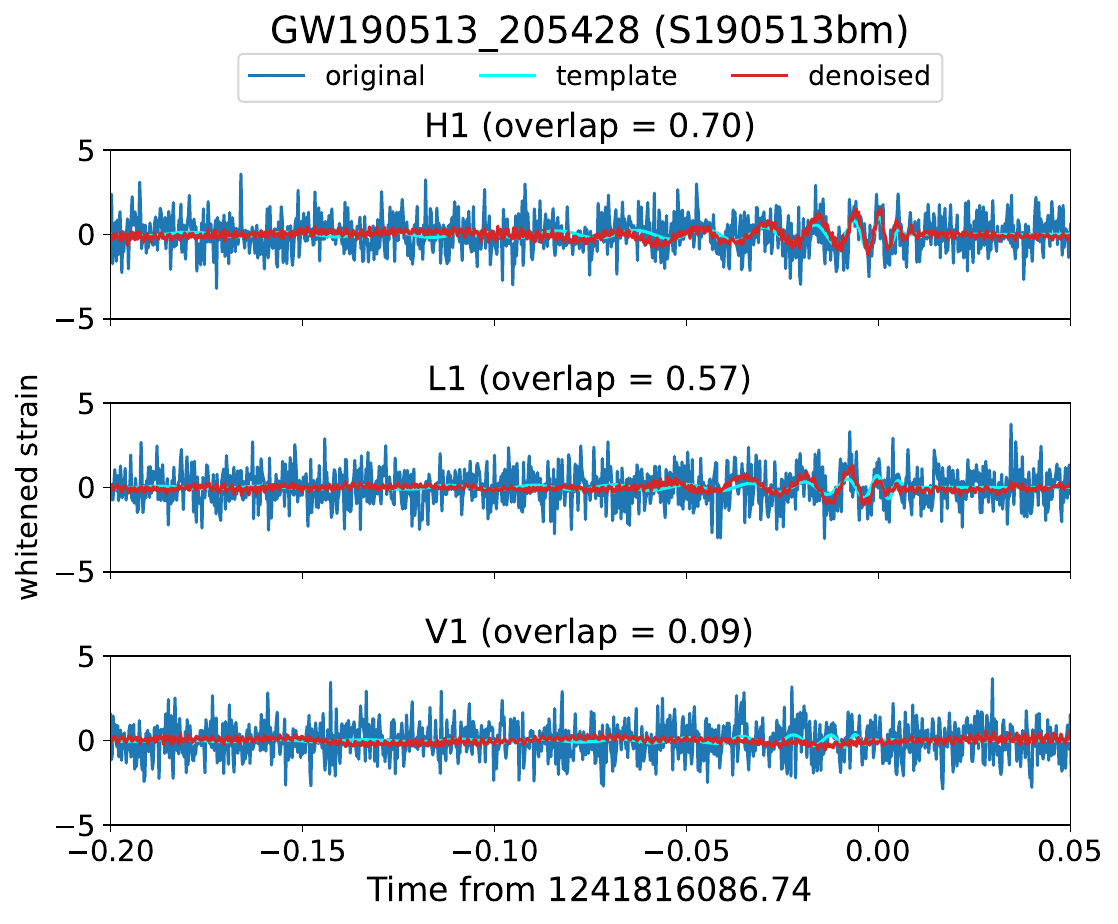}}
    \hfill
    \subfloat[\label{fig:appendix_10_2}]{\includegraphics[width=0.48\linewidth]{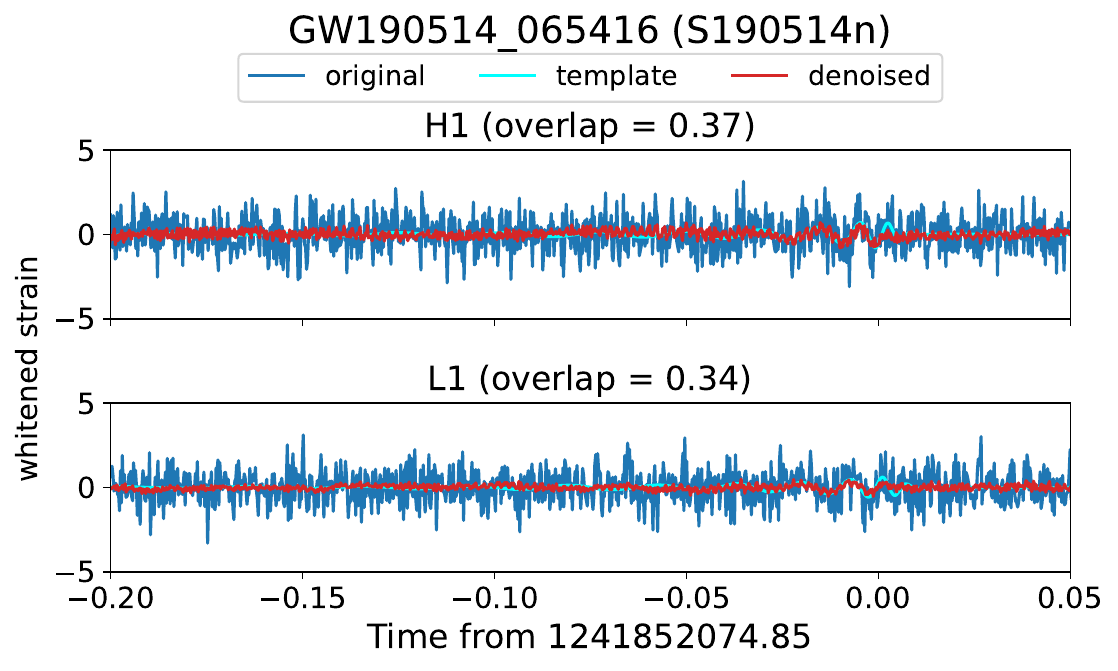}}
    \caption{\label{fig:appendix_10}The $0.25$s windows of the denoised results of the H1, L1 and V1 data around S190513bm and S190514n.}
\end{figure*}

\begin{figure*}[tbp]
    \centering
    \subfloat[\label{fig:appendix_11_1}]{\includegraphics[width=0.48\linewidth]{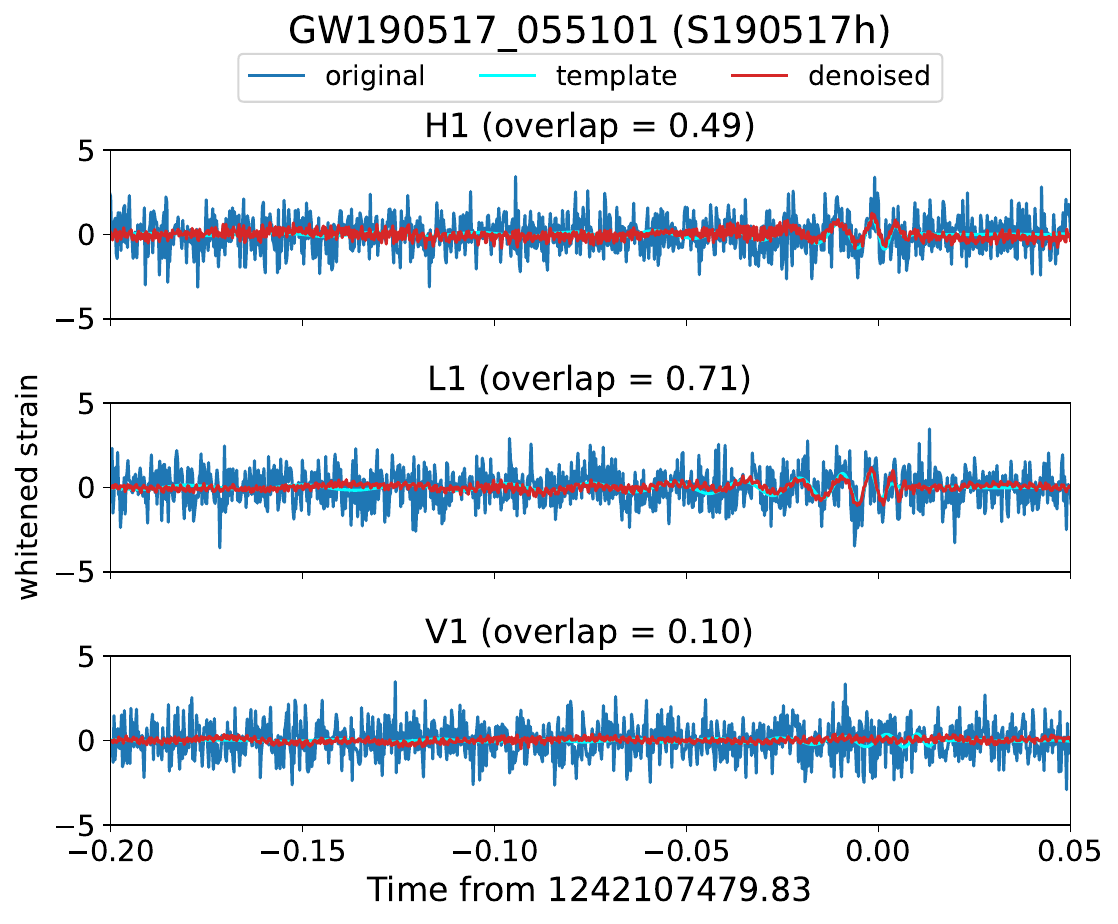}}
    \hfill
    \subfloat[\label{fig:appendix_11_2}]{\includegraphics[width=0.48\linewidth]{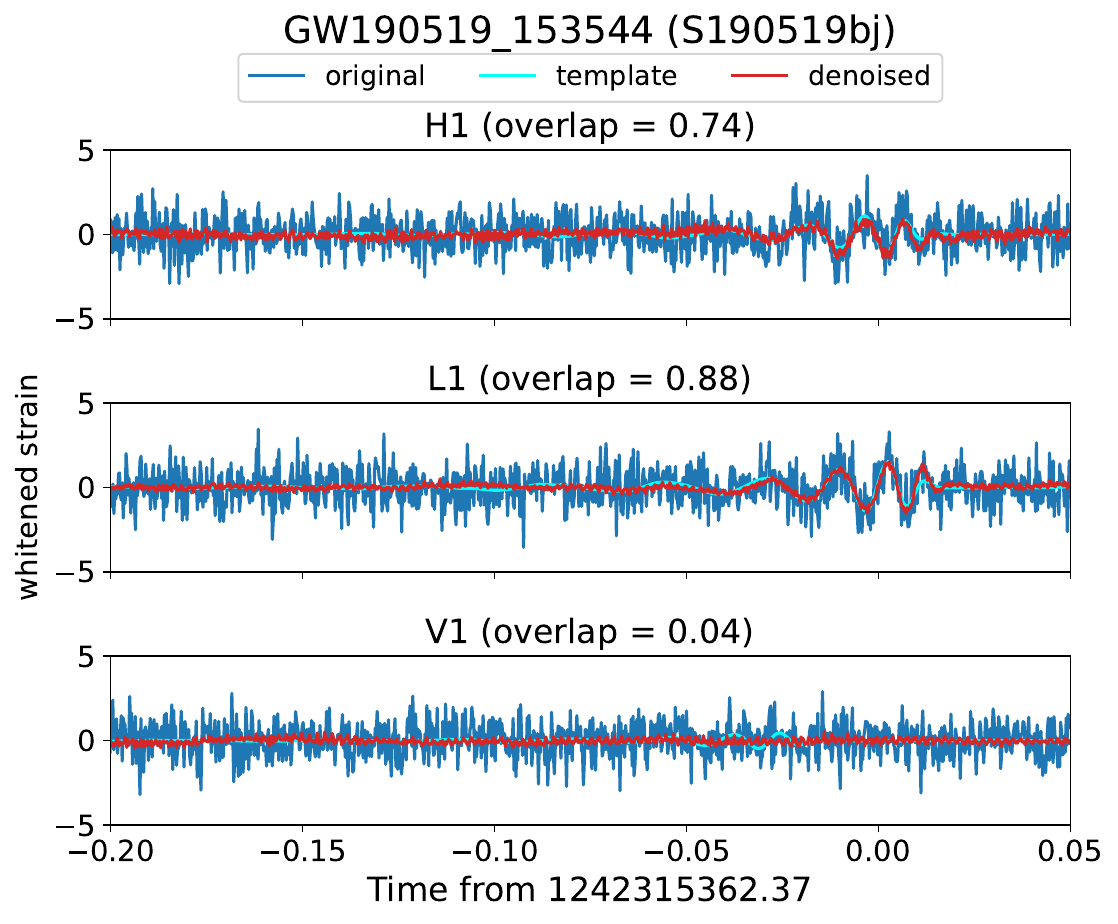}}
    \caption{\label{fig:appendix_11}The $0.25$s windows of the denoised results of the H1, L1 and V1 data around S190517h and S190519bj.}
\end{figure*}

\begin{figure*}[tbp]
    \centering
    \subfloat[\label{fig:appendix_12_1}]{\includegraphics[width=0.48\linewidth]{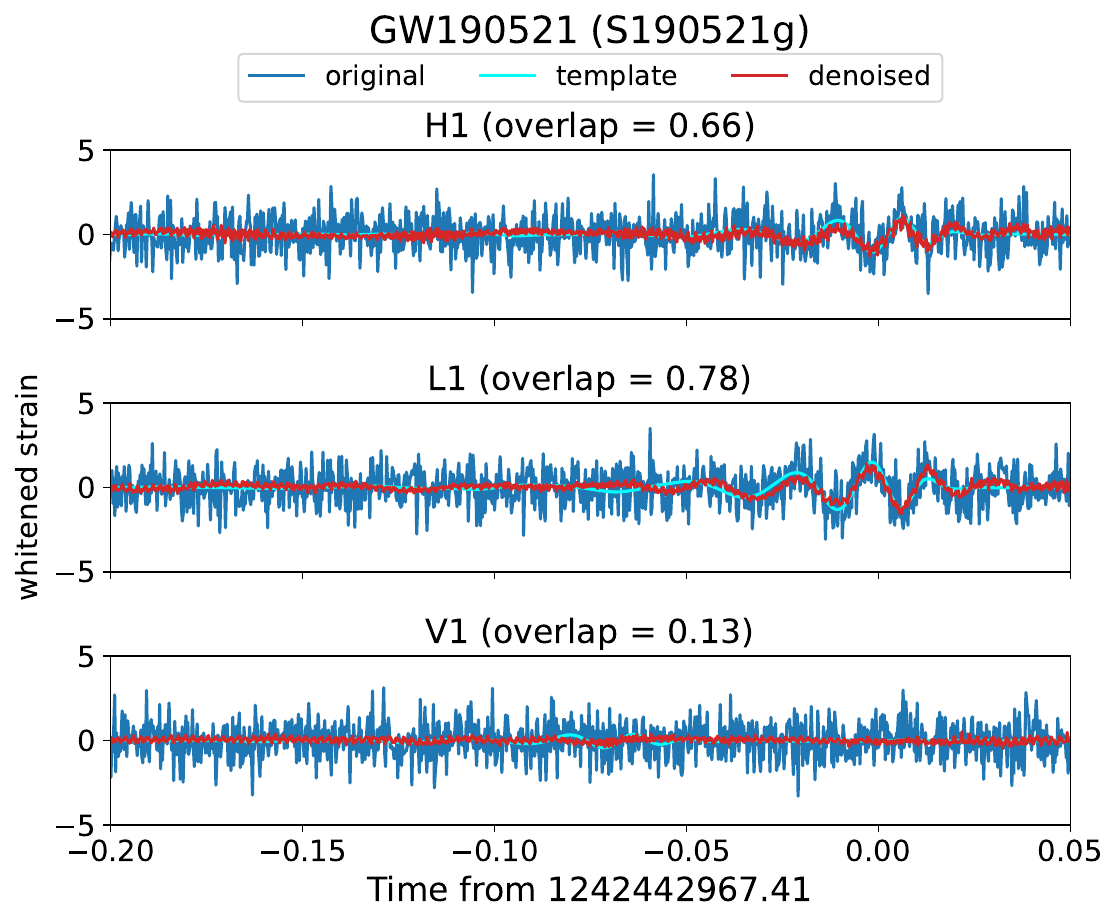}}
    \hfill
    \subfloat[\label{fig:appendix_12_2}]{\includegraphics[width=0.48\linewidth]{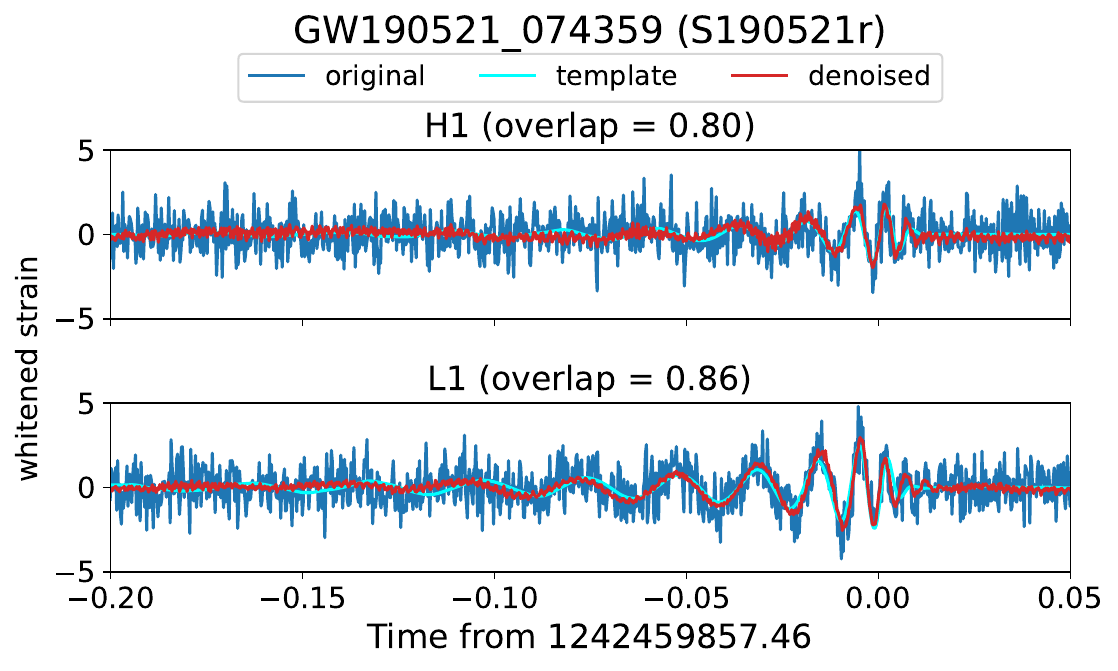}}
    \caption{\label{fig:appendix_12}The $0.25$s windows of the denoised results of the H1, L1 and V1 data around S190521g and S190521r.}
\end{figure*}

\begin{figure*}[tbp]
    \centering
    \subfloat[\label{fig:appendix_13_1}]{\includegraphics[width=0.48\linewidth]{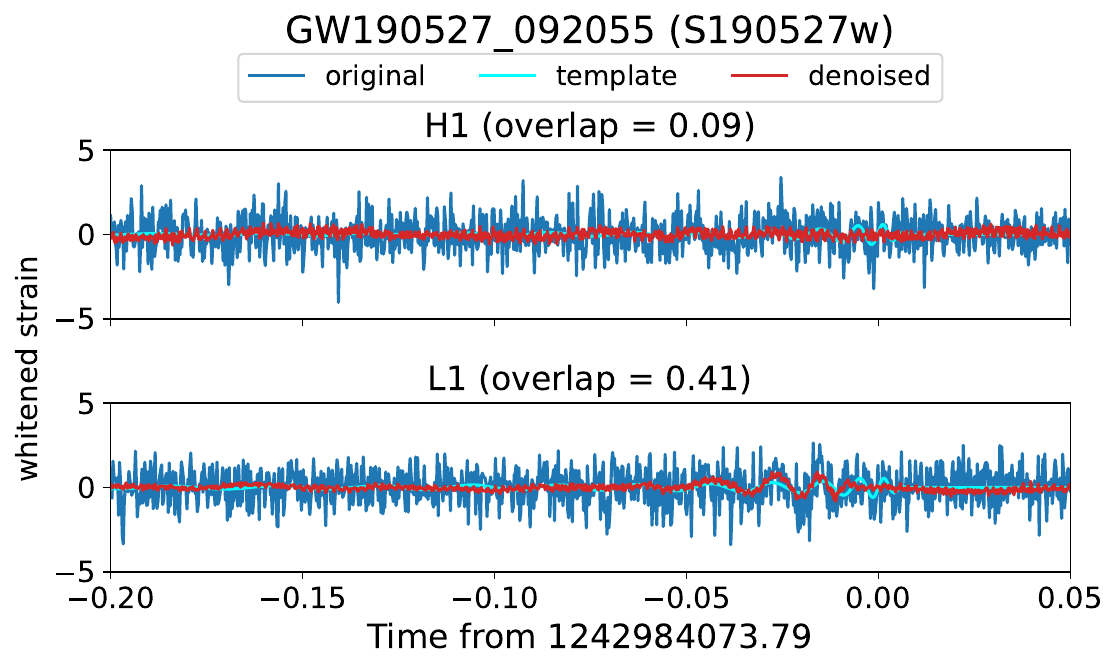}}
    \hfill
    \subfloat[\label{fig:appendix_13_2}]{\includegraphics[width=0.48\linewidth]{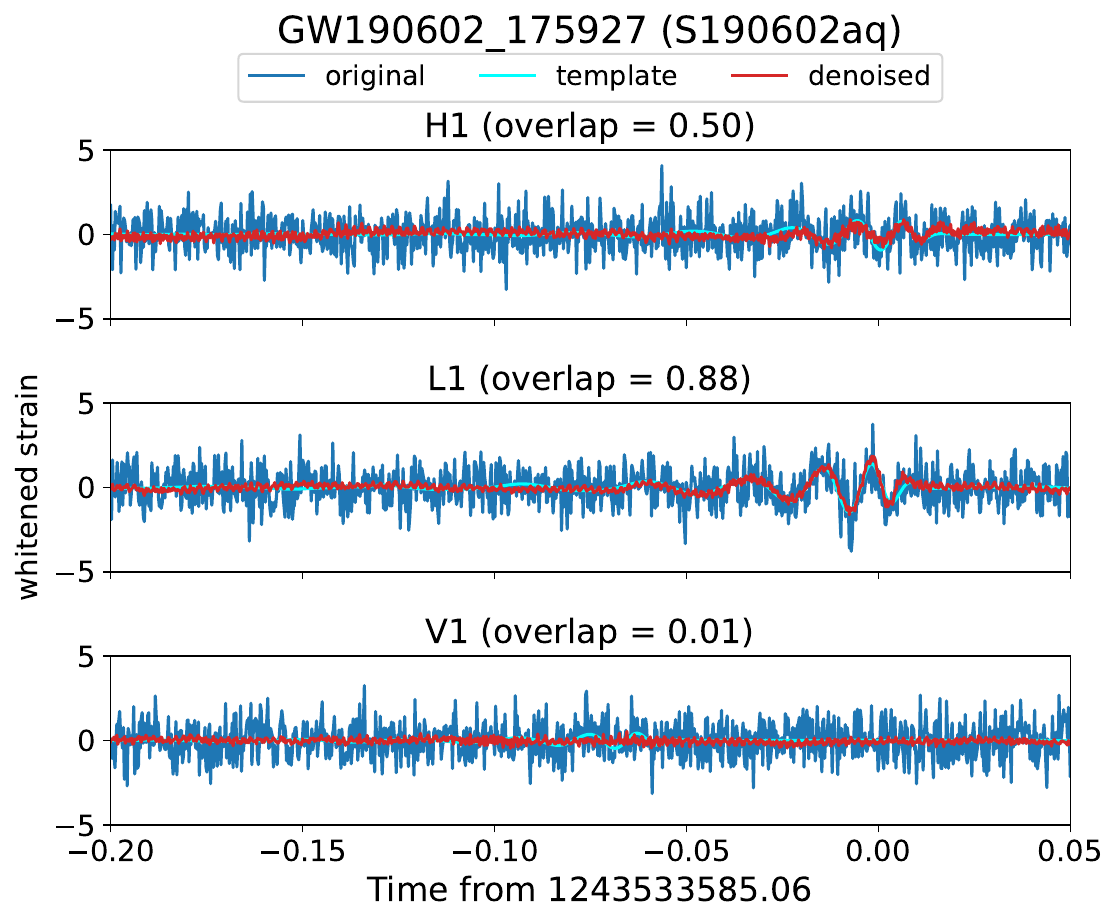}}
    \caption{\label{fig:appendix_13}The $0.25$s windows of the denoised results of the H1, L1 and V1 data around S190527w and S190602aq.}
\end{figure*}

\begin{figure*}[tbp]
    \centering
    \subfloat[\label{fig:appendix_14_1}]{\includegraphics[width=0.48\linewidth]{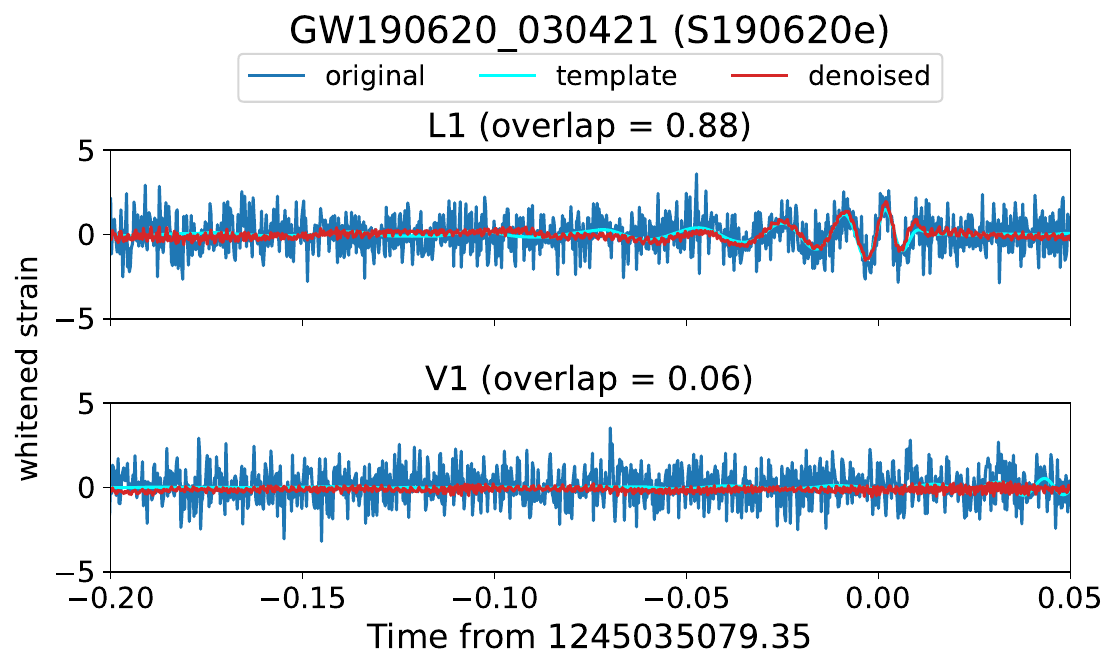}}
    \hfill
    \subfloat[\label{fig:appendix_14_2}]{\includegraphics[width=0.48\linewidth]{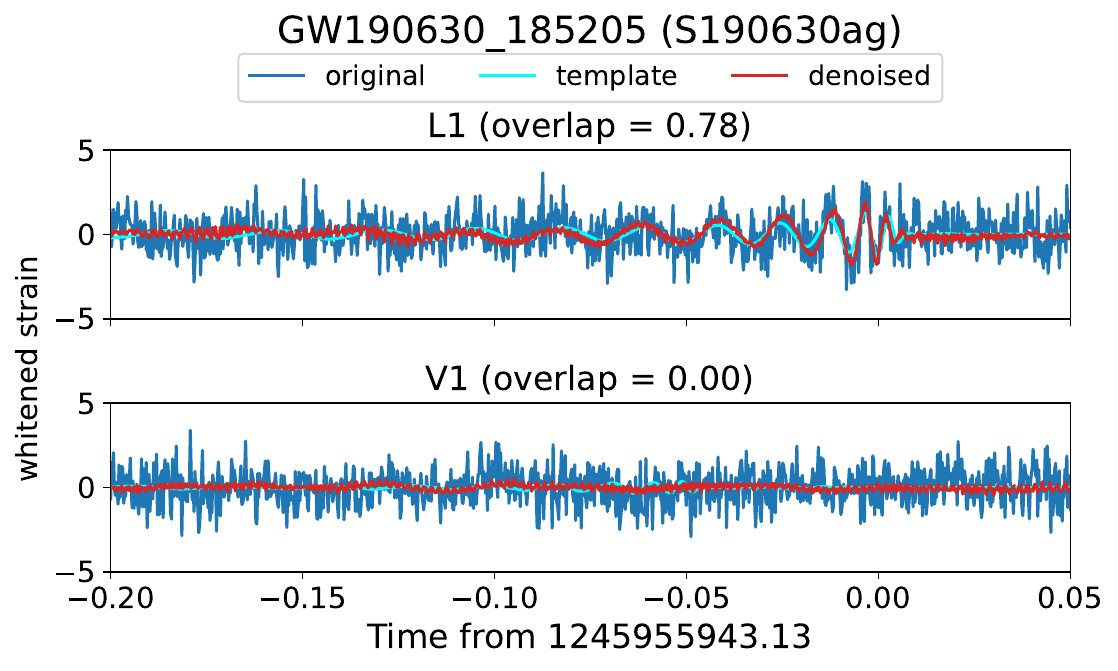}}
    \caption{\label{fig:appendix_14}The $0.25$s windows of the denoised results of the H1, L1 and V1 data around S190620e and S190630ag.}
\end{figure*}

\begin{figure*}[tbp]
    \centering
    \subfloat[\label{fig:appendix_15_1}]{\includegraphics[width=0.48\linewidth]{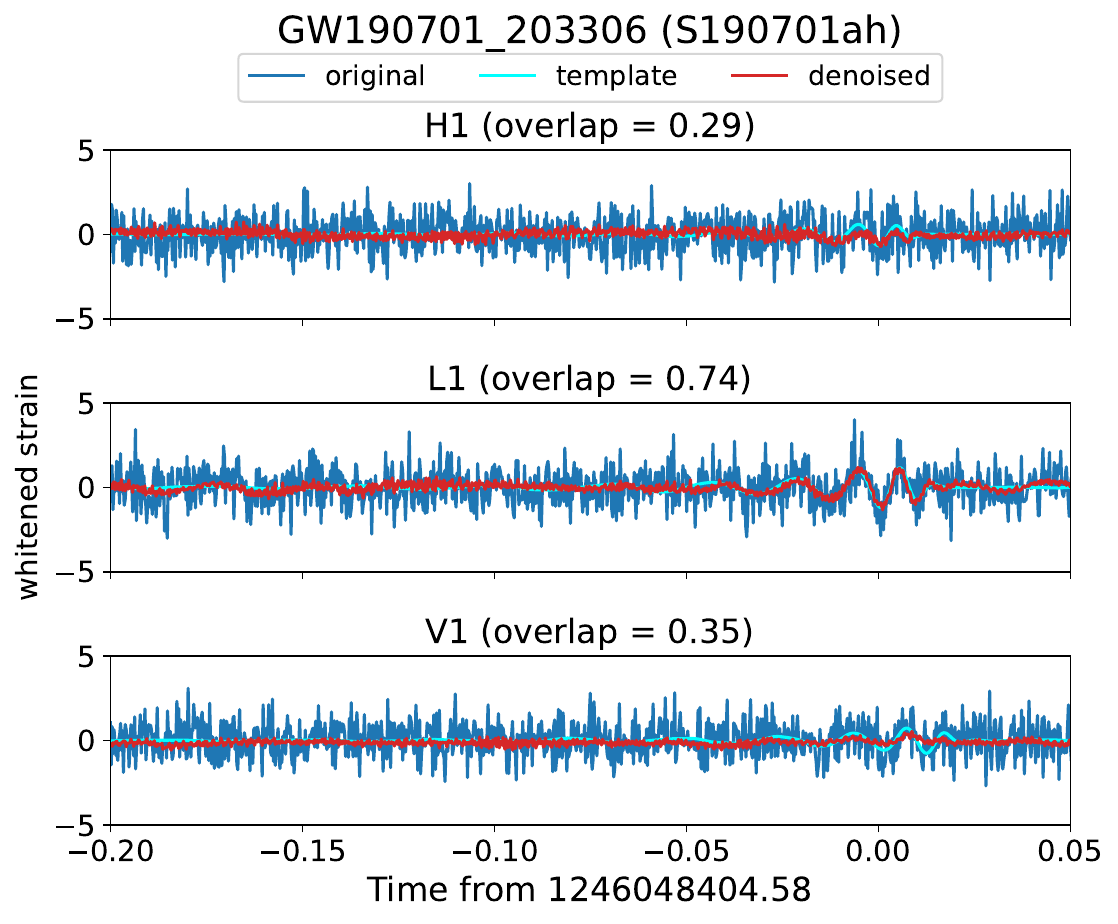}}
    \hfill
    \subfloat[\label{fig:appendix_15_2}]{\includegraphics[width=0.48\linewidth]{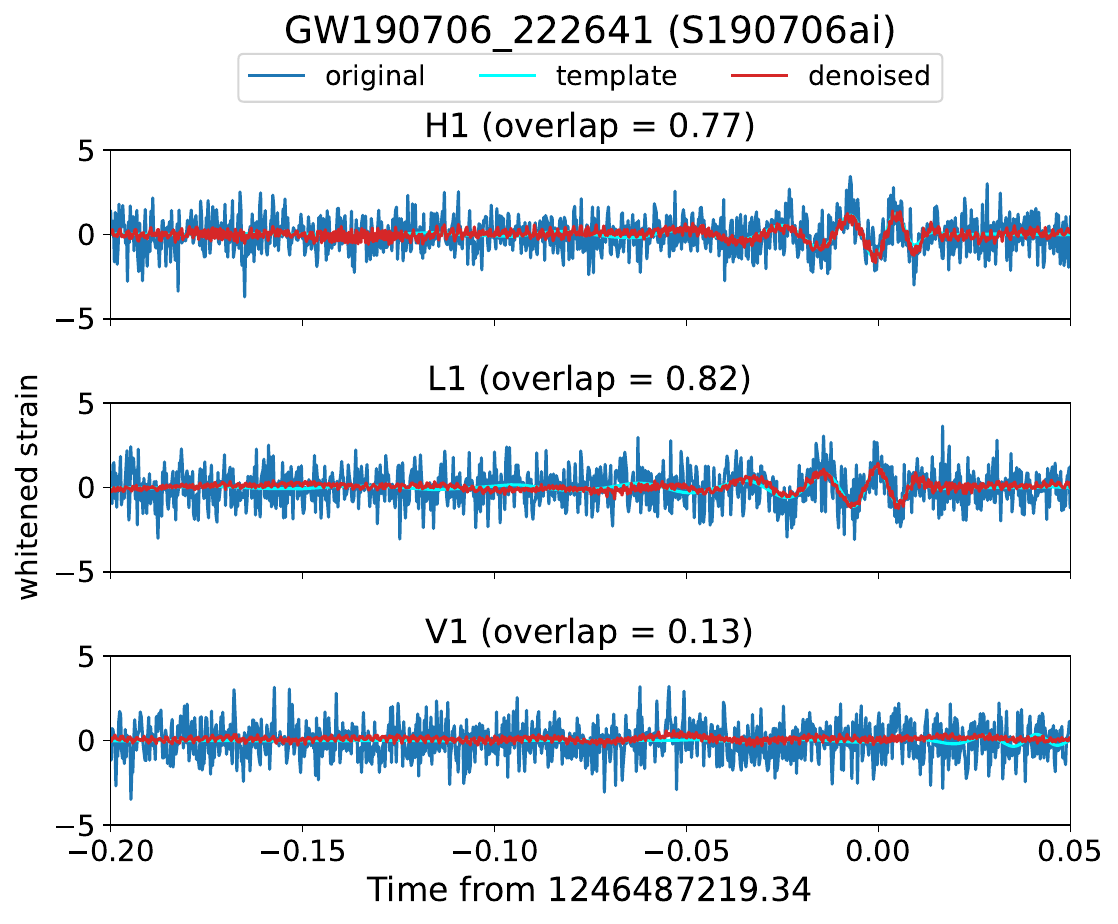}}
    \caption{\label{fig:appendix_15}The $0.25$s windows of the denoised results of the H1, L1 and V1 data around S190701ah and S190706ai.}
\end{figure*}

\begin{figure*}[tbp]
    \centering
    \subfloat[\label{fig:appendix_16_1}]{\includegraphics[width=0.48\linewidth]{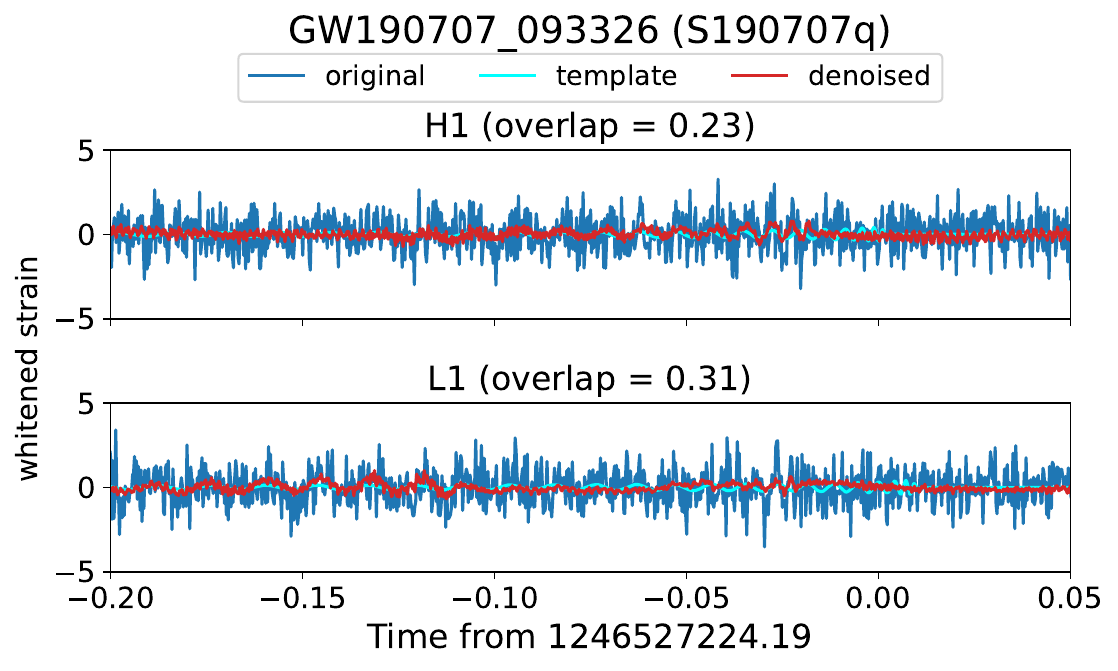}}
    \hfill
    \subfloat[\label{fig:appendix_16_2}]{\includegraphics[width=0.48\linewidth]{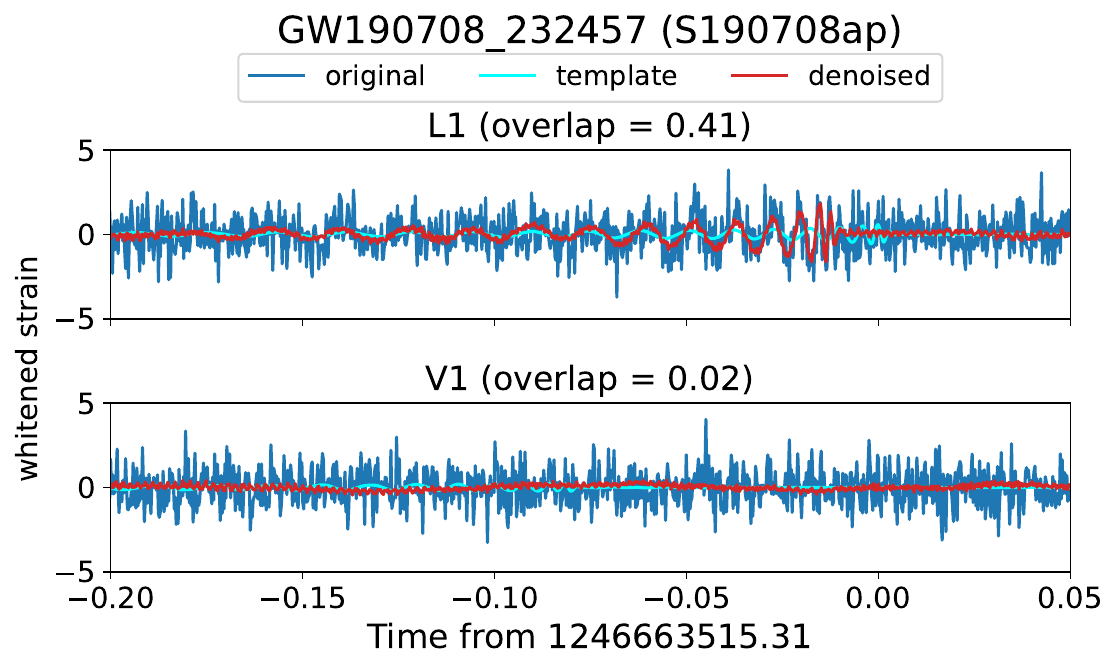}}
    \caption{\label{fig:appendix_16}The $0.25$s windows of the denoised results of the H1, L1 and V1 data around S190707q and S190708ap.}
\end{figure*}

\begin{figure*}[tbp]
    \centering
    \subfloat[\label{fig:appendix_17_1}]{\includegraphics[width=0.48\linewidth]{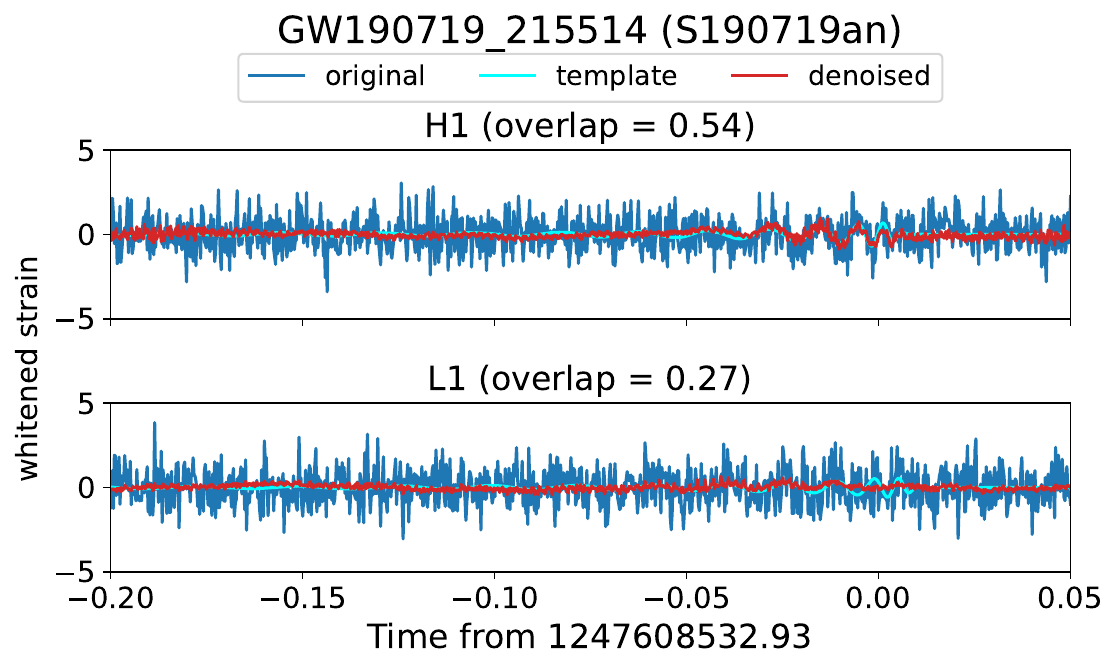}}
    \hfill
    \subfloat[\label{fig:appendix_17_2}]{\includegraphics[width=0.48\linewidth]{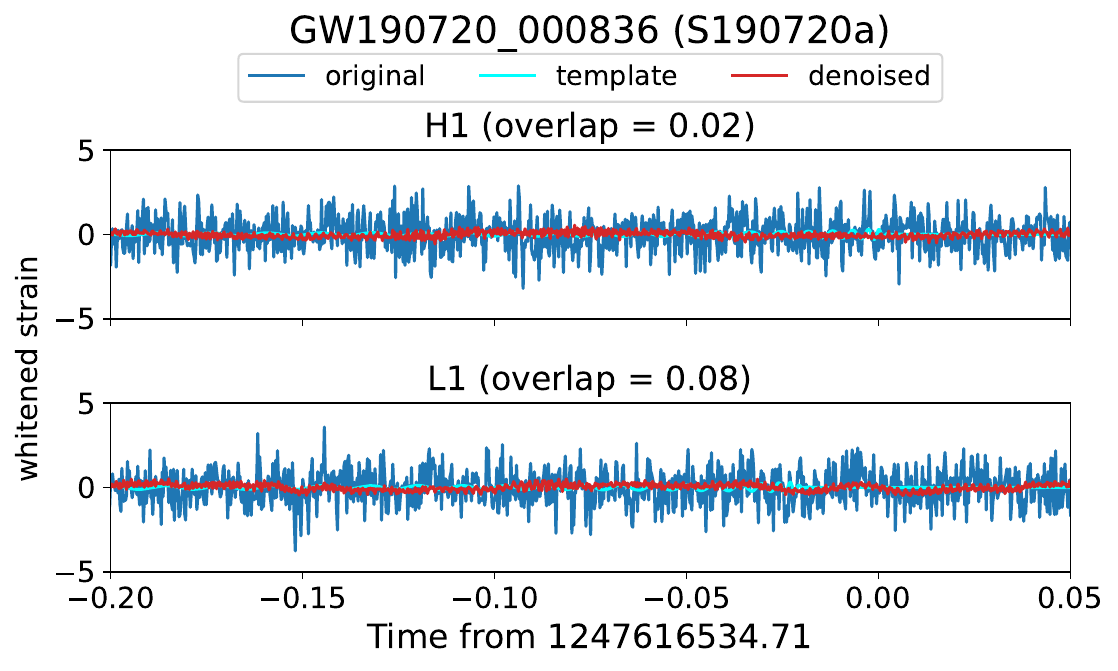}}
    \caption{\label{fig:appendix_17}The $0.25$s windows of the denoised results of the H1, L1 and V1 data around S190719an and S190720a.}
\end{figure*}

\begin{figure*}[tbp]
    \centering
    \subfloat[\label{fig:appendix_18_1}]{\includegraphics[width=0.48\linewidth]{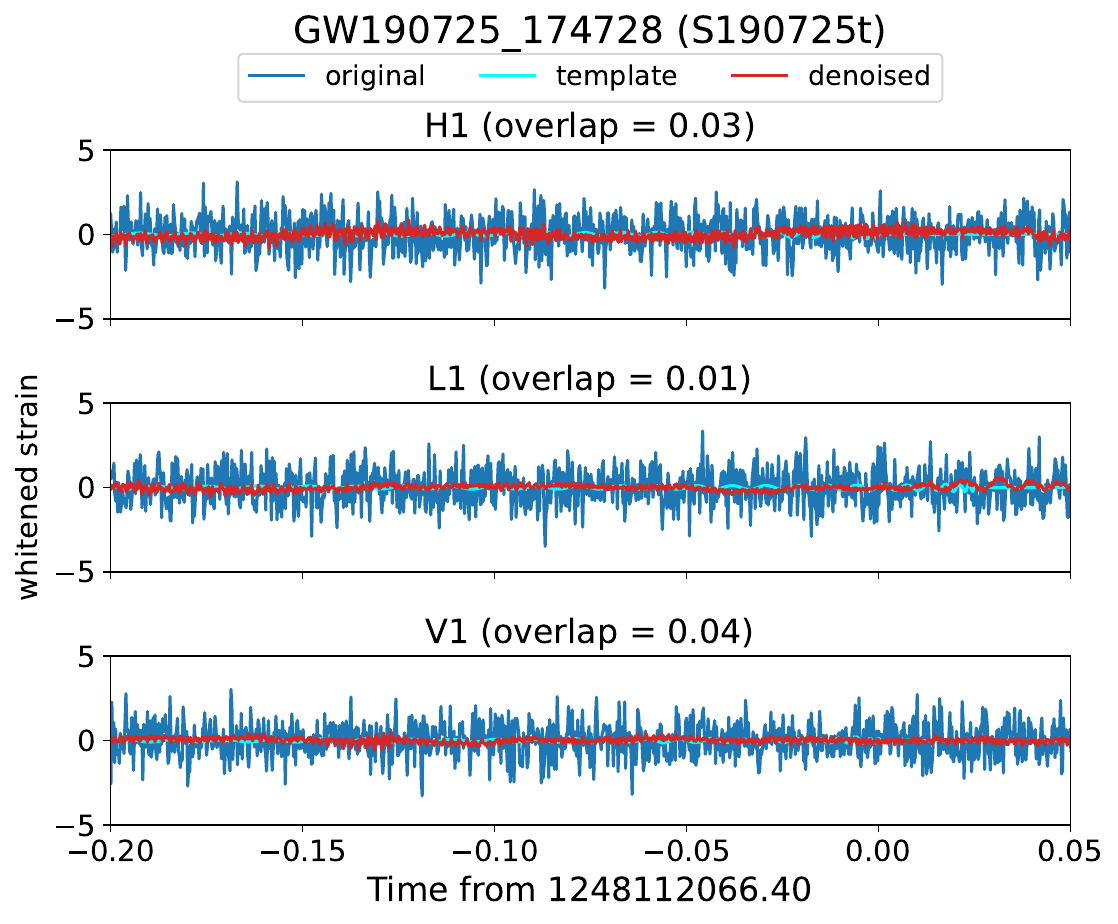}}
    \hfill
    \subfloat[\label{fig:appendix_18_2}]{\includegraphics[width=0.48\linewidth]{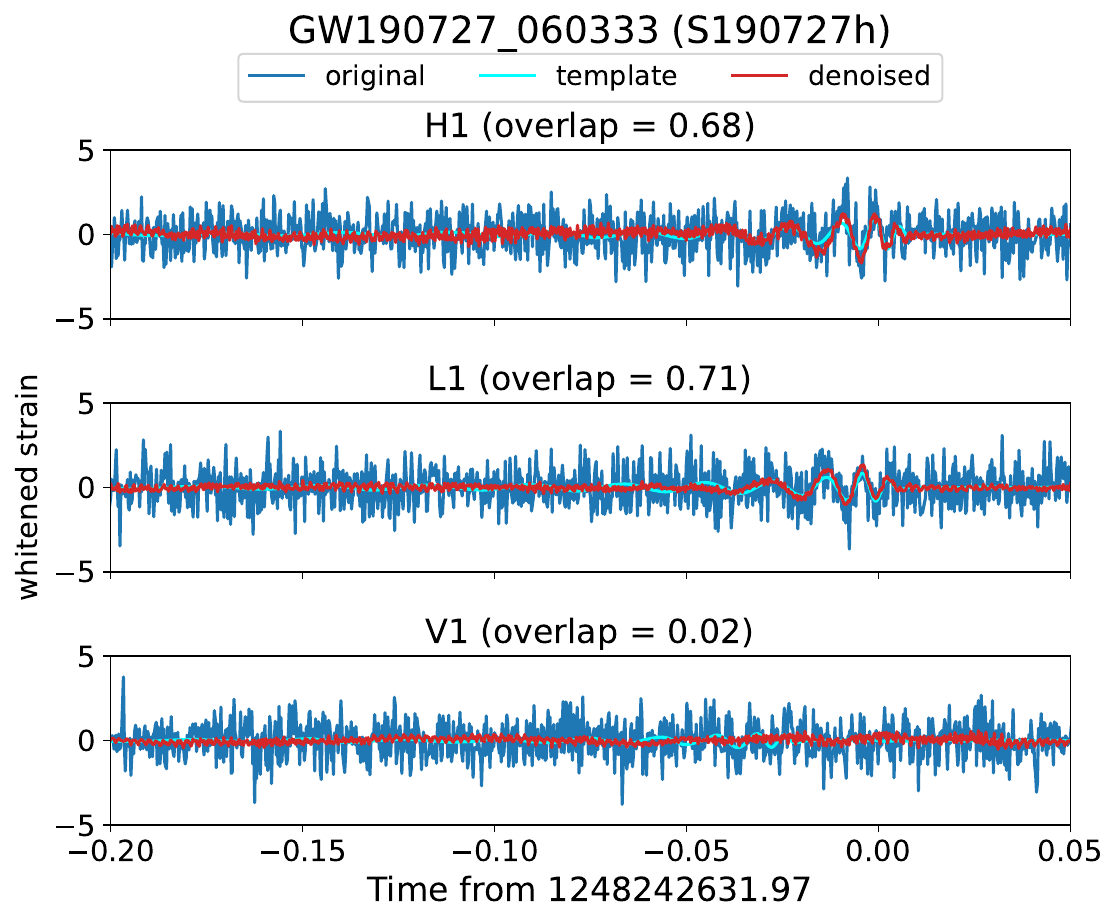}}
    \caption{\label{fig:appendix_18}The $0.25$s windows of the denoised results of the H1, L1 and V1 data around S190725t and S190727h.}
\end{figure*}

\begin{figure*}[tbp]
    \centering
    \subfloat[\label{fig:appendix_19_1}]{\includegraphics[width=0.48\linewidth]{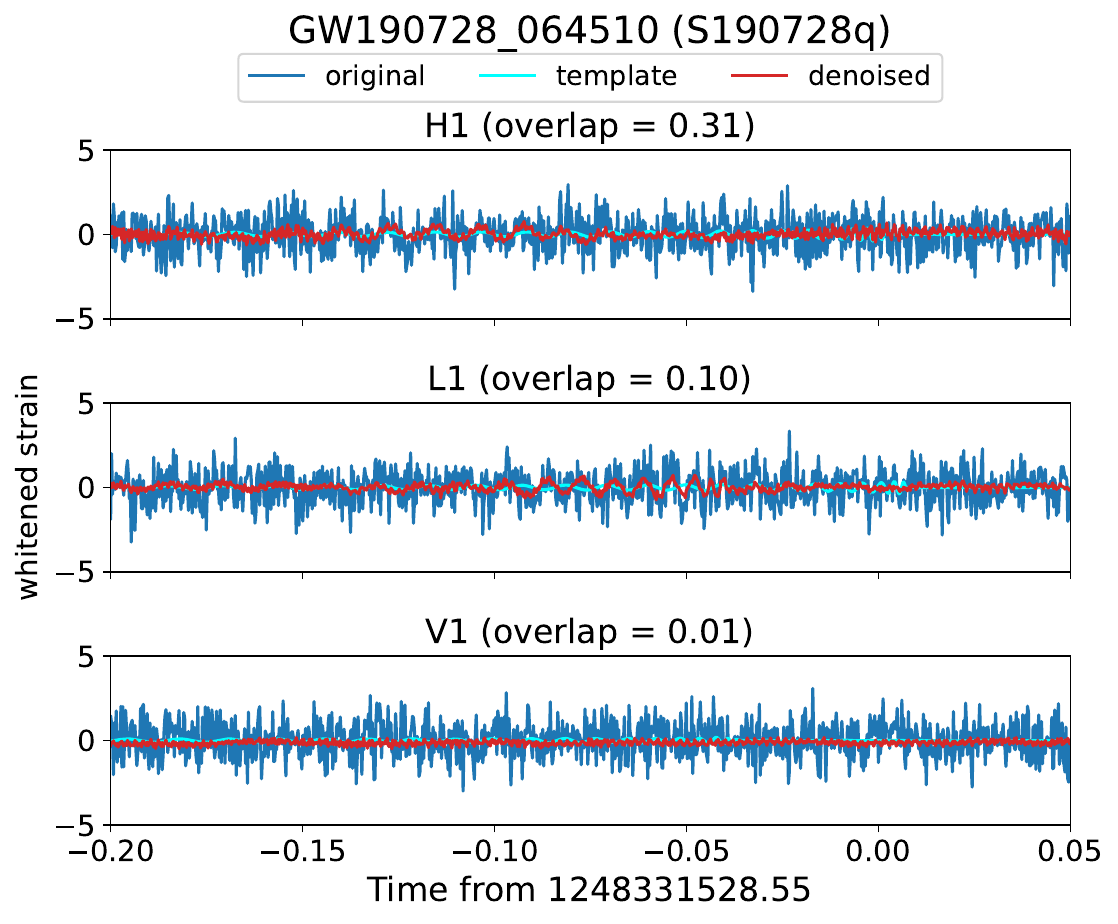}}
    \hfill
    \subfloat[\label{fig:appendix_19_2}]{\includegraphics[width=0.48\linewidth]{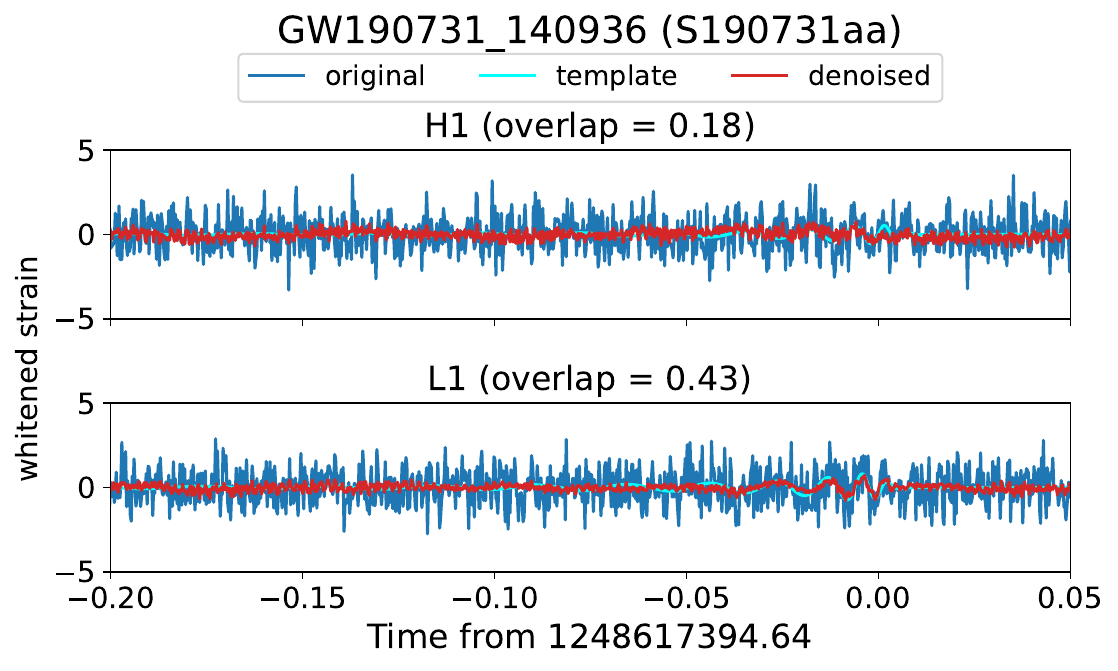}}
    \caption{\label{fig:appendix_19}The $0.25$s windows of the denoised results of the H1, L1 and V1 data around S190728q and S190731aa.}
\end{figure*}

\begin{figure*}[tbp]
    \centering
    \subfloat[\label{fig:appendix_20_1}]{\includegraphics[width=0.48\linewidth]{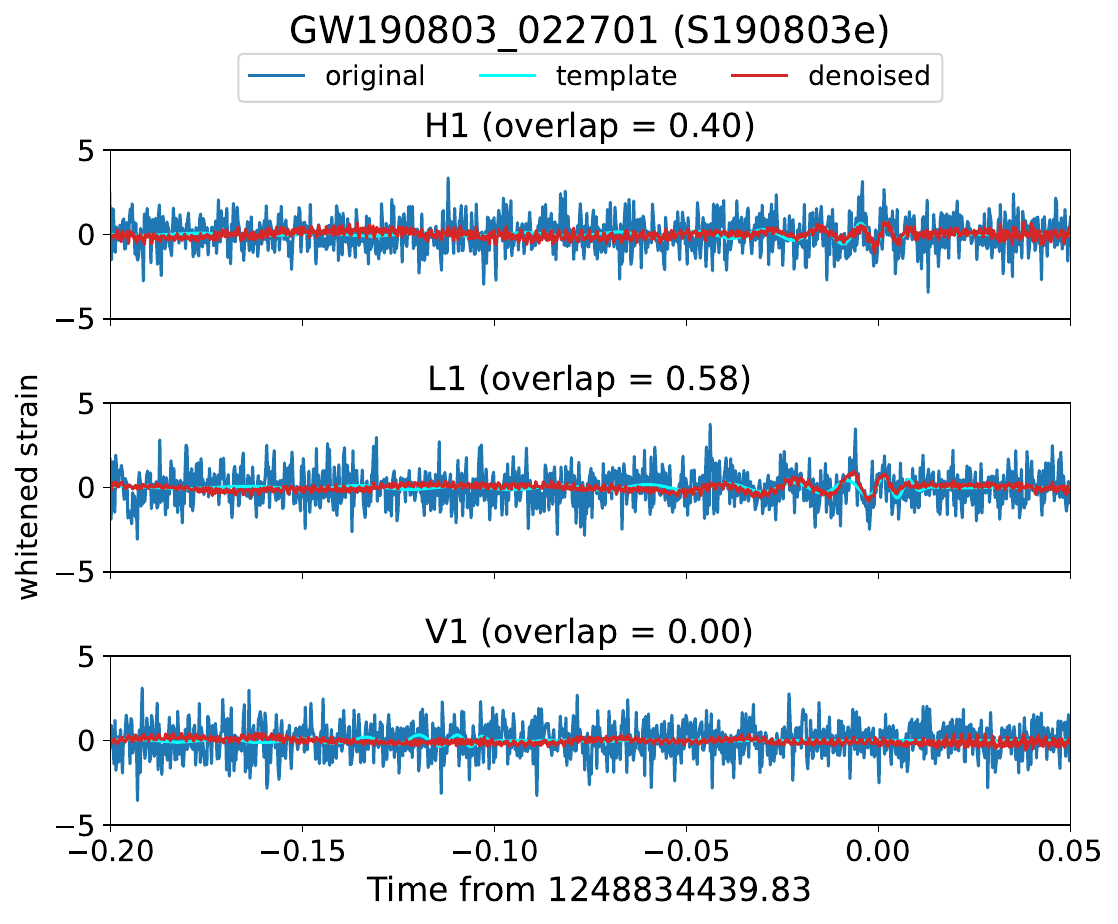}}
    \hfill
    \subfloat[\label{fig:appendix_20_2}]{\includegraphics[width=0.48\linewidth]{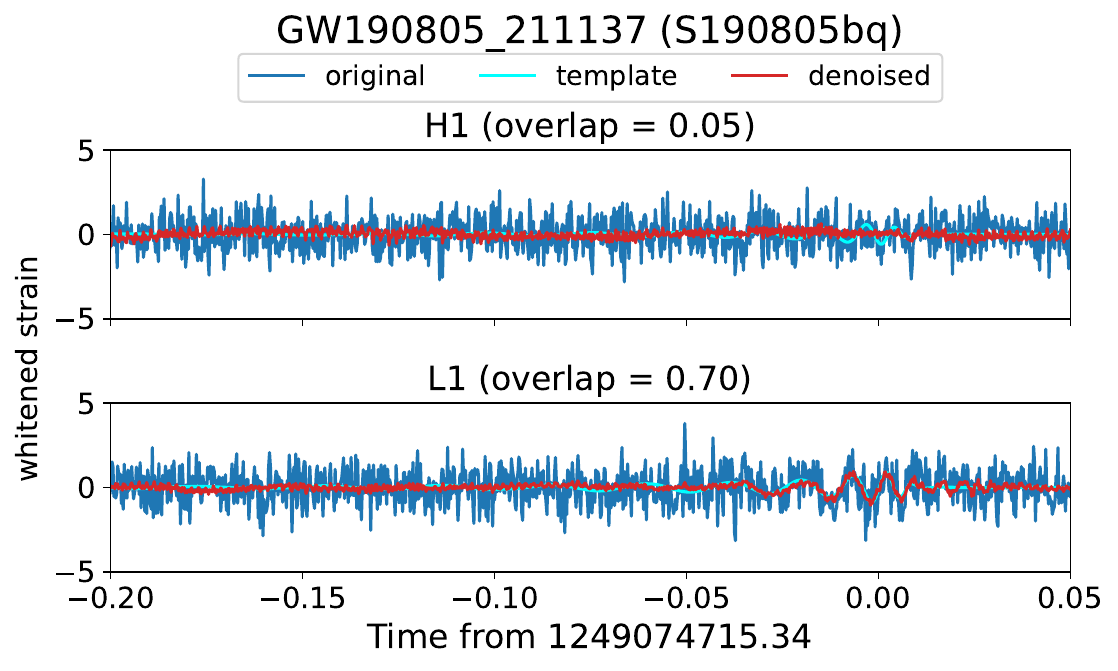}}
    \caption{\label{fig:appendix_20}The $0.25$s windows of the denoised results of the H1, L1 and V1 data around S190803e and S190805bq.}
\end{figure*}

\begin{figure*}[tbp]
    \centering
    \subfloat[\label{fig:appendix_21_1}]{\includegraphics[width=0.48\linewidth]{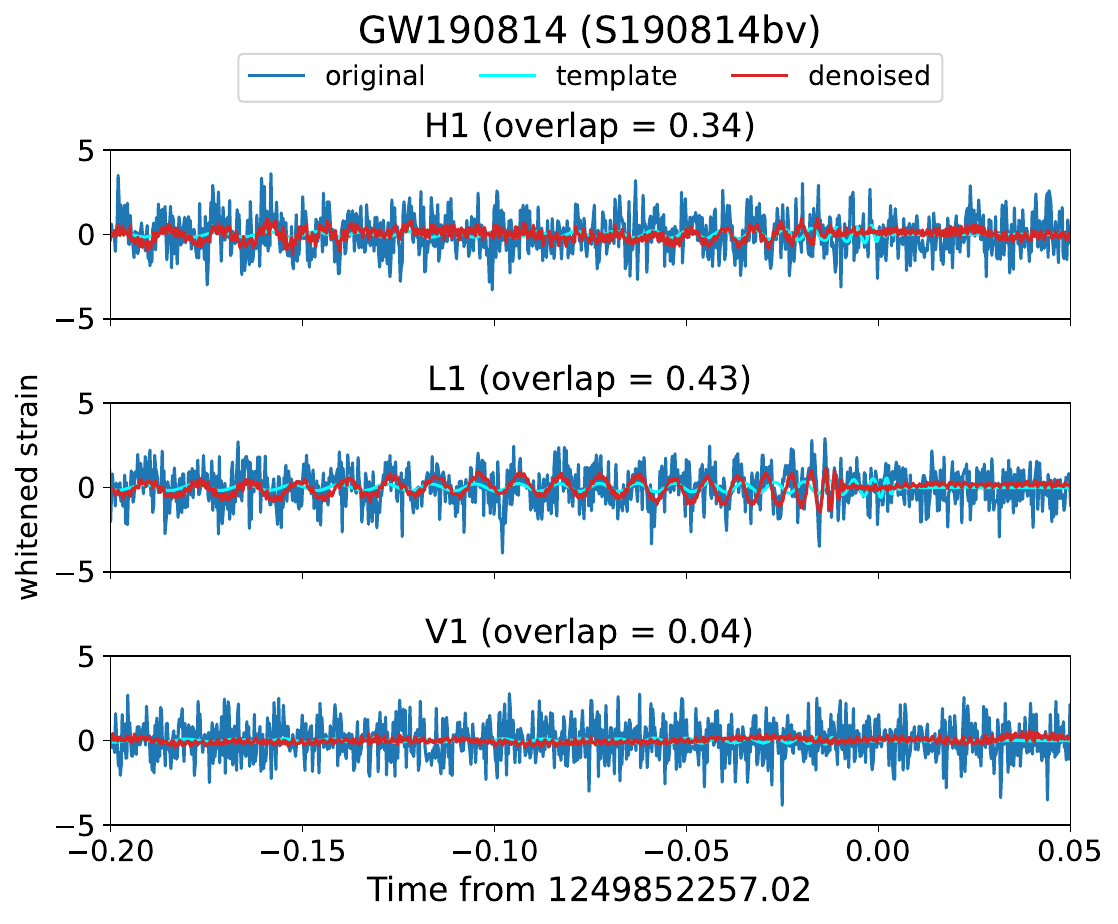}}
    \hfill
    \subfloat[\label{fig:appendix_21_2}]{\includegraphics[width=0.48\linewidth]{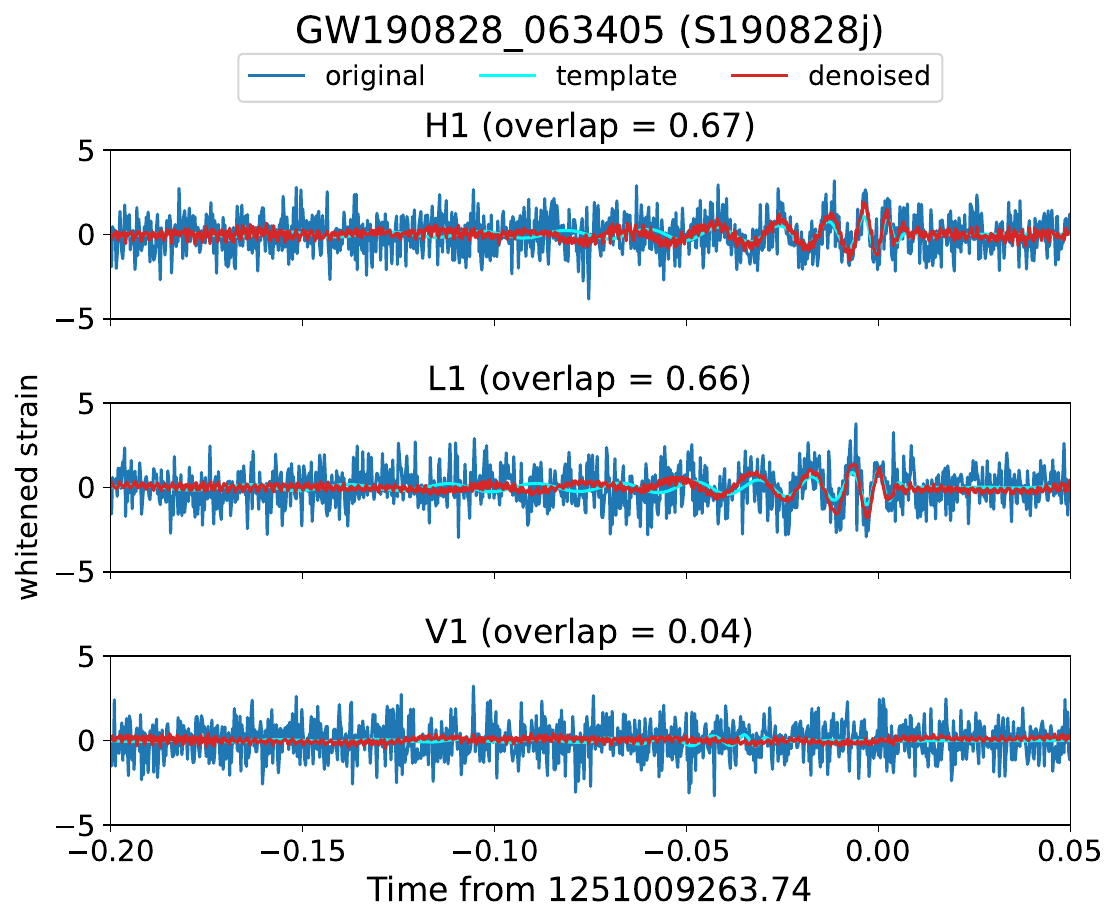}}
    \caption{\label{fig:appendix_21}The $0.25$s windows of the denoised results of the H1, L1 and V1 data around S190814bv and S190828j.}
\end{figure*}

\begin{figure*}[tbp]
    \centering
    \subfloat[\label{fig:appendix_22_1}]{\includegraphics[width=0.48\linewidth]{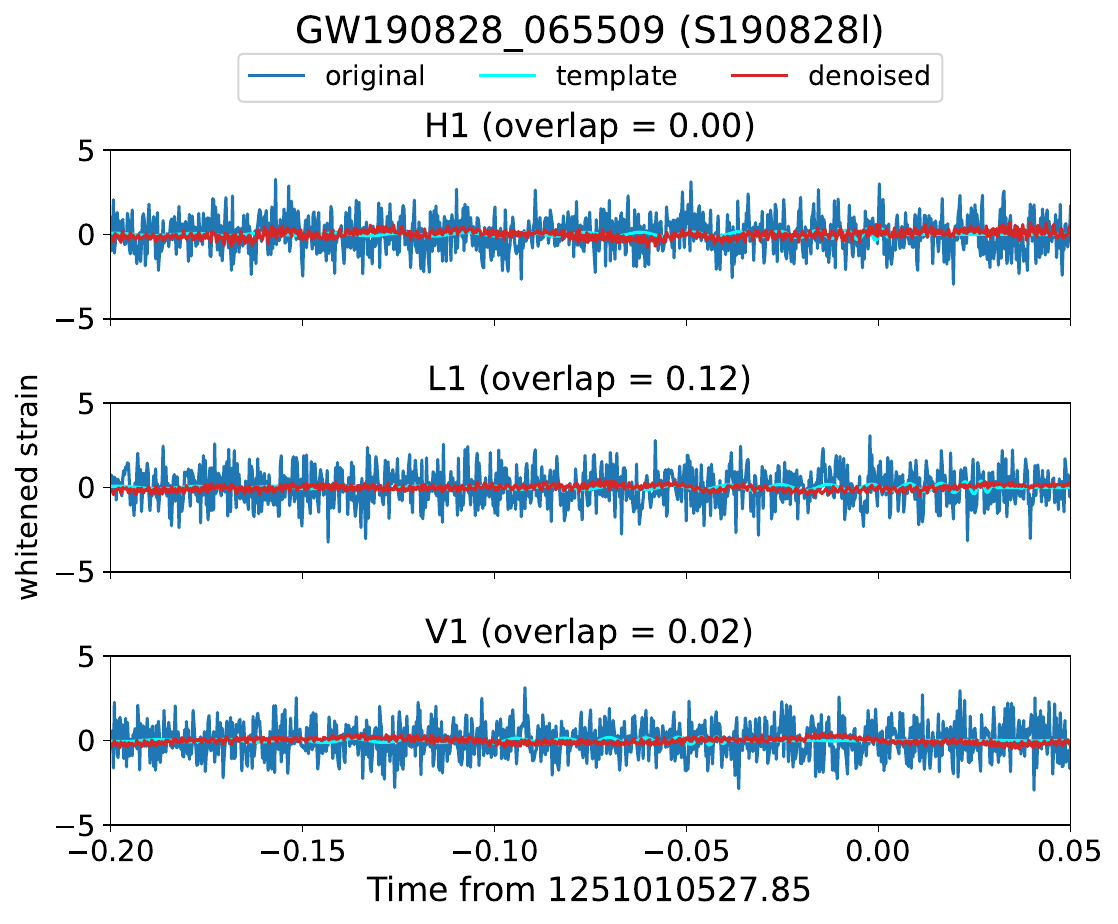}}
    \hfill
    \subfloat[\label{fig:appendix_22_2}]{\includegraphics[width=0.48\linewidth]{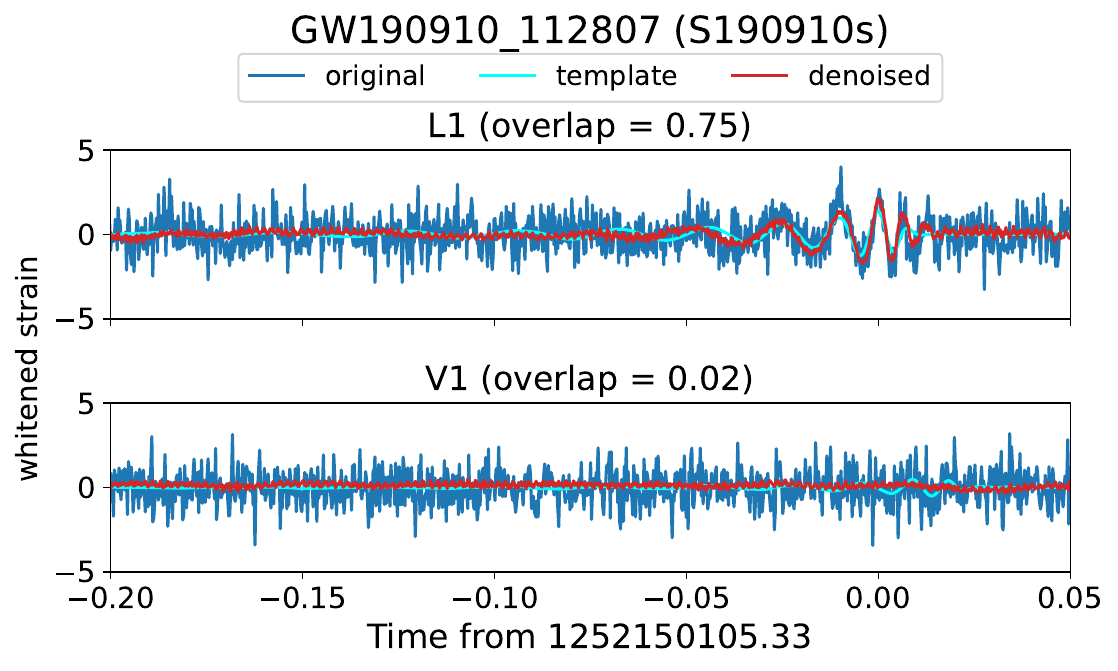}}
    \caption{\label{fig:appendix_22}The $0.25$s windows of the denoised results of the H1, L1 and V1 data around S190828l and S190910s.}
\end{figure*}

\begin{figure*}[tbp]
    \centering
    \subfloat[\label{fig:appendix_23_1}]{\includegraphics[width=0.48\linewidth]{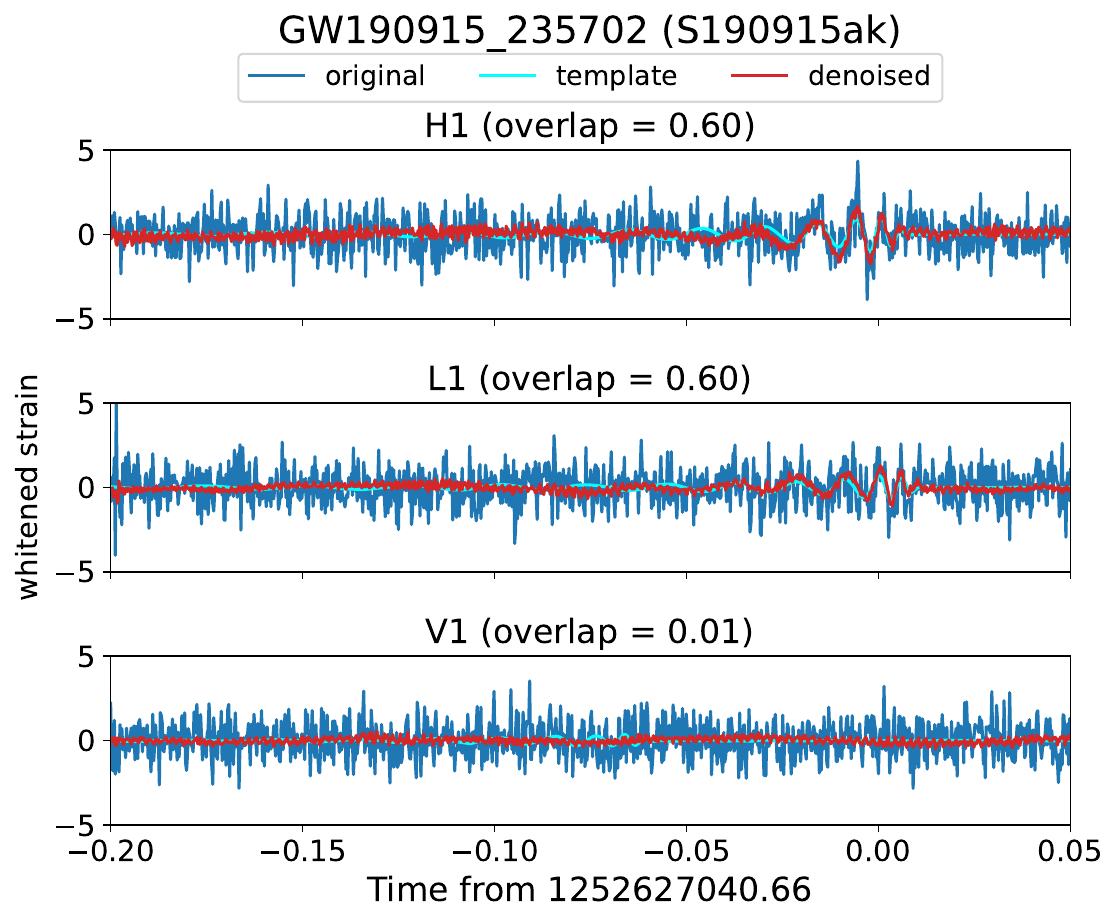}}
    \hfill
    \subfloat[\label{fig:appendix_23_2}]{\includegraphics[width=0.48\linewidth]{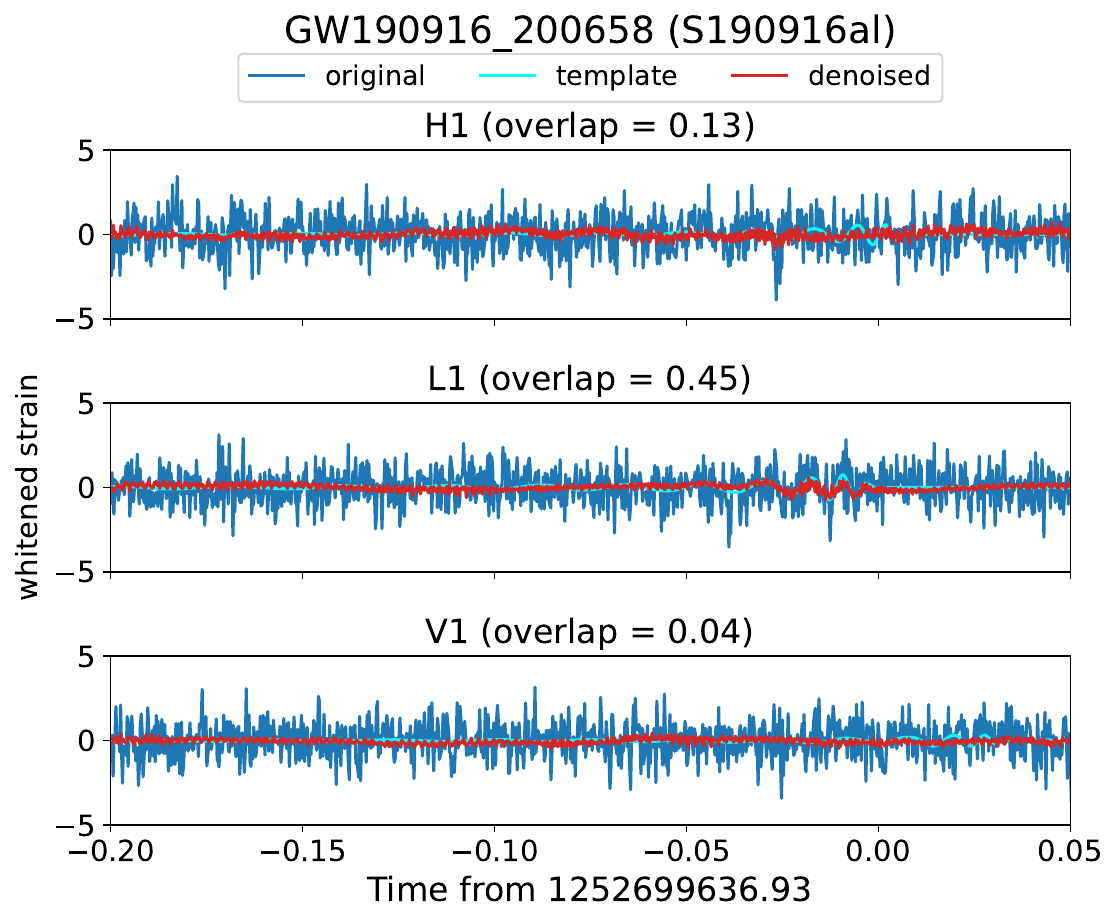}}
    \caption{\label{fig:appendix_23}The $0.25$s windows of the denoised results of the H1, L1 and V1 data around S190915ak and S190916al.}
\end{figure*}

\begin{figure*}[tbp]
    \centering
    \subfloat[\label{fig:appendix_24_1}]{\includegraphics[width=0.48\linewidth]{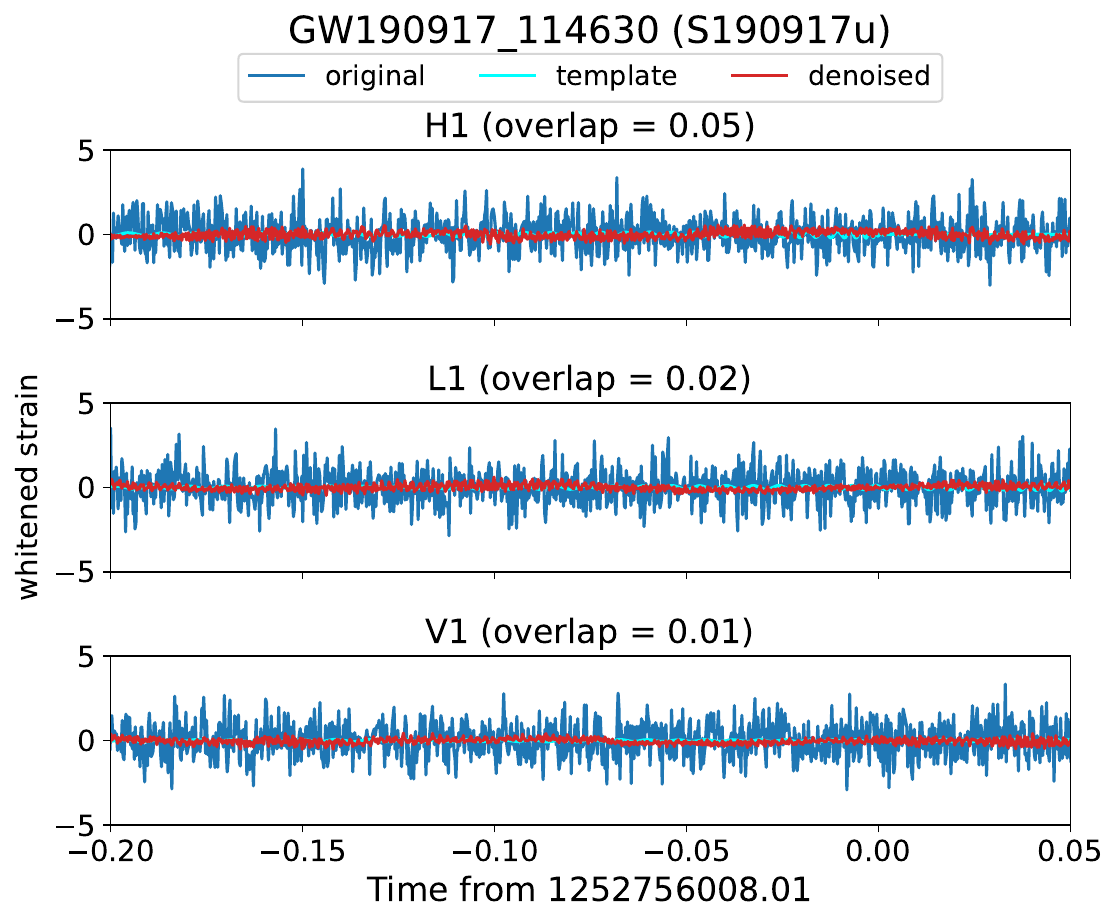}}
    \hfill
    \subfloat[\label{fig:appendix_24_2}]{\includegraphics[width=0.48\linewidth]{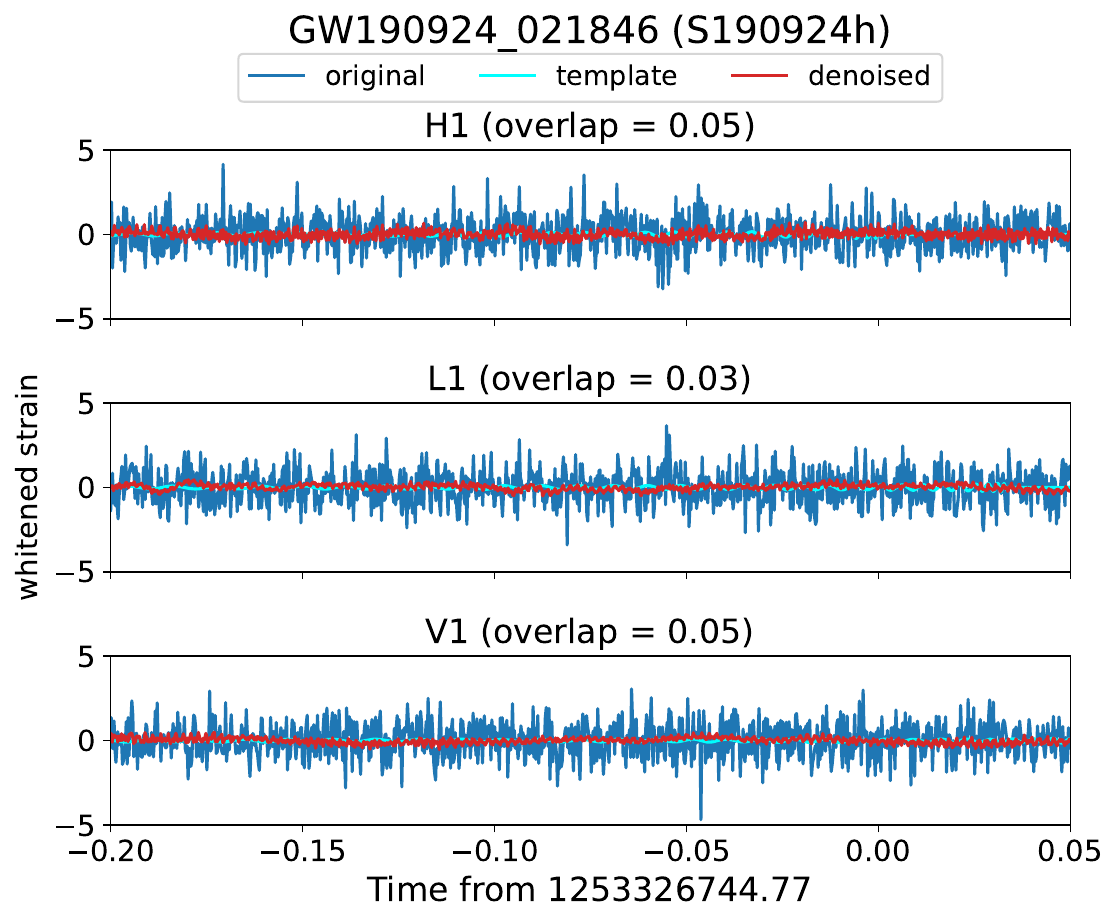}}
    \caption{\label{fig:appendix_24}The $0.25$s windows of the denoised results of the H1, L1 and V1 data around S190917u and S190924h.}
\end{figure*}

\begin{figure*}[tbp]
    \centering
    \subfloat[\label{fig:appendix_25_1}]{\includegraphics[width=0.48\linewidth]{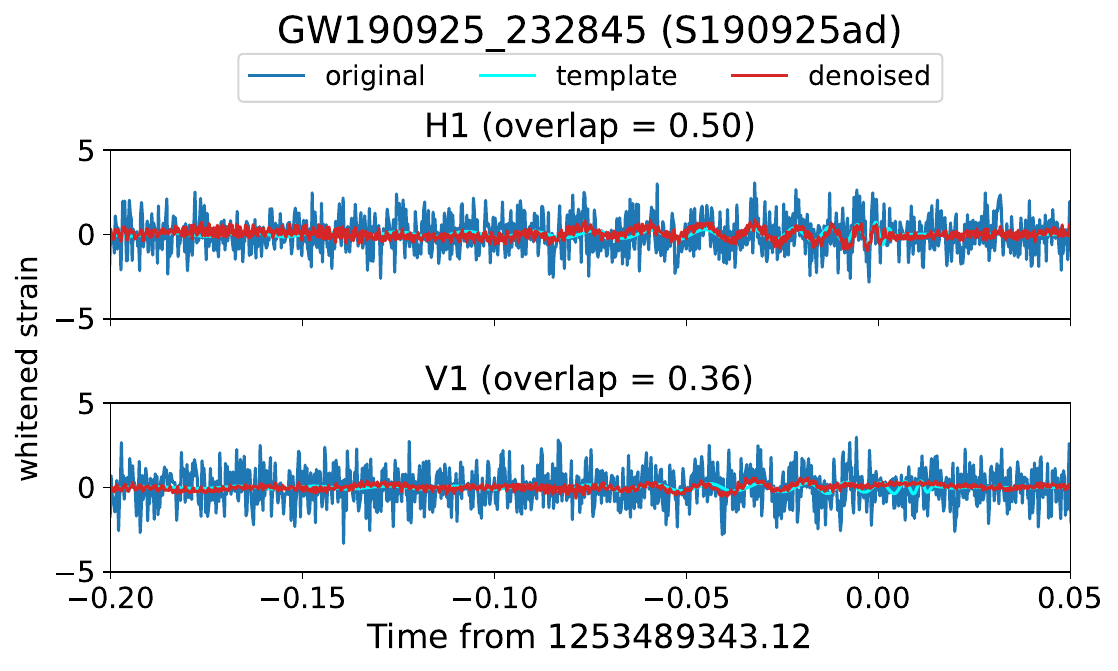}}
    \hfill
    \subfloat[\label{fig:appendix_25_2}]{\includegraphics[width=0.48\linewidth]{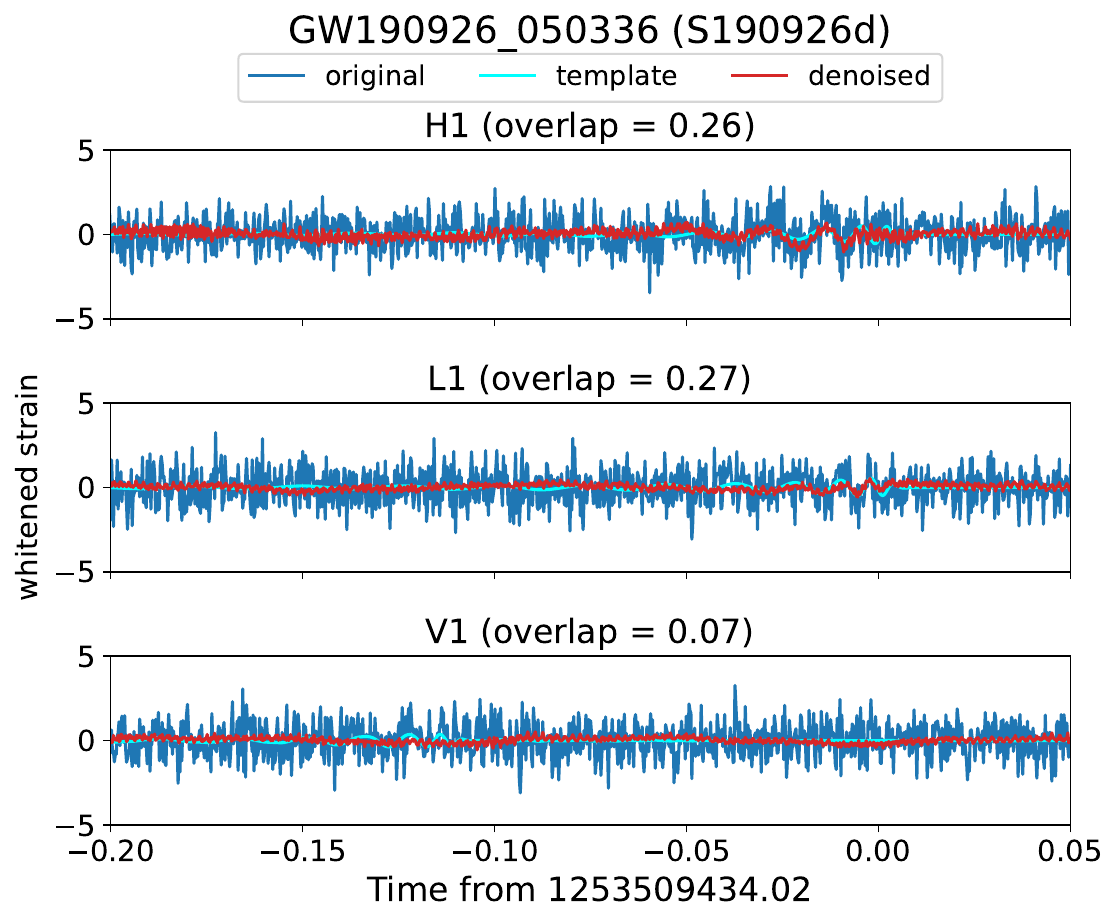}}
    \caption{\label{fig:appendix_25}The $0.25$s windows of the denoised results of the H1, L1 and V1 data around S190925ad and S190926d.}
\end{figure*}

\begin{figure*}[tbp]
    \centering
    \subfloat[\label{fig:appendix_26_1}]{\includegraphics[width=0.48\linewidth]{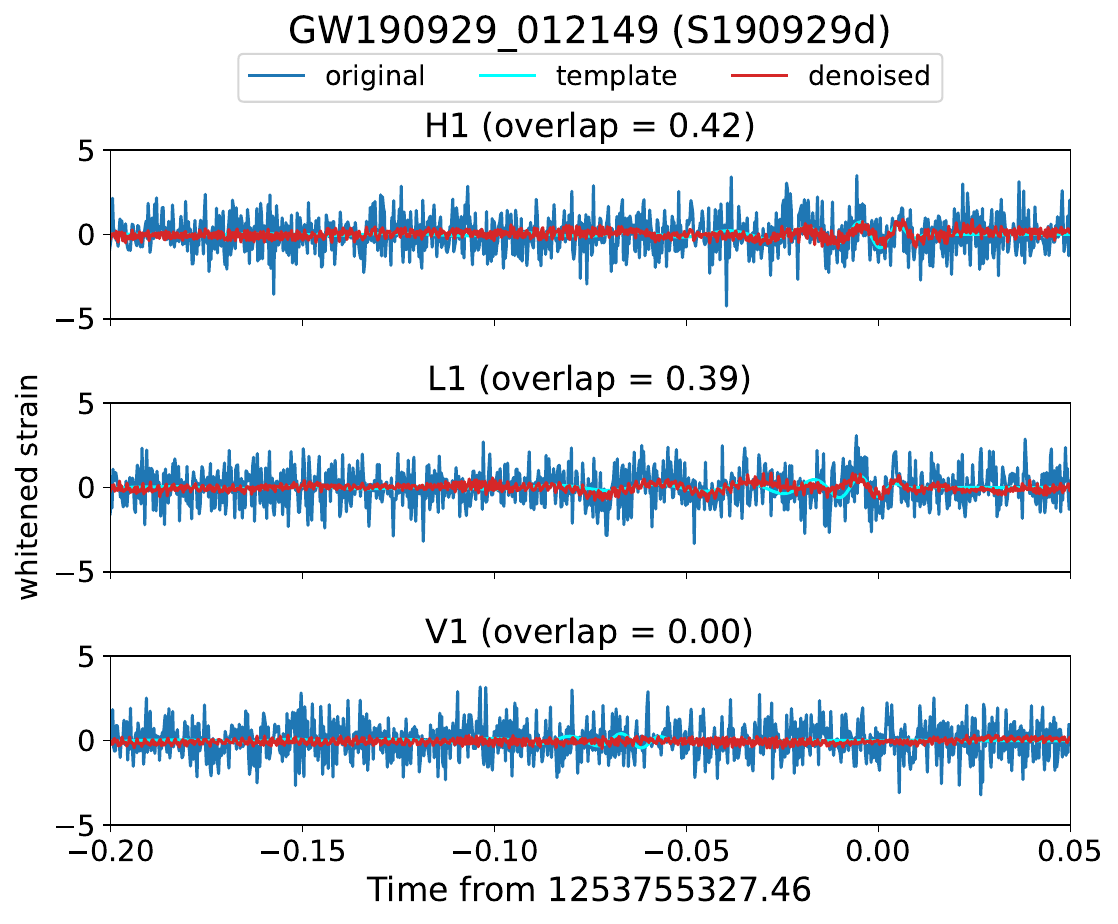}}
    \hfill
    \subfloat[\label{fig:appendix_26_2}]{\includegraphics[width=0.48\linewidth]{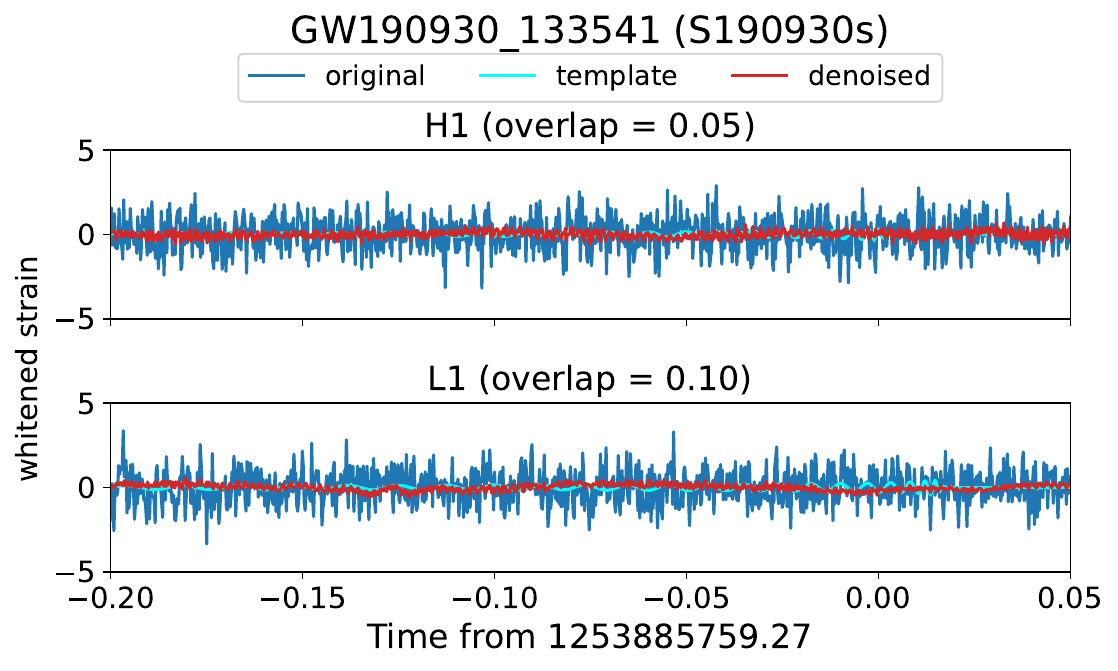}}
    \caption{\label{fig:appendix_26}The $0.25$s windows of the denoised results of the H1, L1 and V1 data around S190929d and S190930s.}
\end{figure*}

\begin{figure*}[tbp]
    \centering
    \subfloat[\label{fig:appendix_27_1}]{\includegraphics[width=0.48\linewidth]{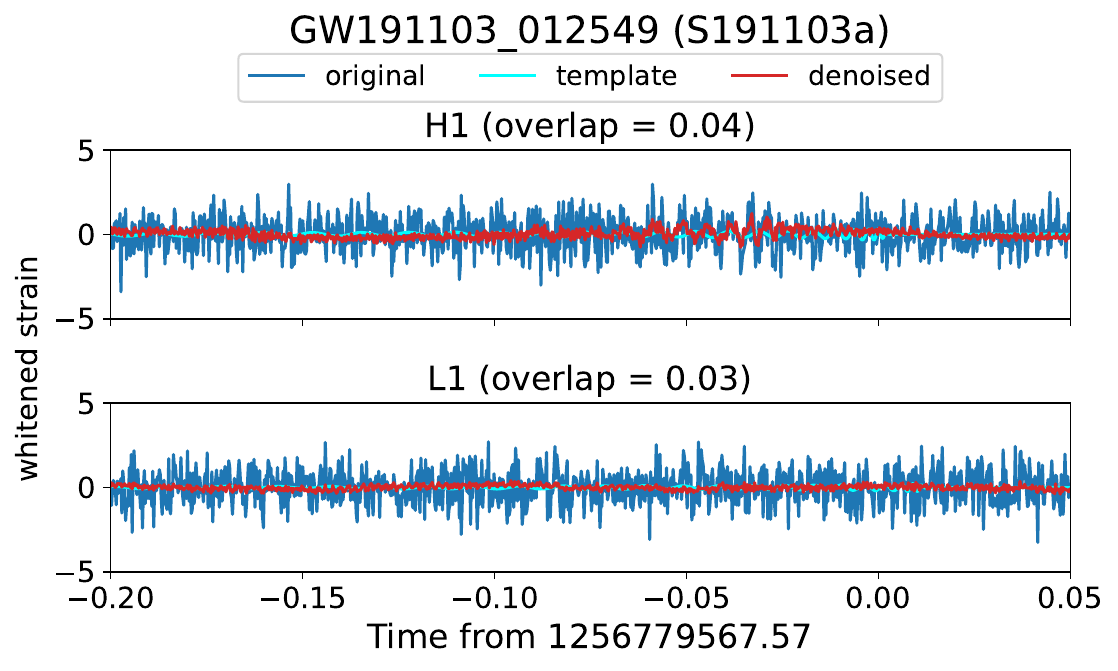}}
    \hfill
    \subfloat[\label{fig:appendix_27_2}]{\includegraphics[width=0.48\linewidth]{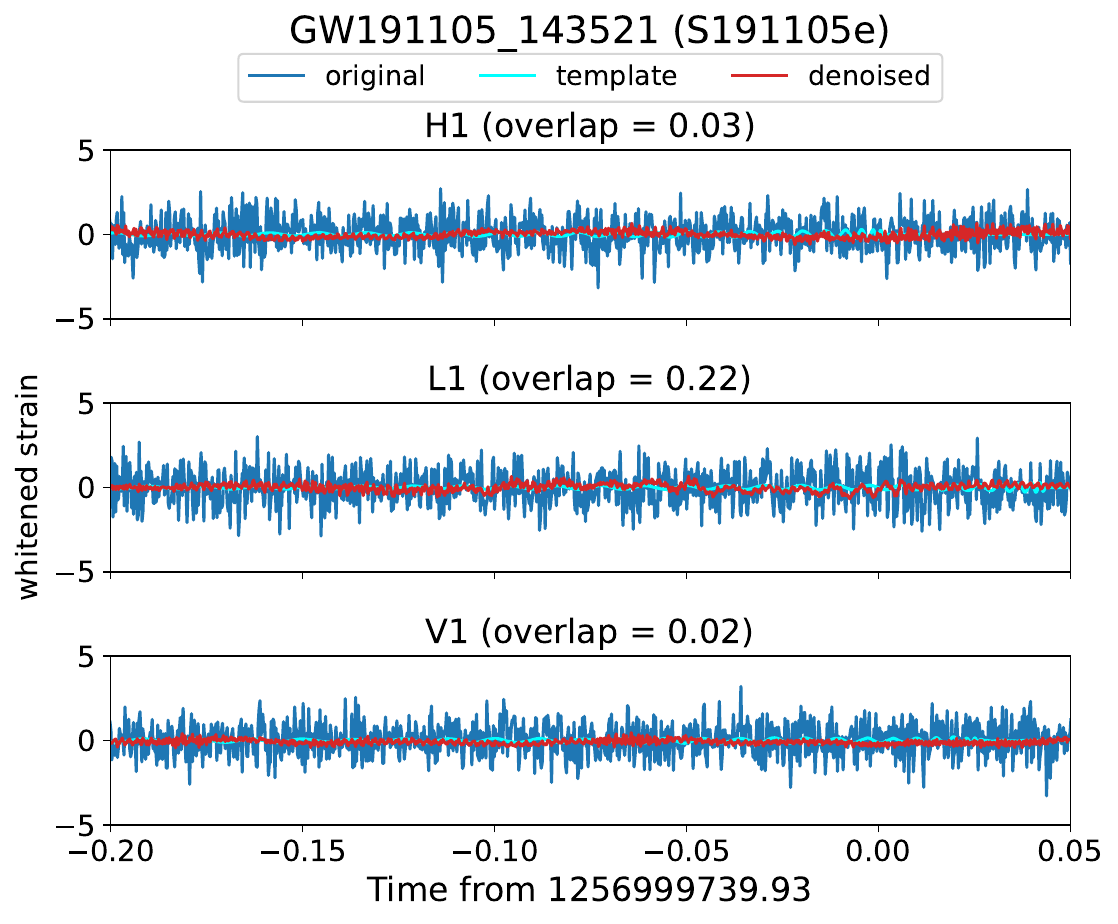}}
    \caption{\label{fig:appendix_27}The $0.25$s windows of the denoised results of the H1, L1 and V1 data around S191103a and S191105e.}
\end{figure*}

\begin{figure*}[tbp]
    \centering
    \subfloat[\label{fig:appendix_28_1}]{\includegraphics[width=0.48\linewidth]{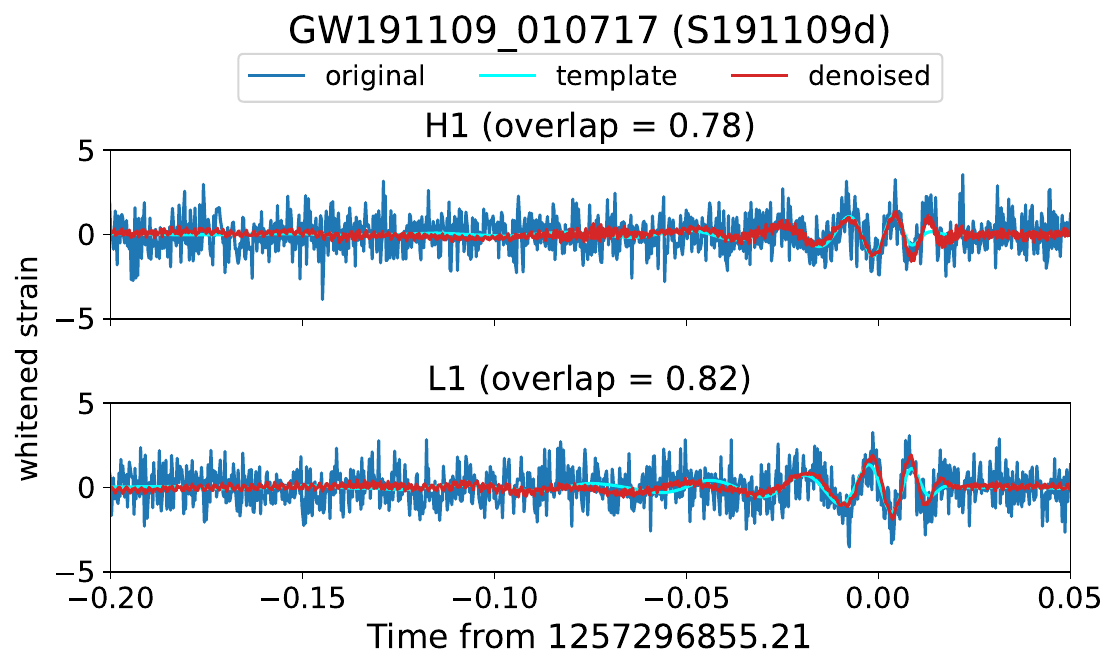}}
    \hfill
    \subfloat[\label{fig:appendix_28_2}]{\includegraphics[width=0.48\linewidth]{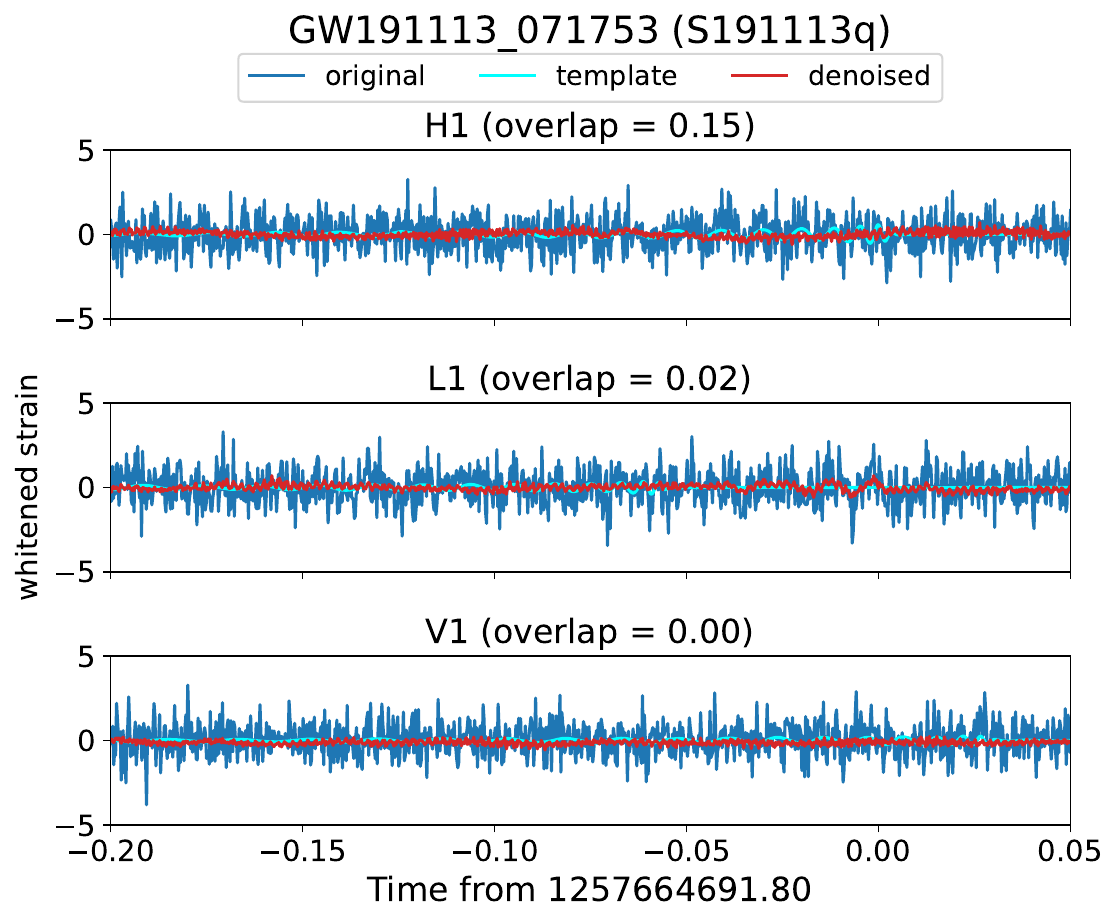}}
    \caption{\label{fig:appendix_28}The $0.25$s windows of the denoised results of the H1, L1 and V1 data around S191109d and S191113q.}
\end{figure*}

\begin{figure*}[tbp]
    \centering
    \subfloat[\label{fig:appendix_29_1}]{\includegraphics[width=0.48\linewidth]{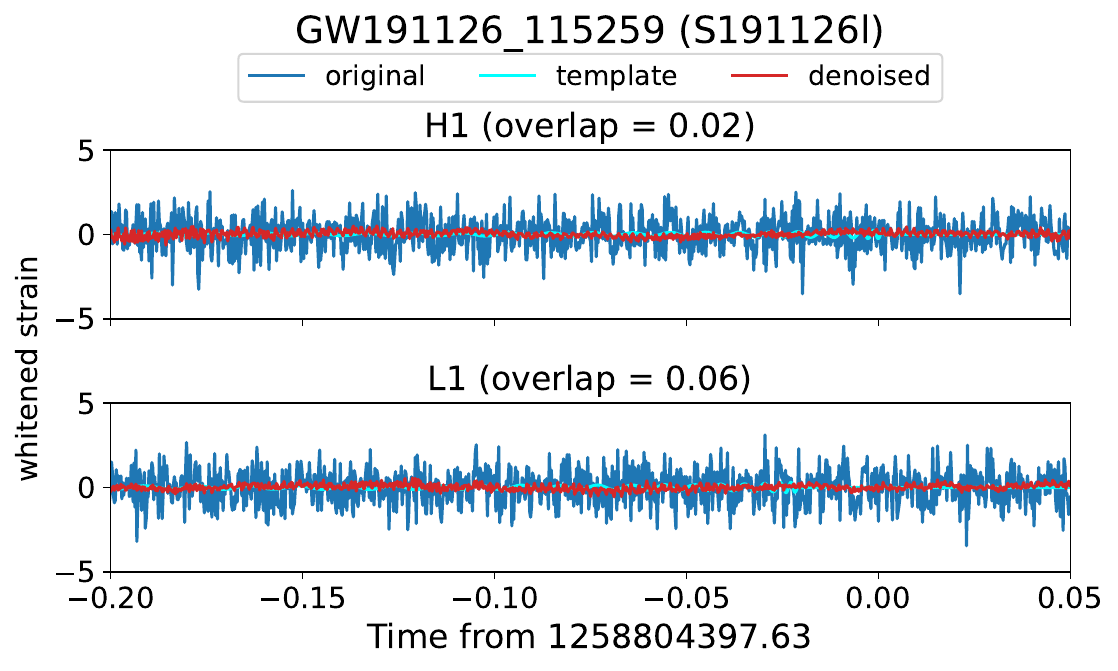}}
    \hfill
    \subfloat[\label{fig:appendix_29_2}]{\includegraphics[width=0.48\linewidth]{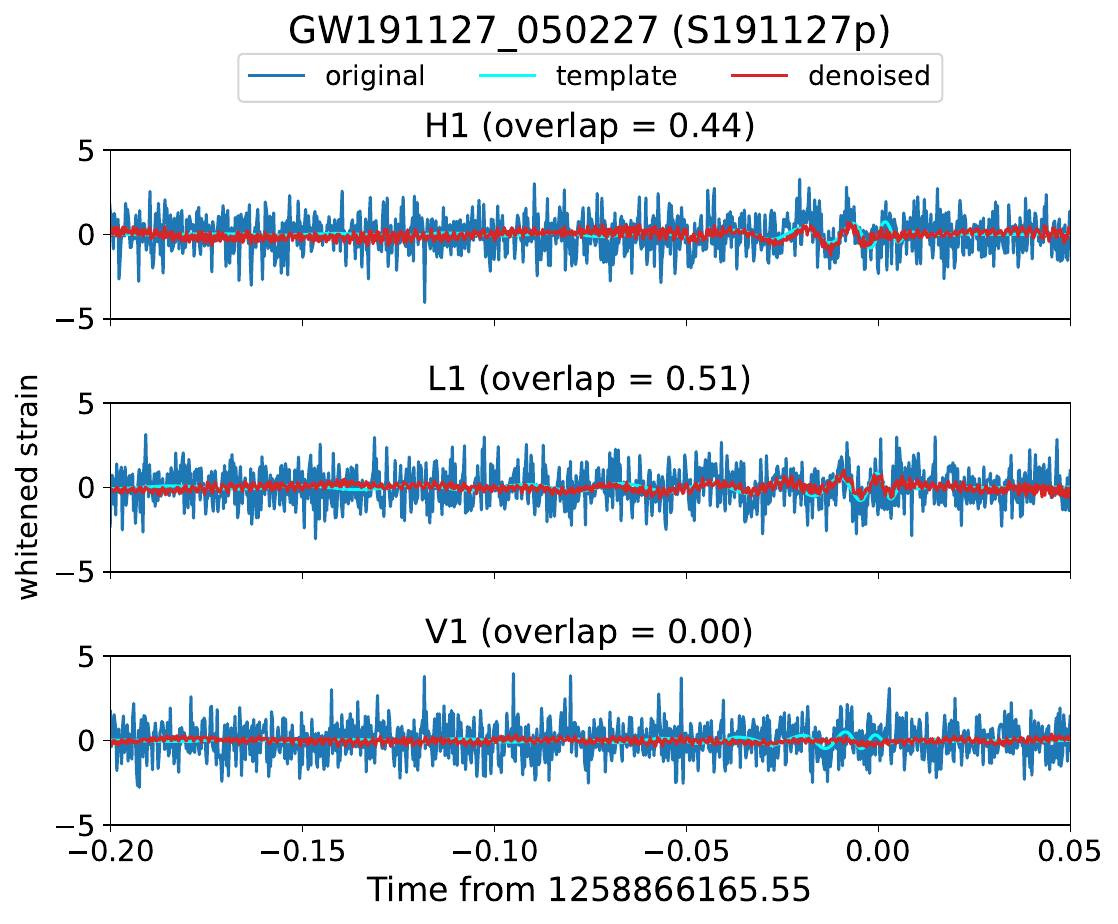}}
    \caption{\label{fig:appendix_29}The $0.25$s windows of the denoised results of the H1, L1 and V1 data around S191126l and S191127p.}
\end{figure*}

\begin{figure*}[tbp]
    \centering
    \subfloat[\label{fig:appendix_30_1}]{\includegraphics[width=0.48\linewidth]{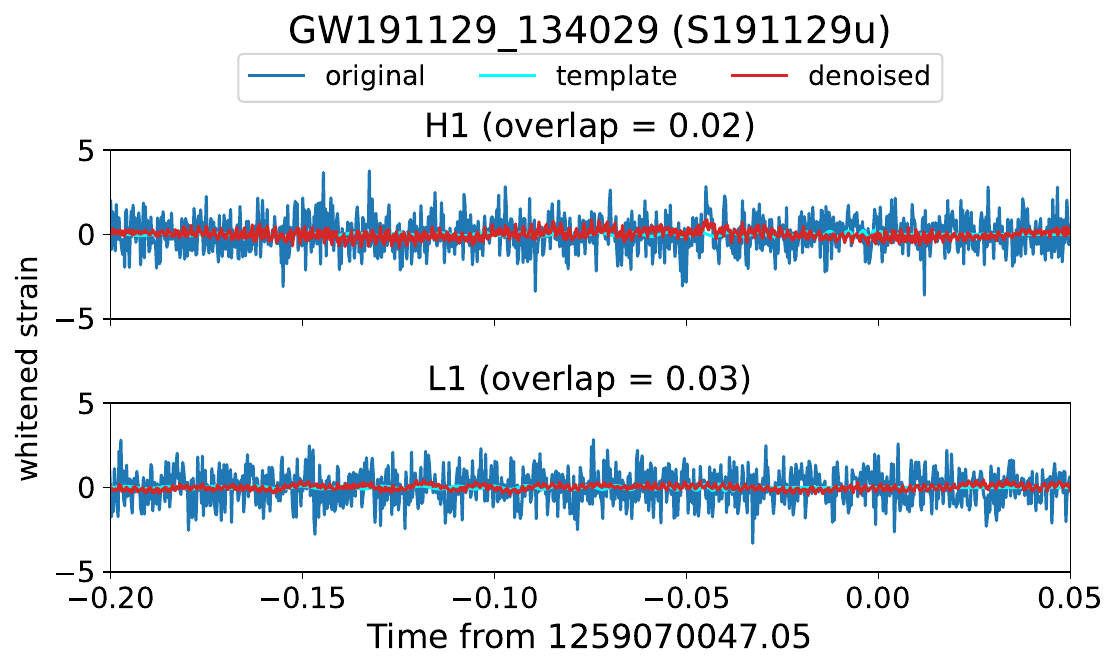}}
    \hfill
    \subfloat[\label{fig:appendix_30_2}]{\includegraphics[width=0.48\linewidth]{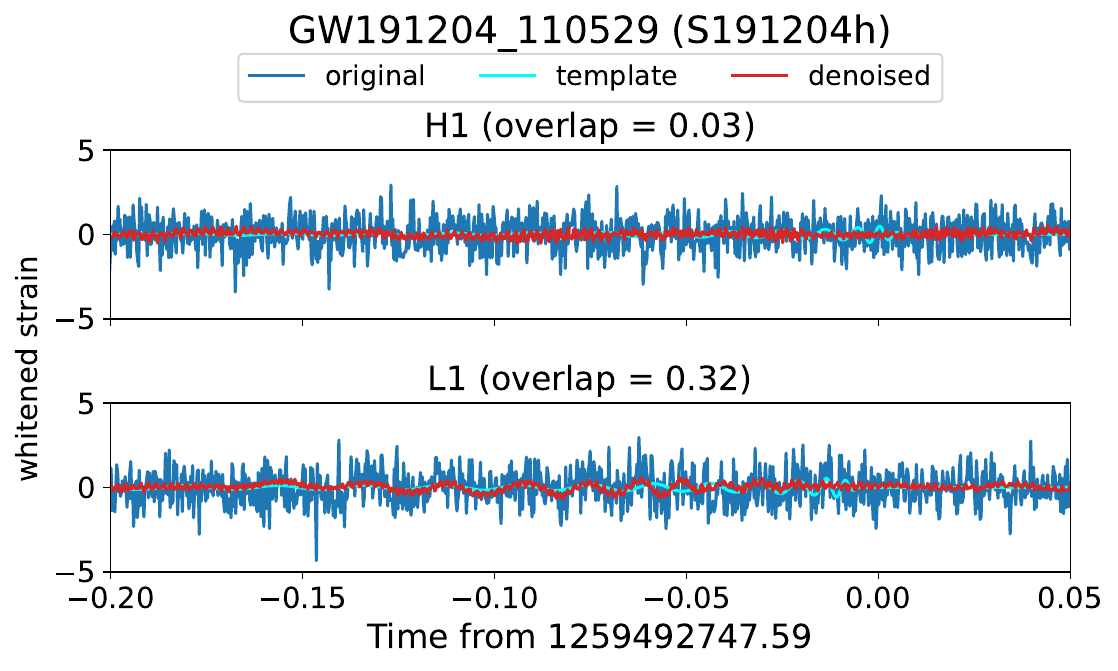}}
    \caption{\label{fig:appendix_30}The $0.25$s windows of the denoised results of the H1, L1 and V1 data around S191129u and S191204h.}
\end{figure*}

\begin{figure*}[tbp]
    \centering
    \subfloat[\label{fig:appendix_31_1}]{\includegraphics[width=0.48\linewidth]{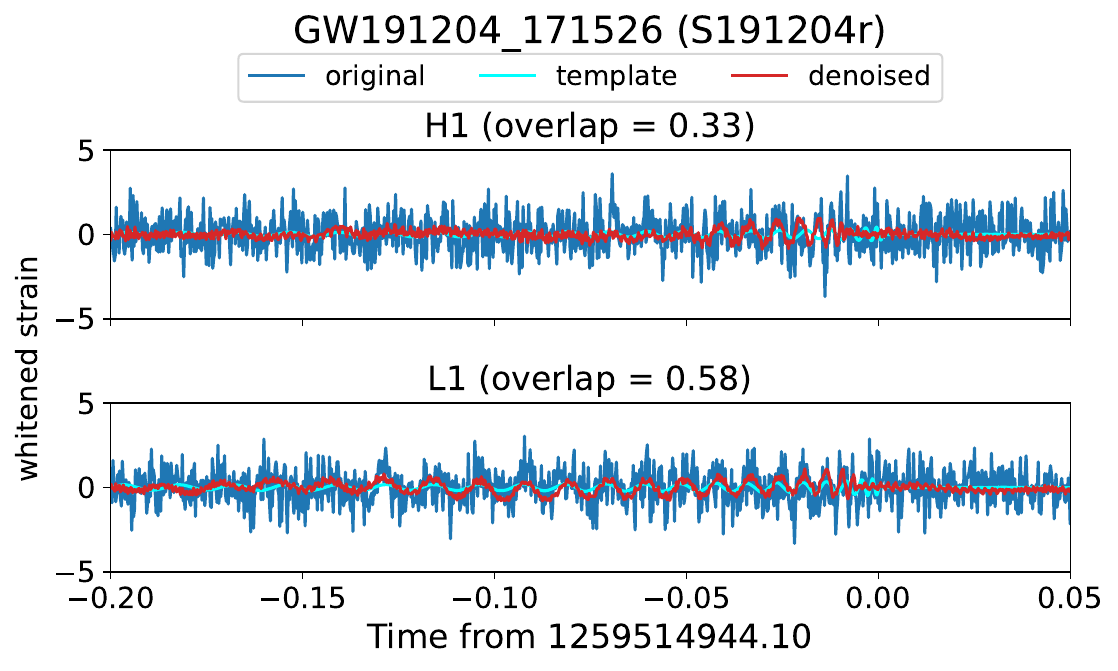}}
    \hfill
    \subfloat[\label{fig:appendix_31_2}]{\includegraphics[width=0.48\linewidth]{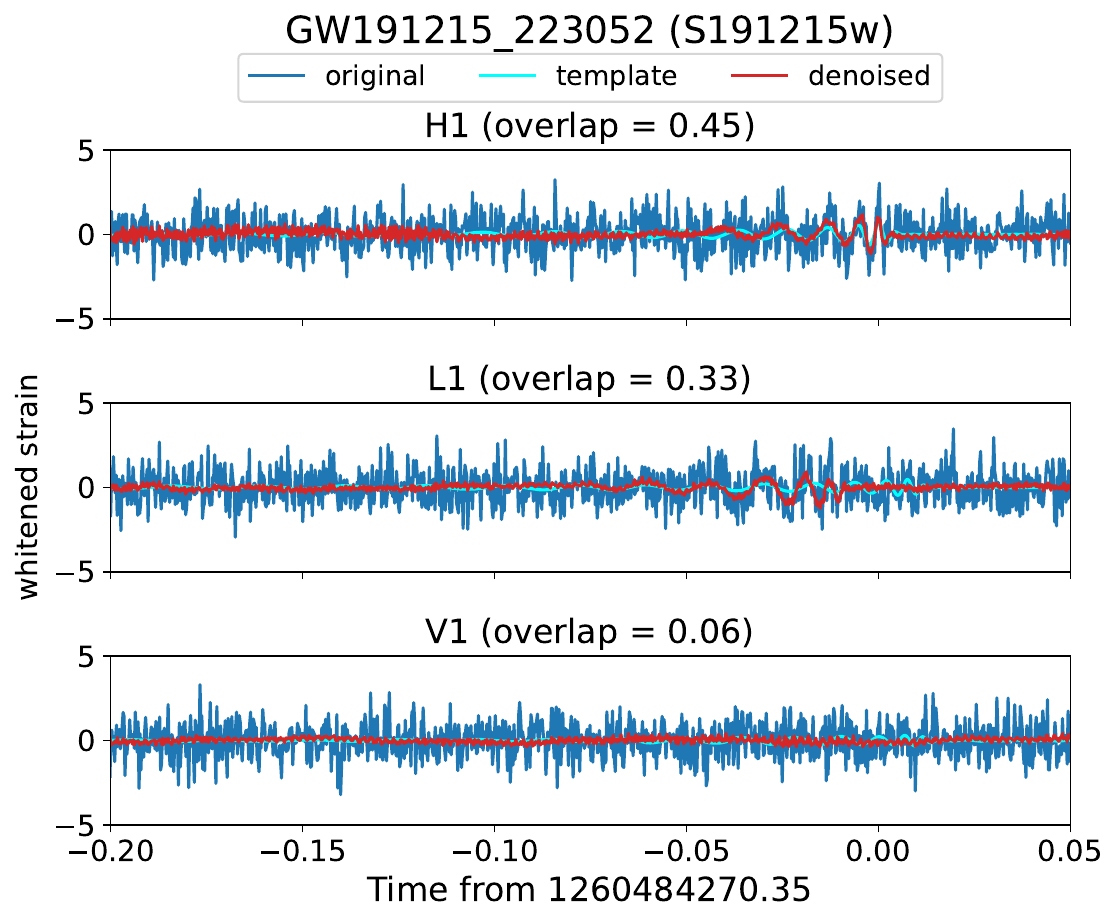}}
    \caption{\label{fig:appendix_31}The $0.25$s windows of the denoised results of the H1, L1 and V1 data around S191204r and S191215w.}
\end{figure*}

\begin{figure*}[tbp]
    \centering
    \subfloat[\label{fig:appendix_32_1}]{\includegraphics[width=0.48\linewidth]{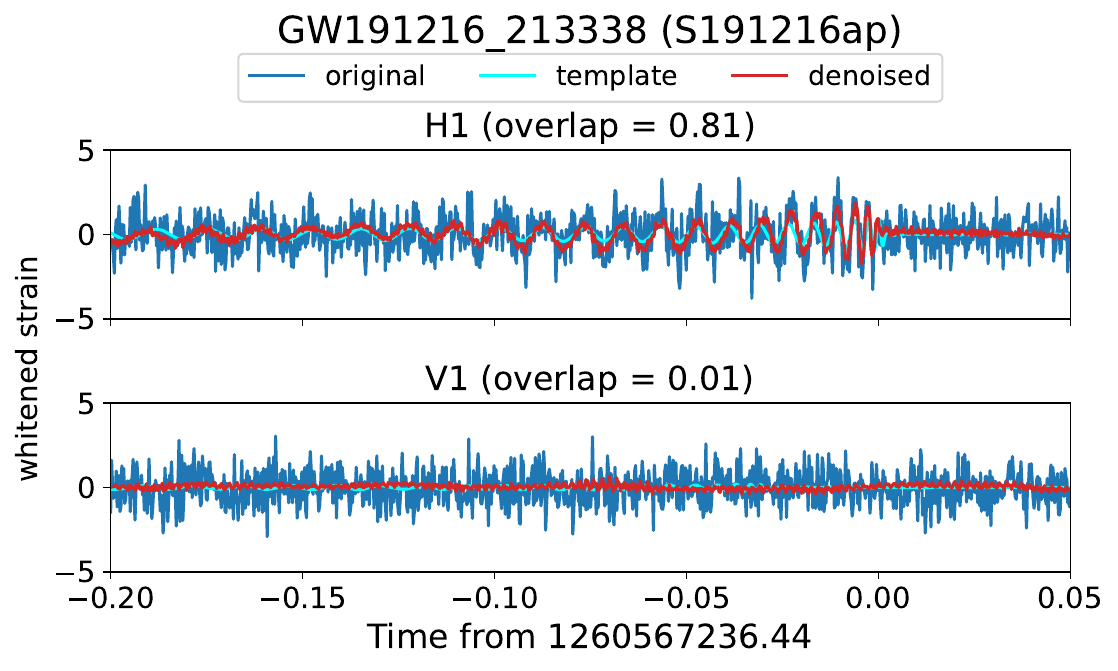}}
    \hfill
    \subfloat[\label{fig:appendix_32_2}]{\includegraphics[width=0.48\linewidth]{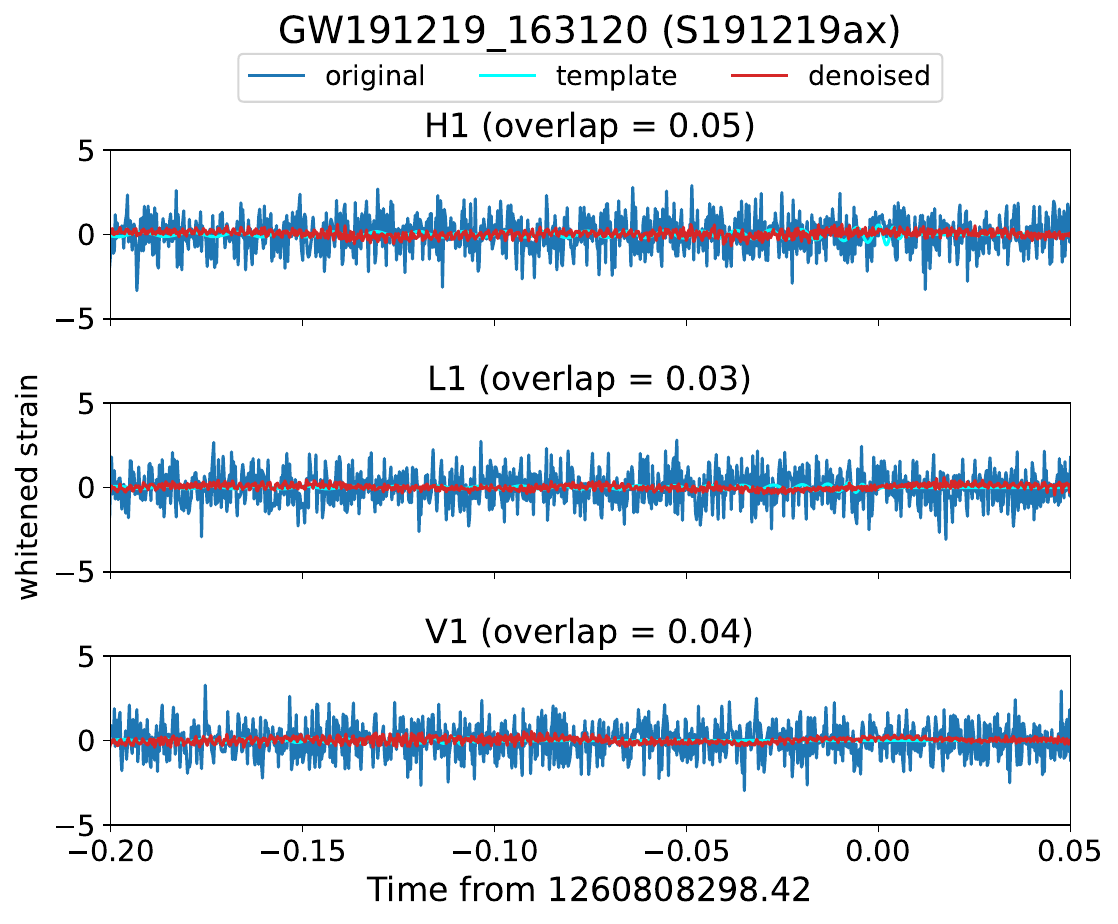}}
    \caption{\label{fig:appendix_32}The $0.25$s windows of the denoised results of the H1, L1 and V1 data around S191216ap and S191219ax.}
\end{figure*}

\begin{figure*}[tbp]
    \centering
    \subfloat[\label{fig:appendix_33_1}]{\includegraphics[width=0.48\linewidth]{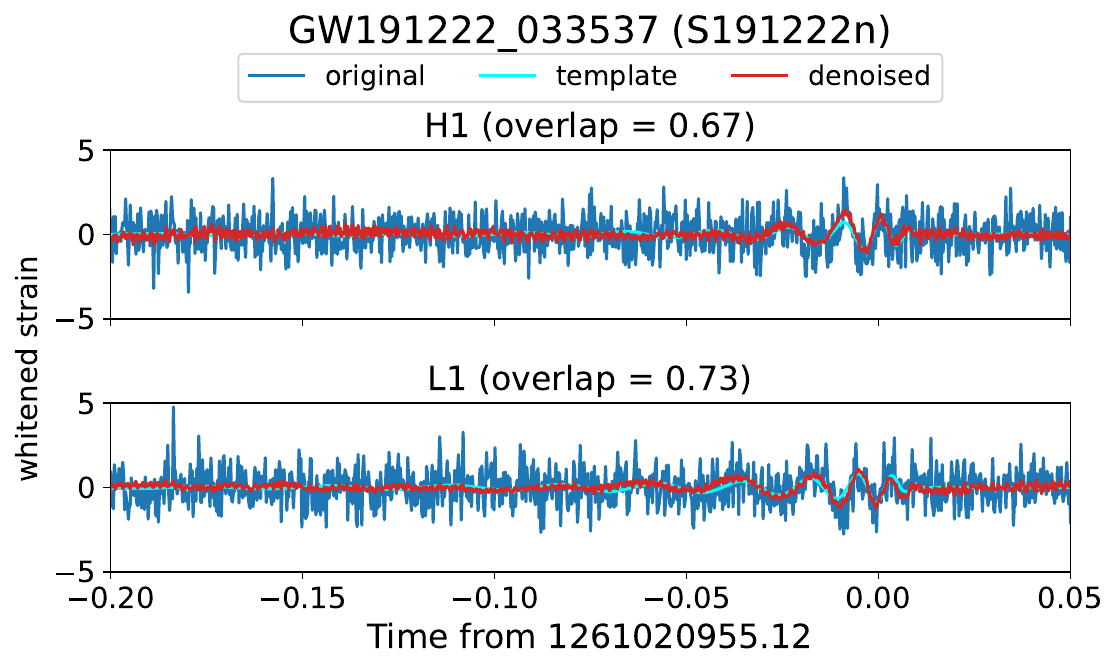}}
    \hfill
    \subfloat[\label{fig:appendix_33_2}]{\includegraphics[width=0.48\linewidth]{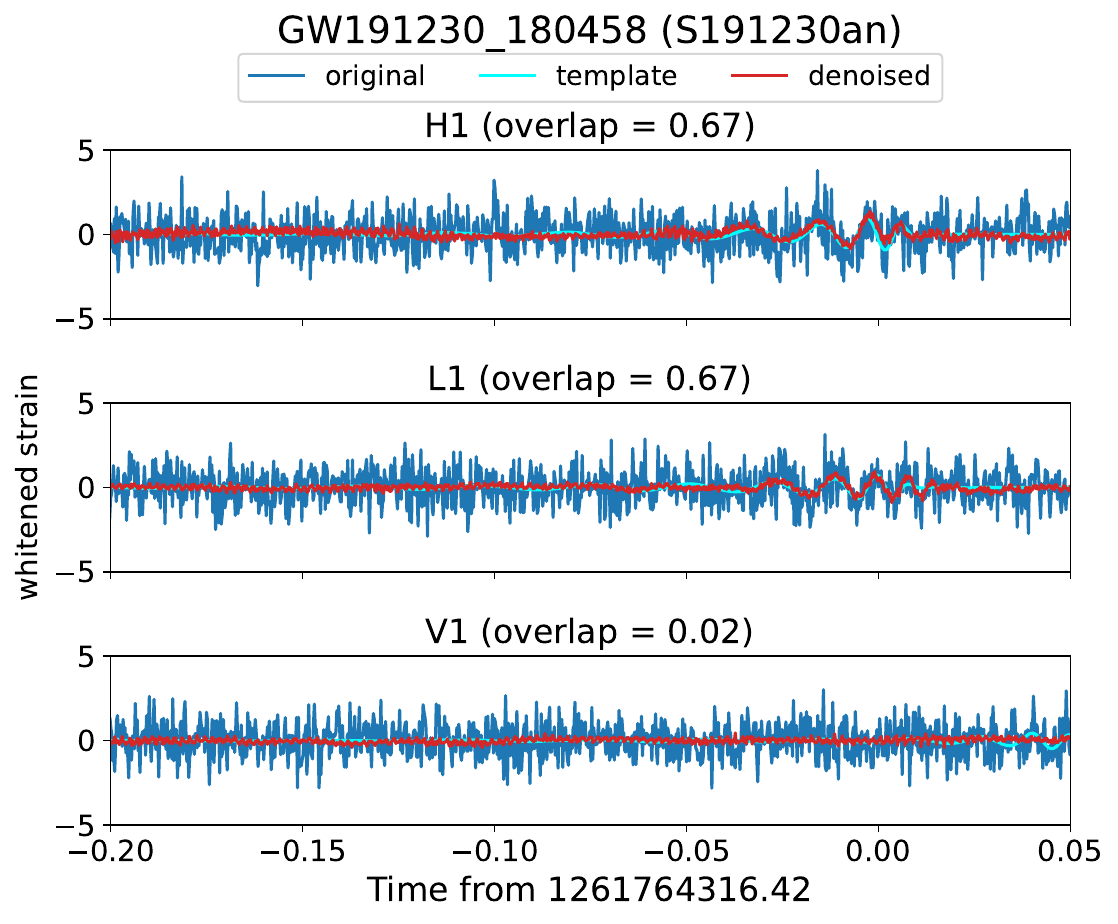}}
    \caption{\label{fig:appendix_33}The $0.25$s windows of the denoised results of the H1, L1 and V1 data around S191222n and S191230an.}
\end{figure*}

\begin{figure*}[tbp]
    \centering
    \subfloat[\label{fig:appendix_34_1}]{\includegraphics[width=0.48\linewidth]{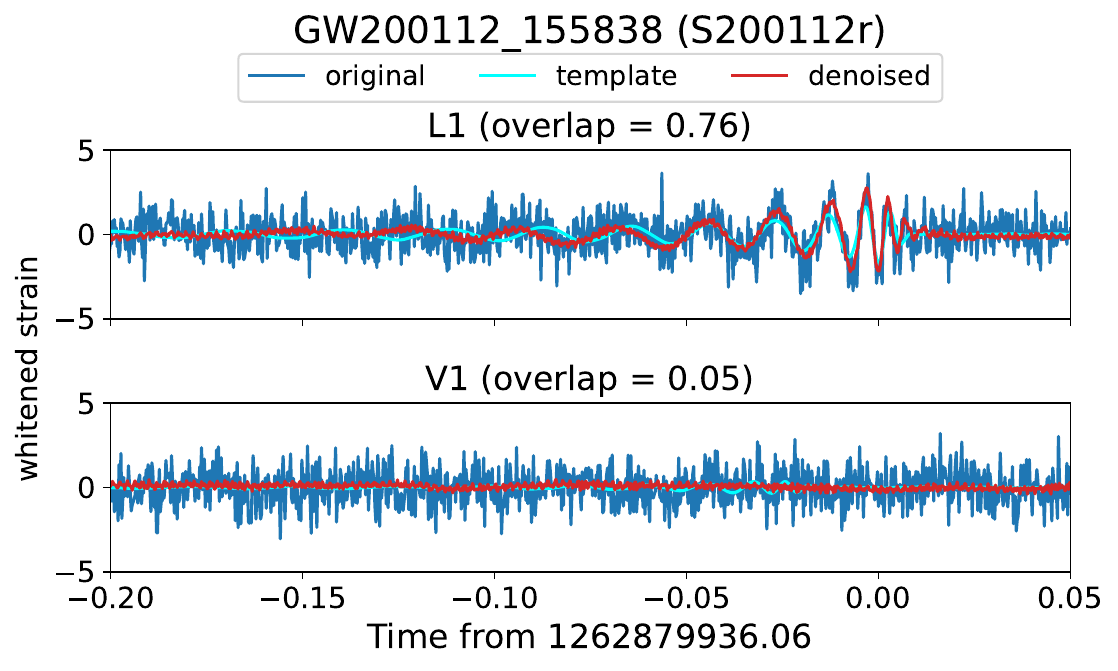}}
    \hfill
    \subfloat[\label{fig:appendix_34_2}]{\includegraphics[width=0.48\linewidth]{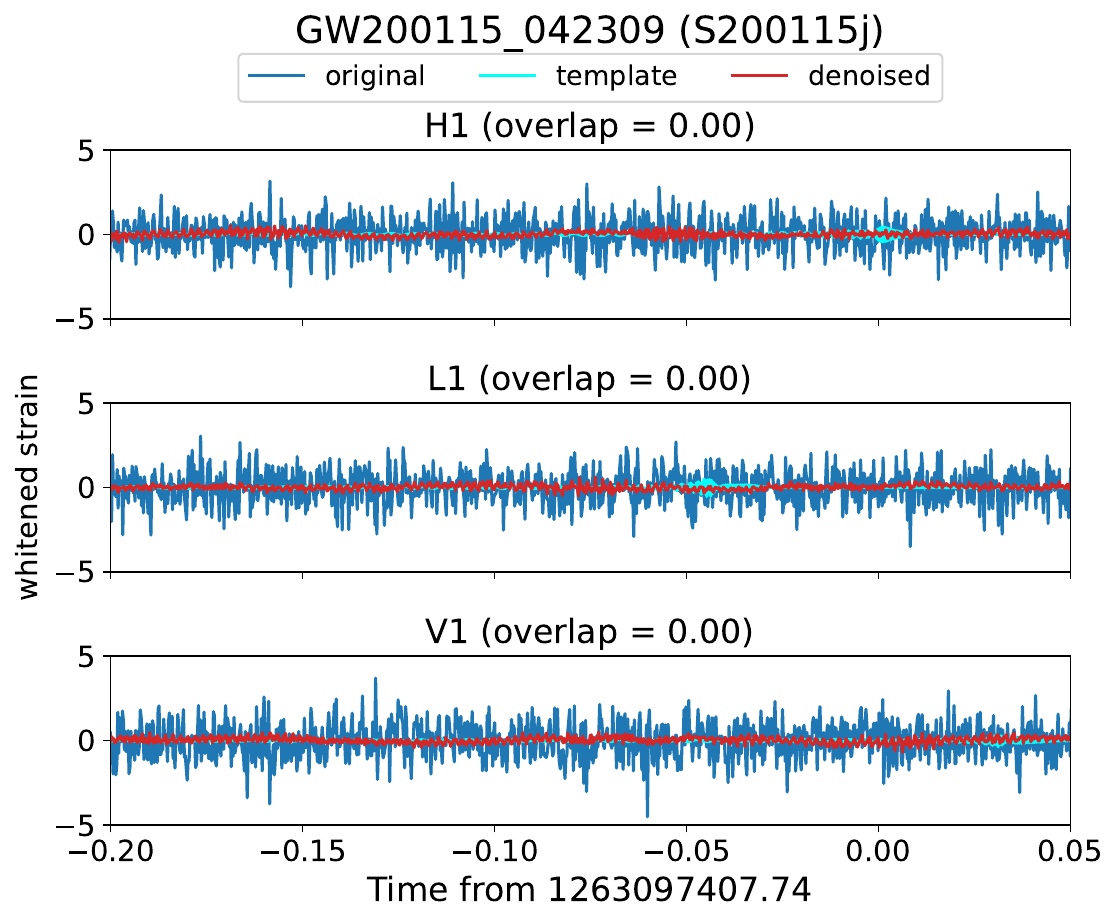}}
    \caption{\label{fig:appendix_34}The $0.25$s windows of the denoised results of the H1, L1 and V1 data around S200112r and S200115j.}
\end{figure*}

\begin{figure*}[tbp]
    \centering
    \subfloat[\label{fig:appendix_35_1}]{\includegraphics[width=0.48\linewidth]{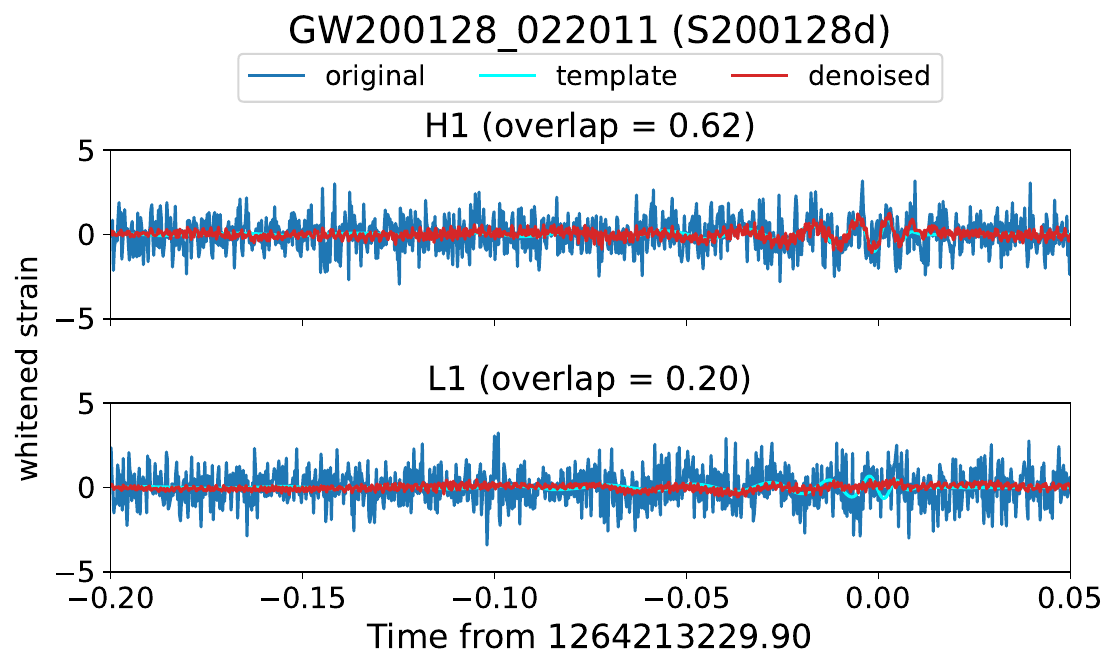}}
    \hfill
    \subfloat[\label{fig:appendix_35_2}]{\includegraphics[width=0.48\linewidth]{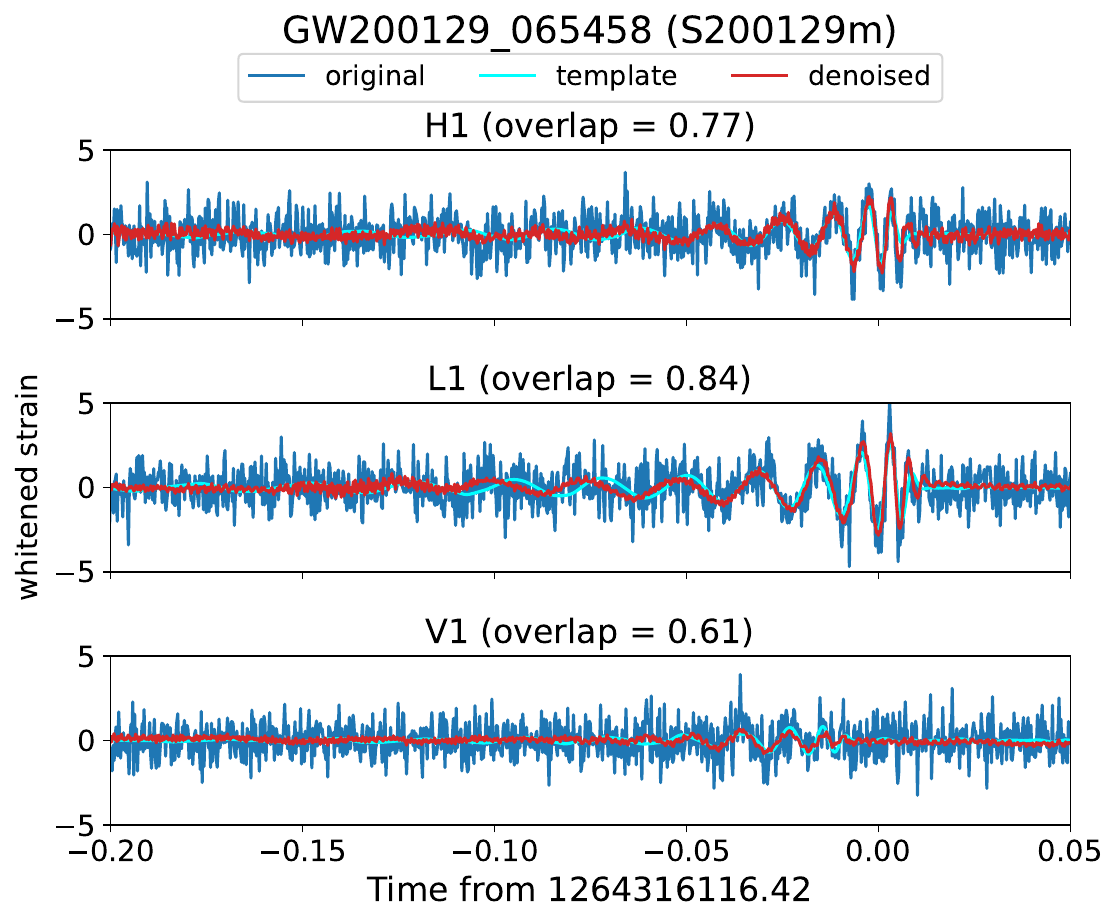}}
    \caption{\label{fig:appendix_35}The $0.25$s windows of the denoised results of the H1, L1 and V1 data around S200128d and S200129m.}
\end{figure*}

\begin{figure*}[tbp]
    \centering
    \subfloat[\label{fig:appendix_36_1}]{\includegraphics[width=0.48\linewidth]{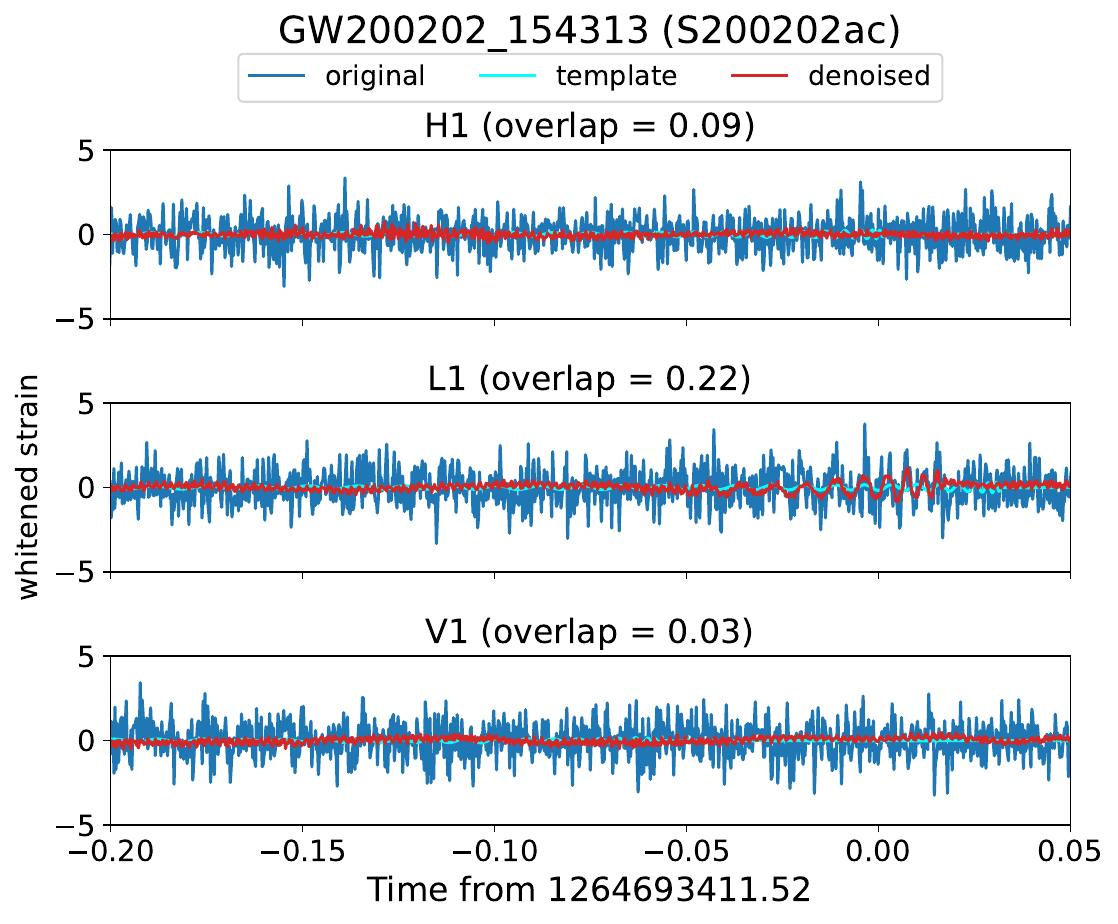}}
    \hfill
    \subfloat[\label{fig:appendix_36_2}]{\includegraphics[width=0.48\linewidth]{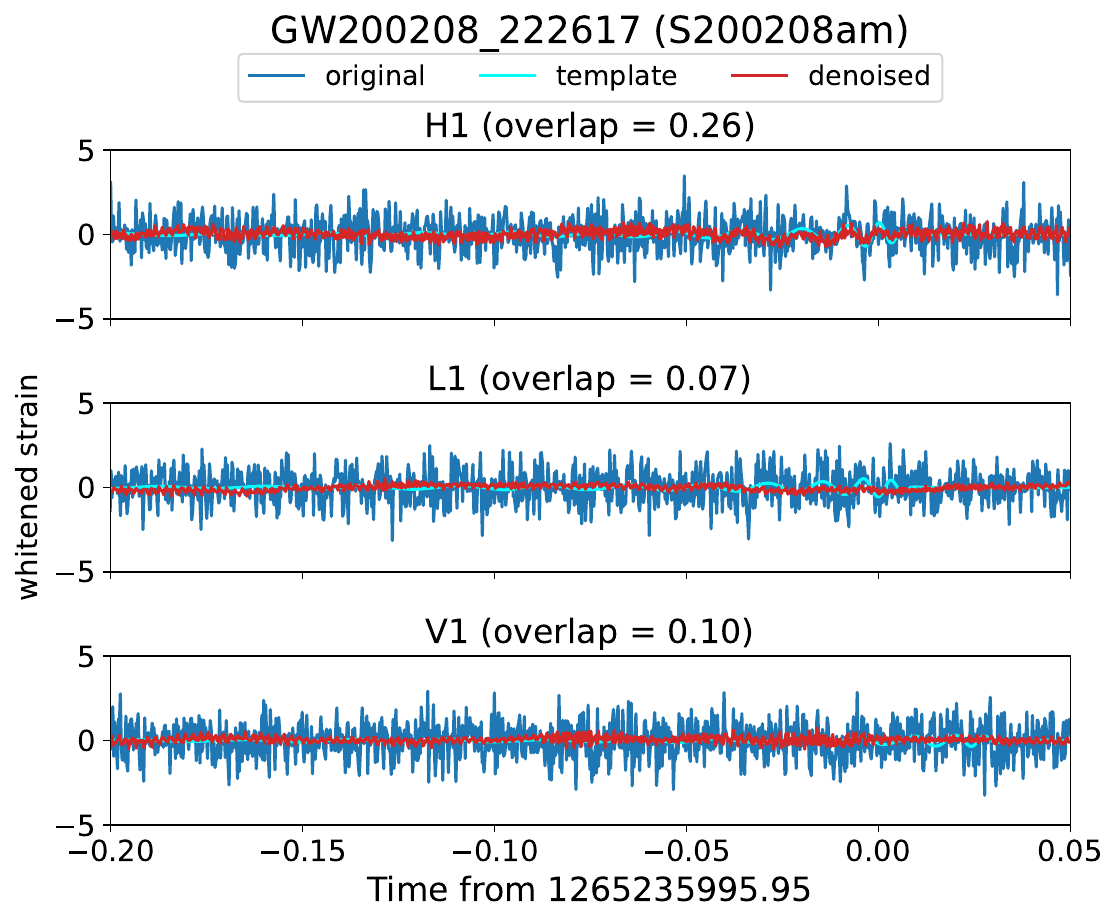}}
    \caption{\label{fig:appendix_36}The $0.25$s windows of the denoised results of the H1, L1 and V1 data around S200202ac and S200208am.}
\end{figure*}

\begin{figure*}[tbp]
    \centering
    \subfloat[\label{fig:appendix_37_1}]{\includegraphics[width=0.48\linewidth]{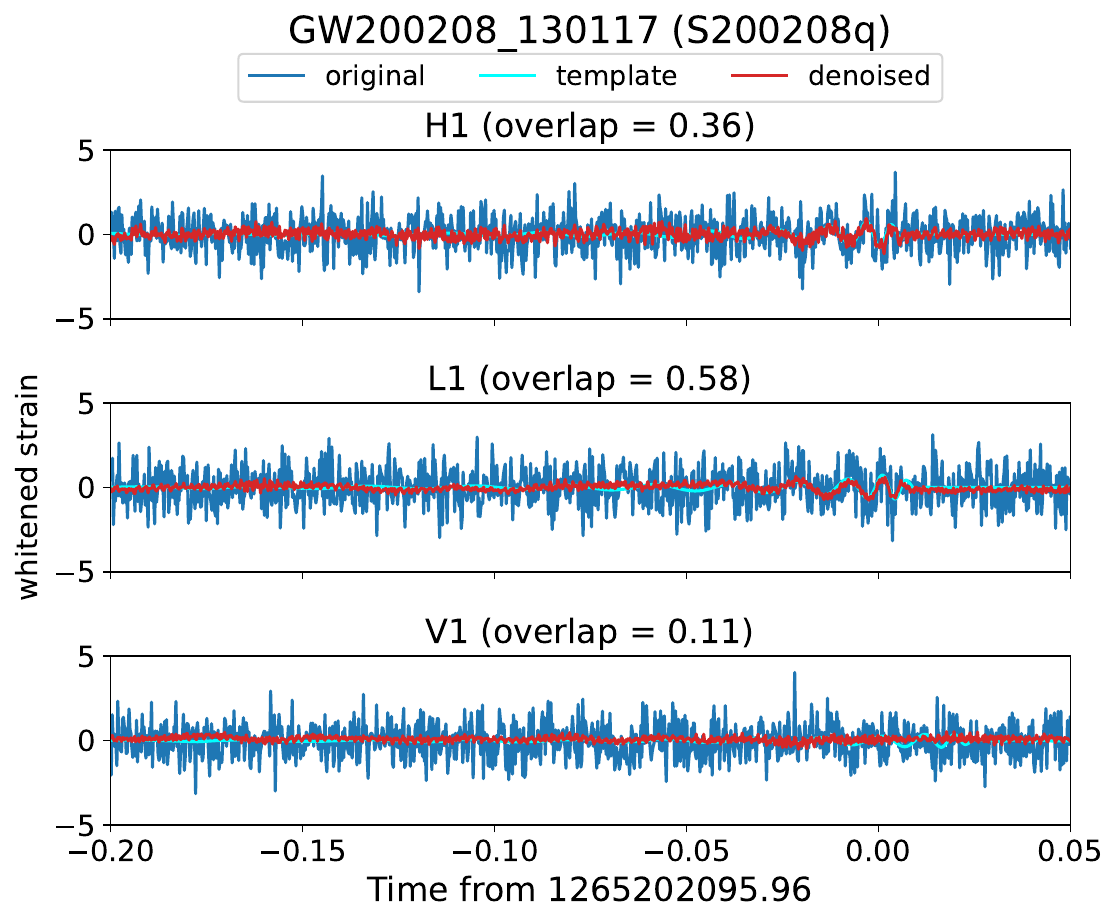}}
    \hfill
    \subfloat[\label{fig:appendix_37_2}]{\includegraphics[width=0.48\linewidth]{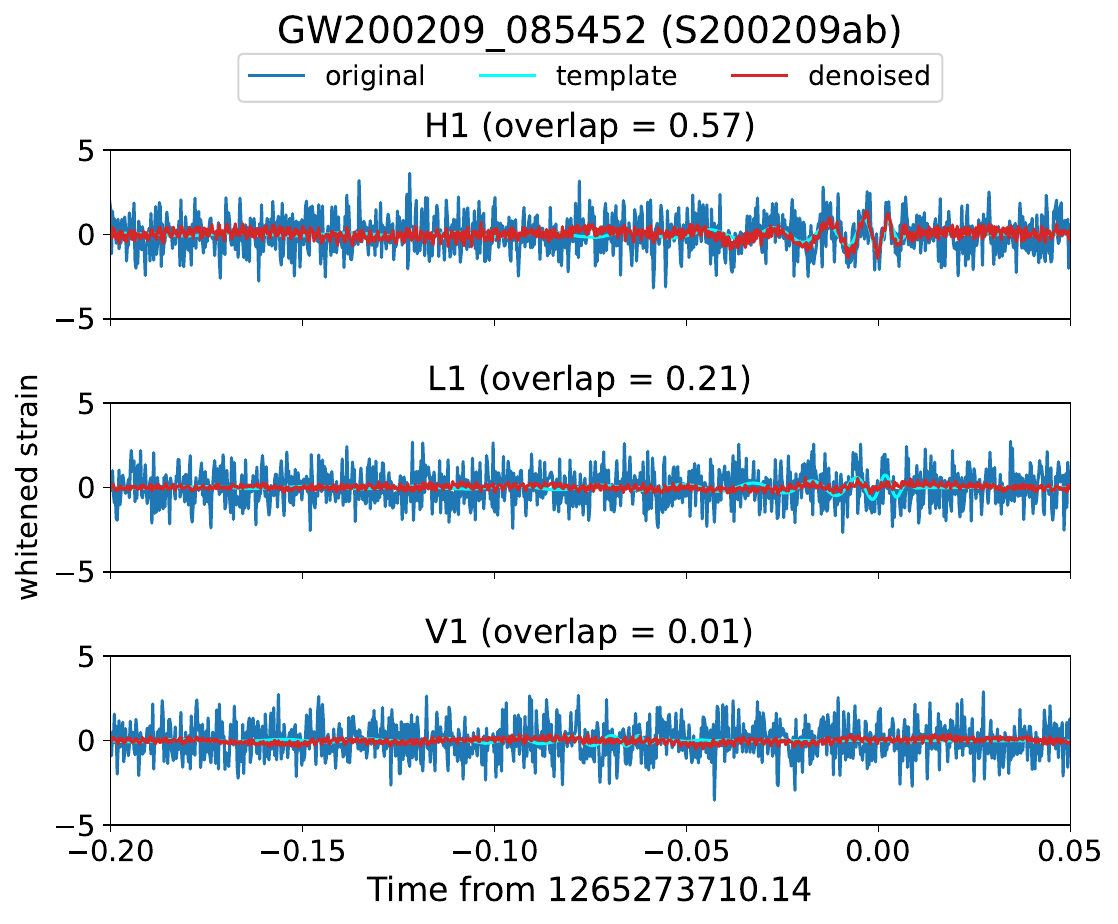}}
    \caption{\label{fig:appendix_37}The $0.25$s windows of the denoised results of the H1, L1 and V1 data around S200208q and S200209ab.}
\end{figure*}

\begin{figure*}[tbp]
    \centering
    \subfloat[\label{fig:appendix_38_1}]{\includegraphics[width=0.48\linewidth]{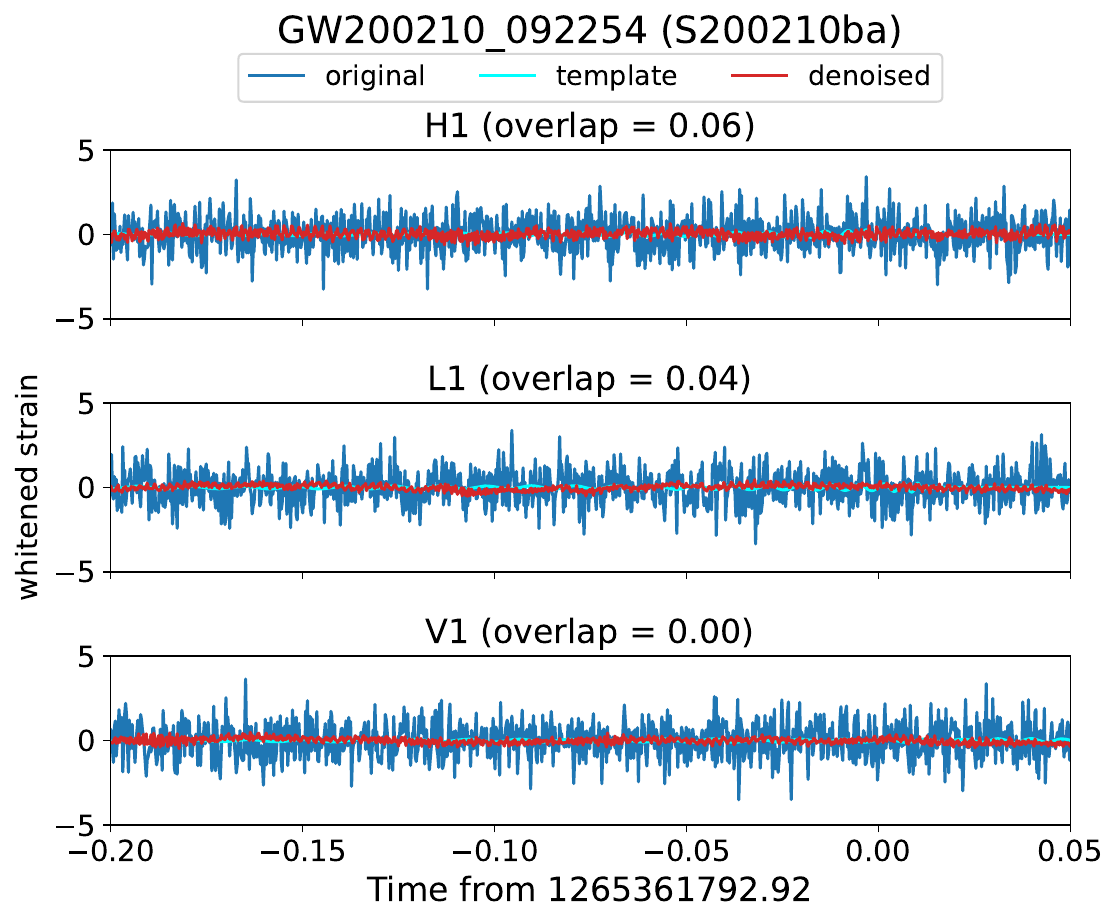}}
    \hfill
    \subfloat[\label{fig:appendix_38_2}]{\includegraphics[width=0.48\linewidth]{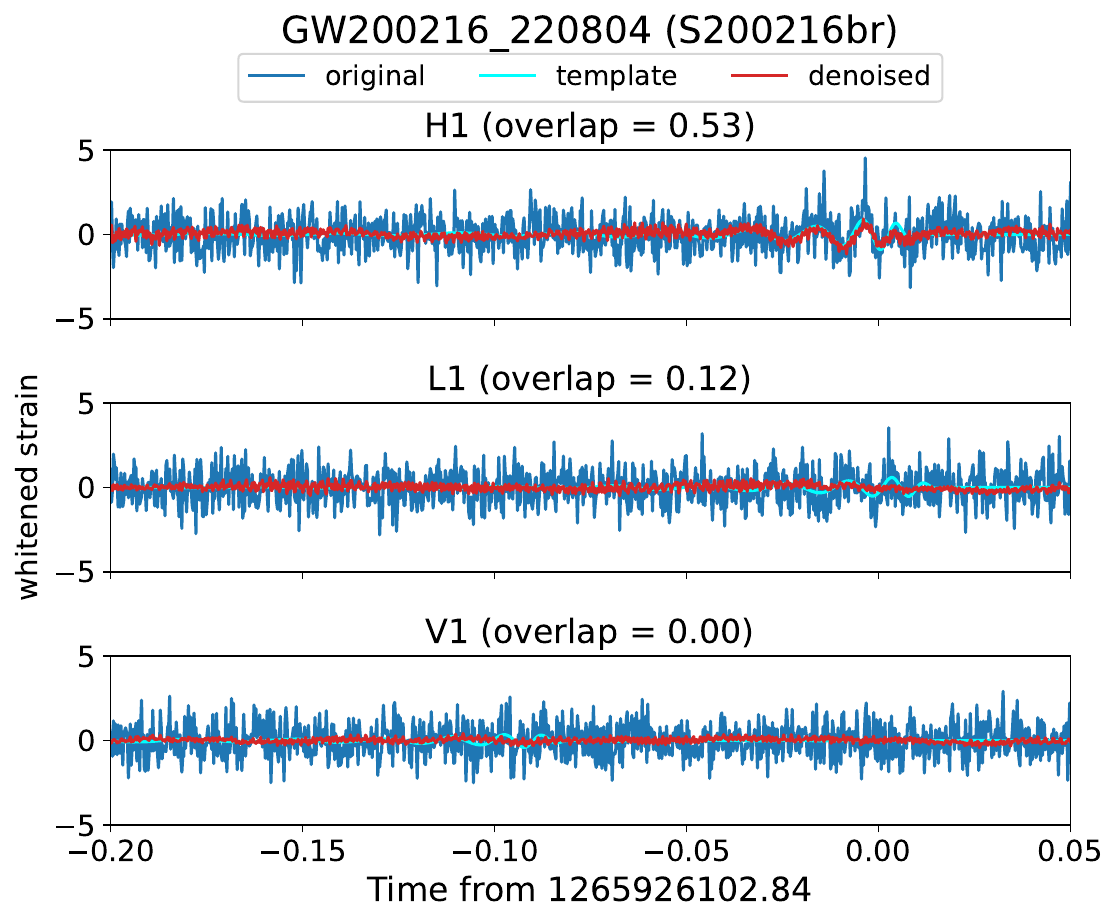}}
    \caption{\label{fig:appendix_38}The $0.25$s windows of the denoised results of the H1, L1 and V1 data around S200210ba and S200216br.}
\end{figure*}

\begin{figure*}[tbp]
    \centering
    \subfloat[\label{fig:appendix_39_1}]{\includegraphics[width=0.48\linewidth]{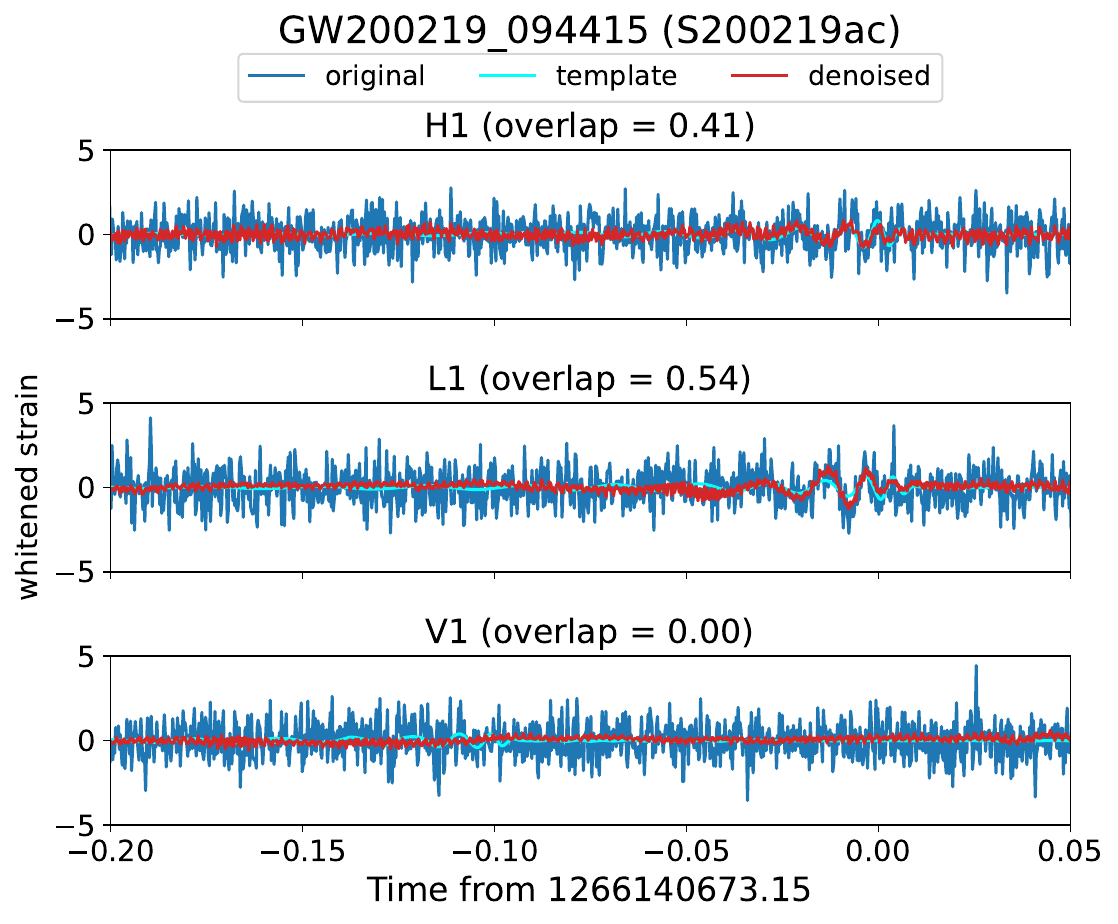}}
    \hfill
    \subfloat[\label{fig:appendix_39_2}]{\includegraphics[width=0.48\linewidth]{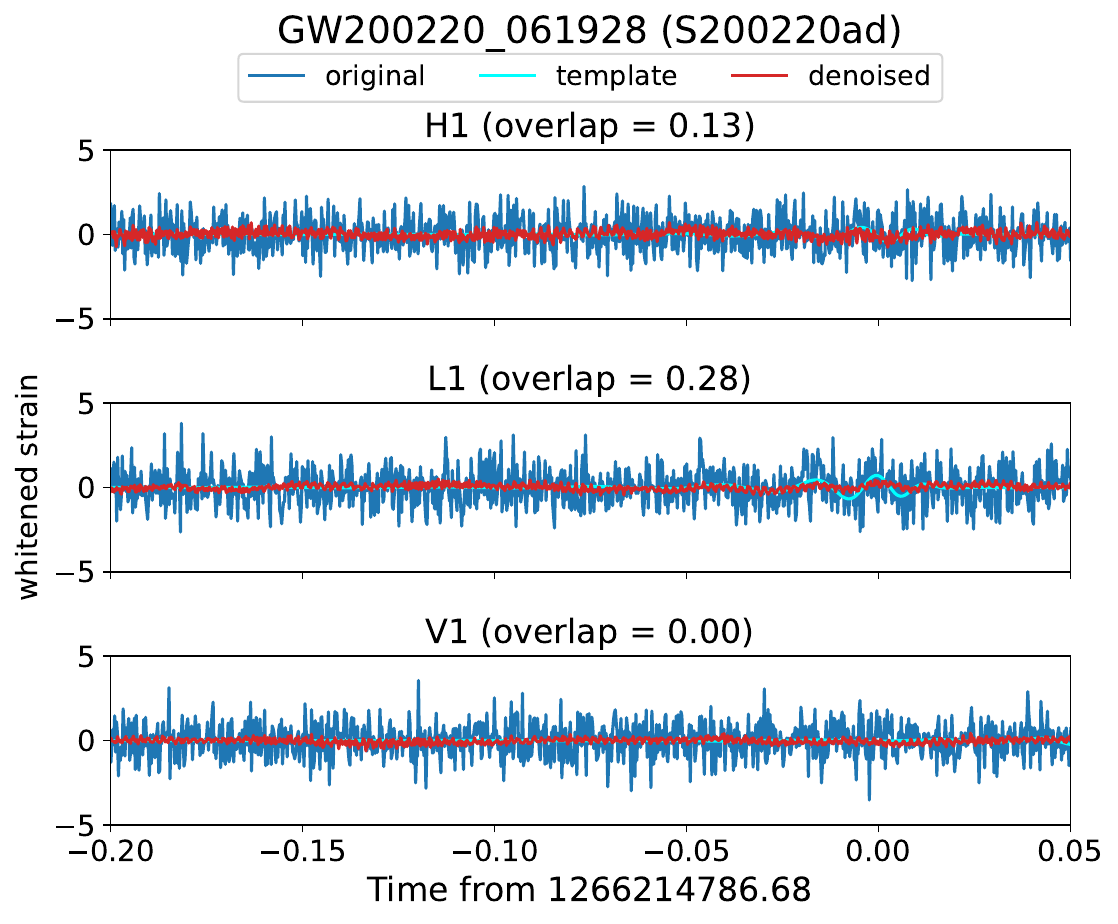}}
    \caption{\label{fig:appendix_39}The $0.25$s windows of the denoised results of the H1, L1 and V1 data around S200219ac and S200220ad.}
\end{figure*}

\begin{figure*}[tbp]
    \centering
    \subfloat[\label{fig:appendix_40_1}]{\includegraphics[width=0.48\linewidth]{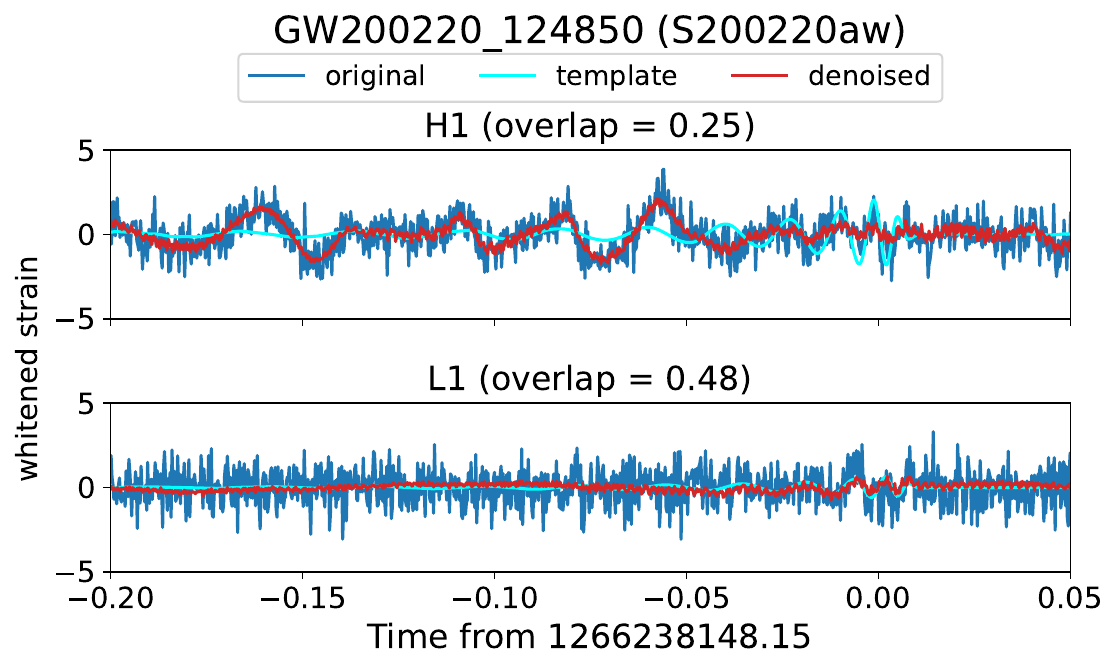}}
    \hfill
    \subfloat[\label{fig:appendix_40_2}]{\includegraphics[width=0.48\linewidth]{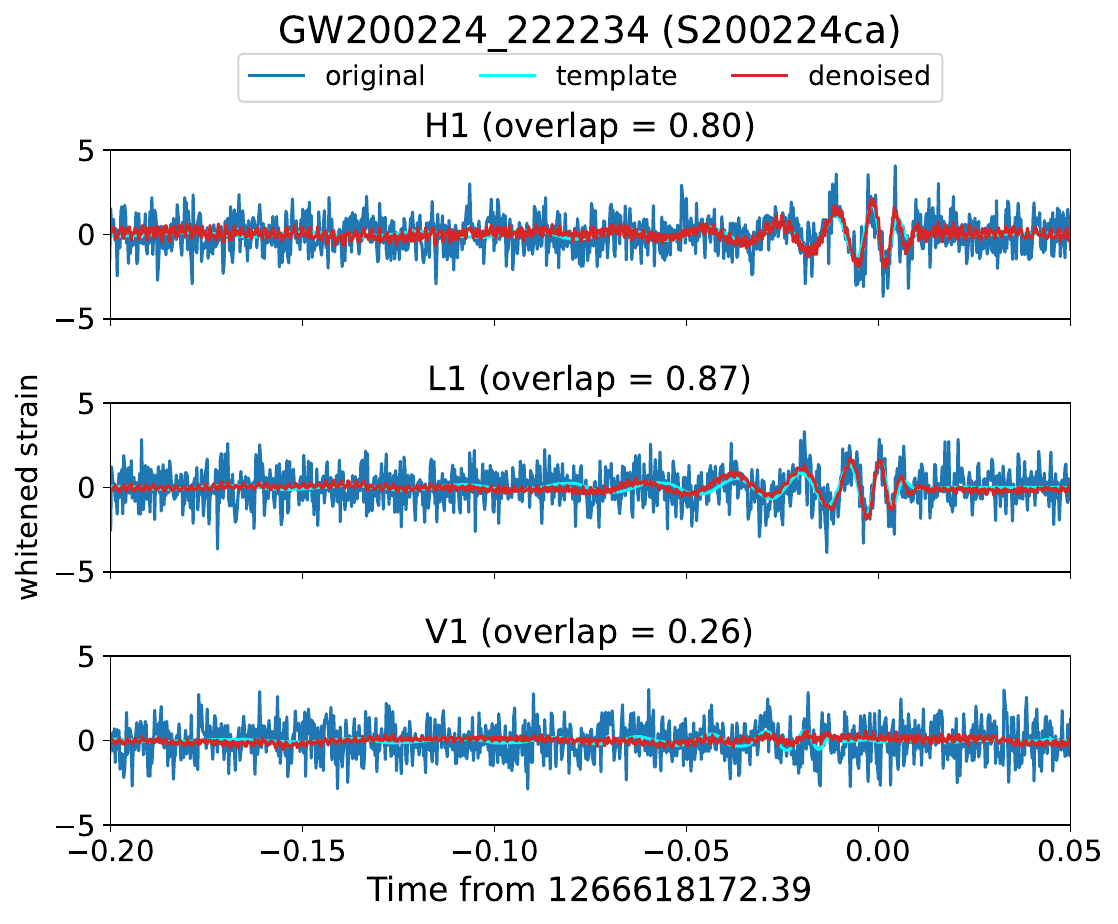}}
    \caption{\label{fig:appendix_40}The $0.25$s windows of the denoised results of the H1, L1 and V1 data around S200220aw and S200224ca.}
\end{figure*}

\begin{figure*}[tbp]
    \centering
    \subfloat[\label{fig:appendix_41_1}]{\includegraphics[width=0.48\linewidth]{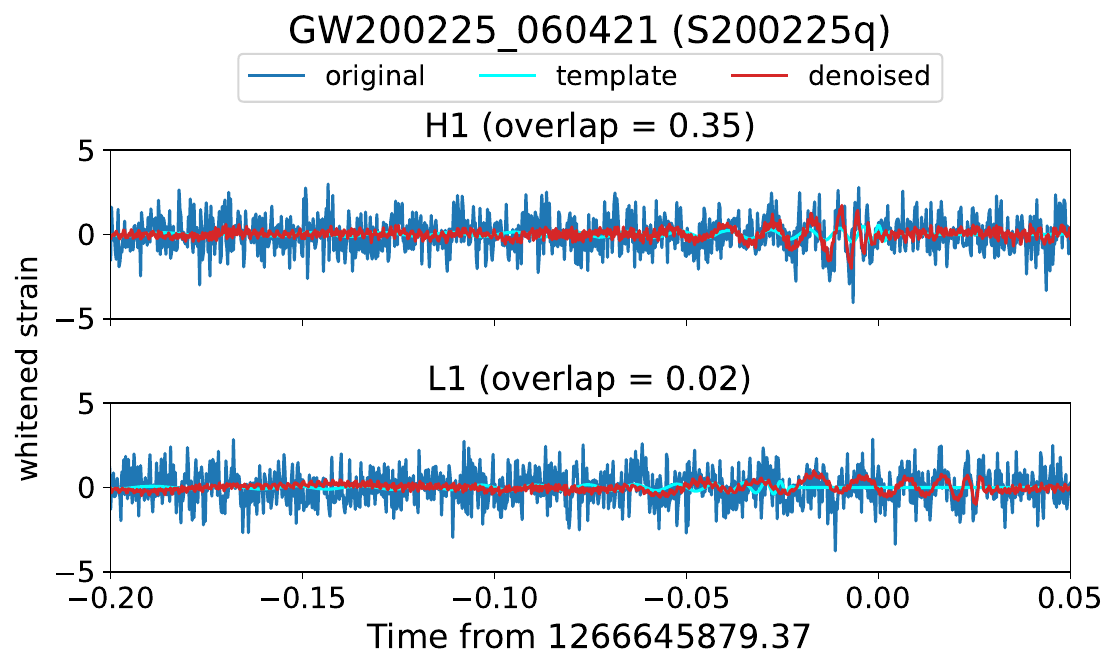}}
    \hfill
    \subfloat[\label{fig:appendix_41_2}]{\includegraphics[width=0.48\linewidth]{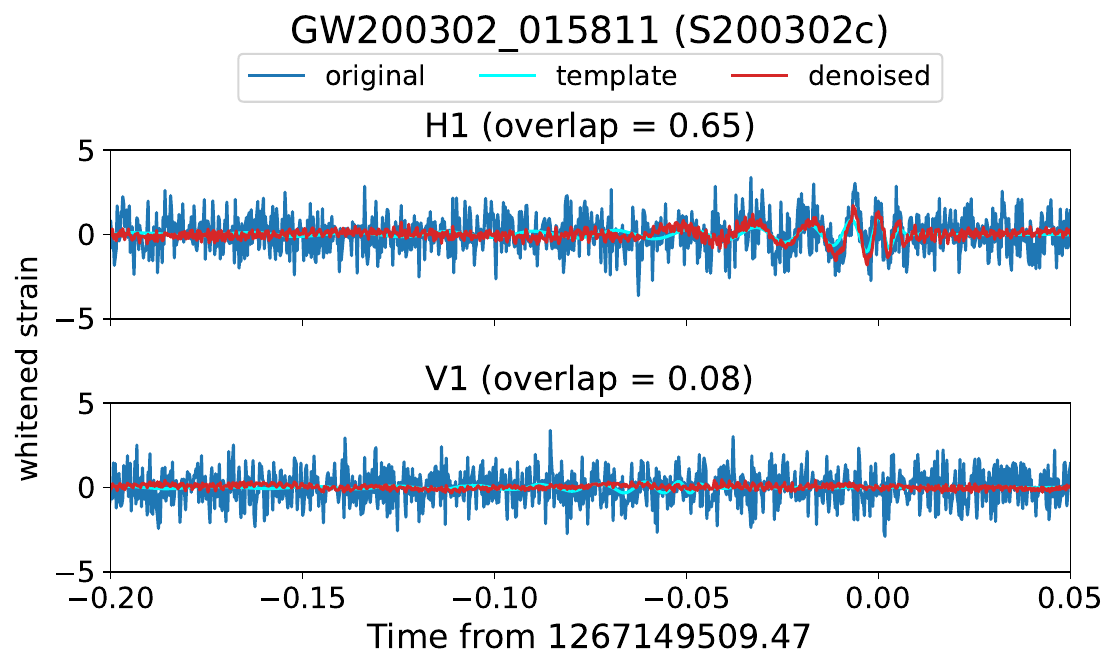}}
    \caption{\label{fig:appendix_41}The $0.25$s windows of the denoised results of the H1, L1 and V1 data around S200115j and S200128d.}
\end{figure*}

\begin{figure*}[tbp]
    \centering
    \subfloat[\label{fig:appendix_42_1}]{\includegraphics[width=0.48\linewidth]{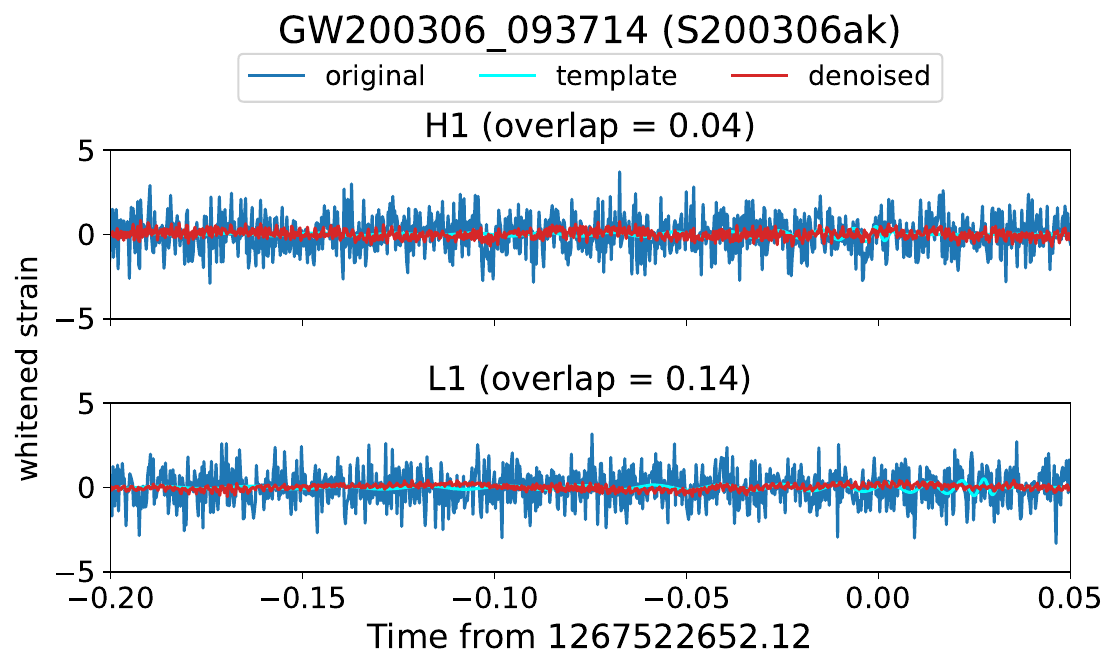}}
    \hfill
    \subfloat[\label{fig:appendix_42_2}]{\includegraphics[width=0.48\linewidth]{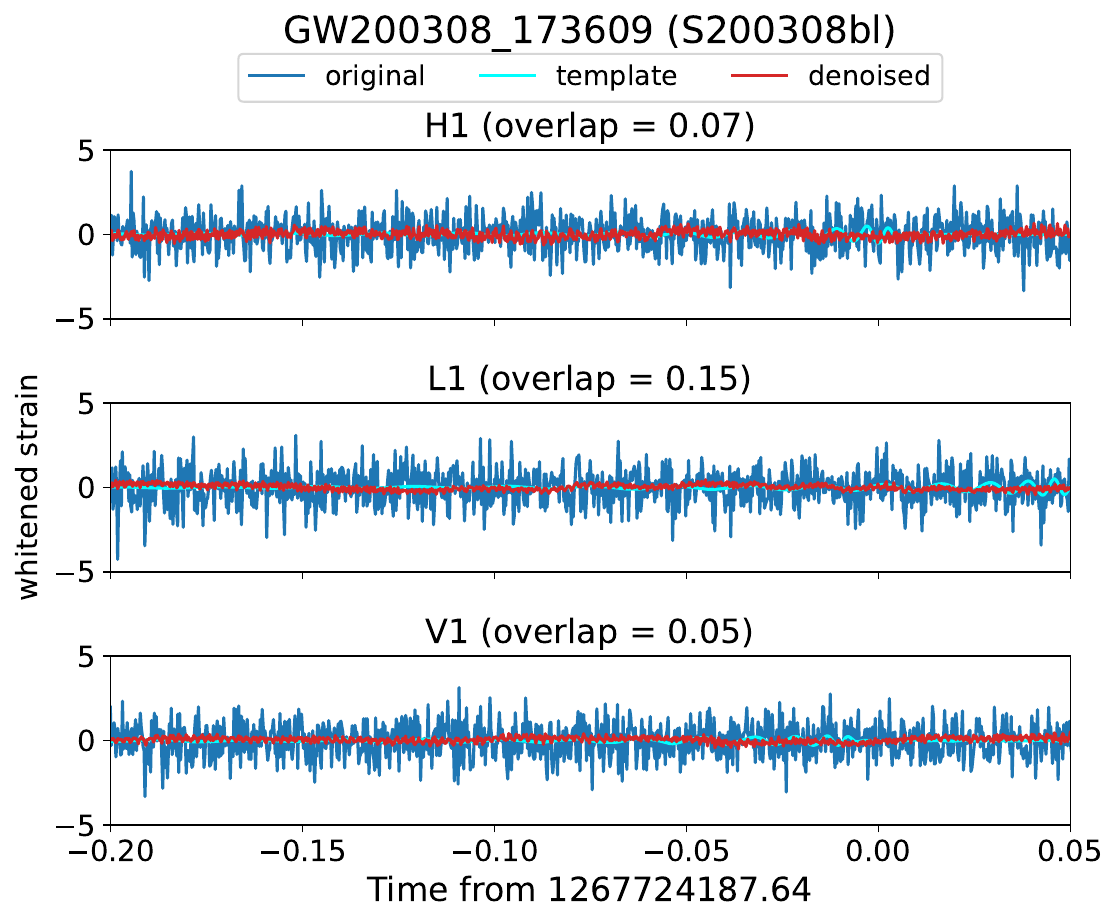}}
    \caption{\label{fig:appendix_42}The $0.25$s windows of the denoised results of the H1, L1 and V1 data around S200306ak and S200308bl.}
\end{figure*}

\begin{figure*}[tbp]
    \centering
    \subfloat[\label{fig:appendix_43_1}]{\includegraphics[width=0.48\linewidth]{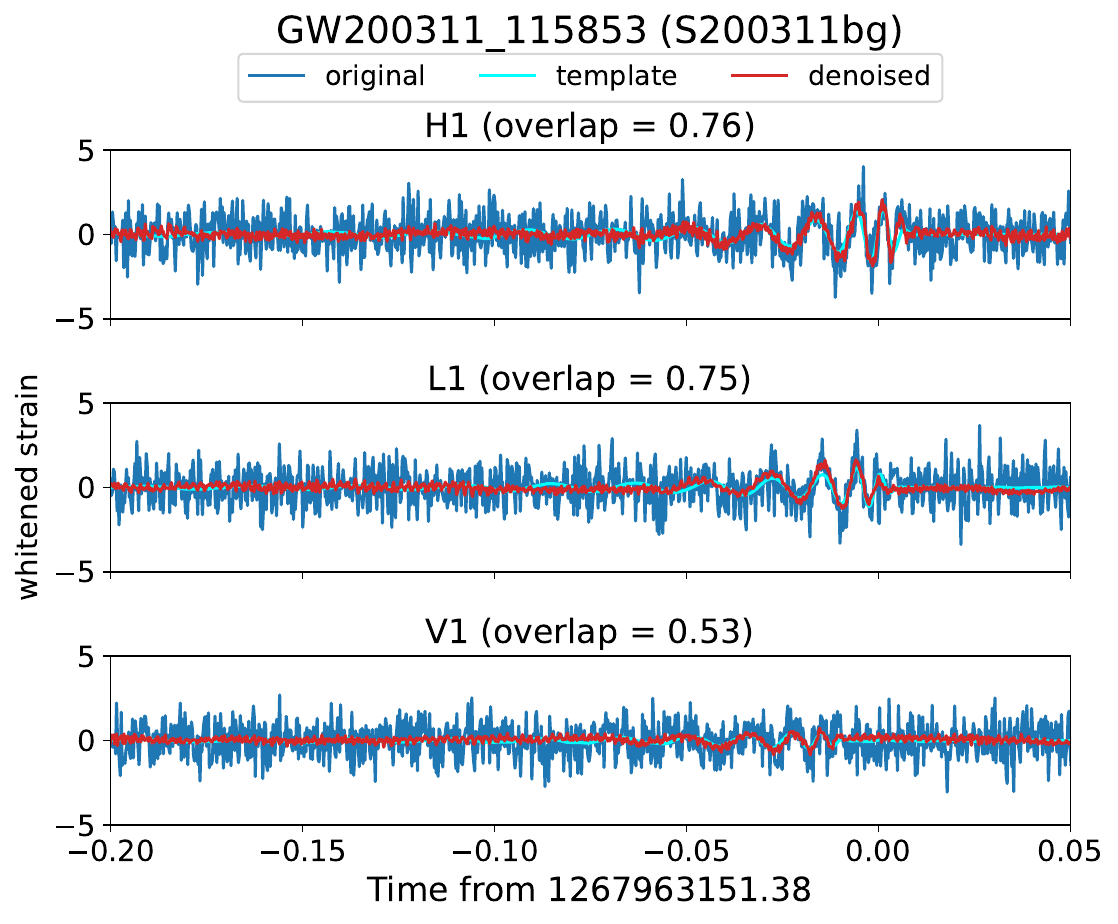}}
    \hfill
    \subfloat[\label{fig:appendix_43_2}]{\includegraphics[width=0.48\linewidth]{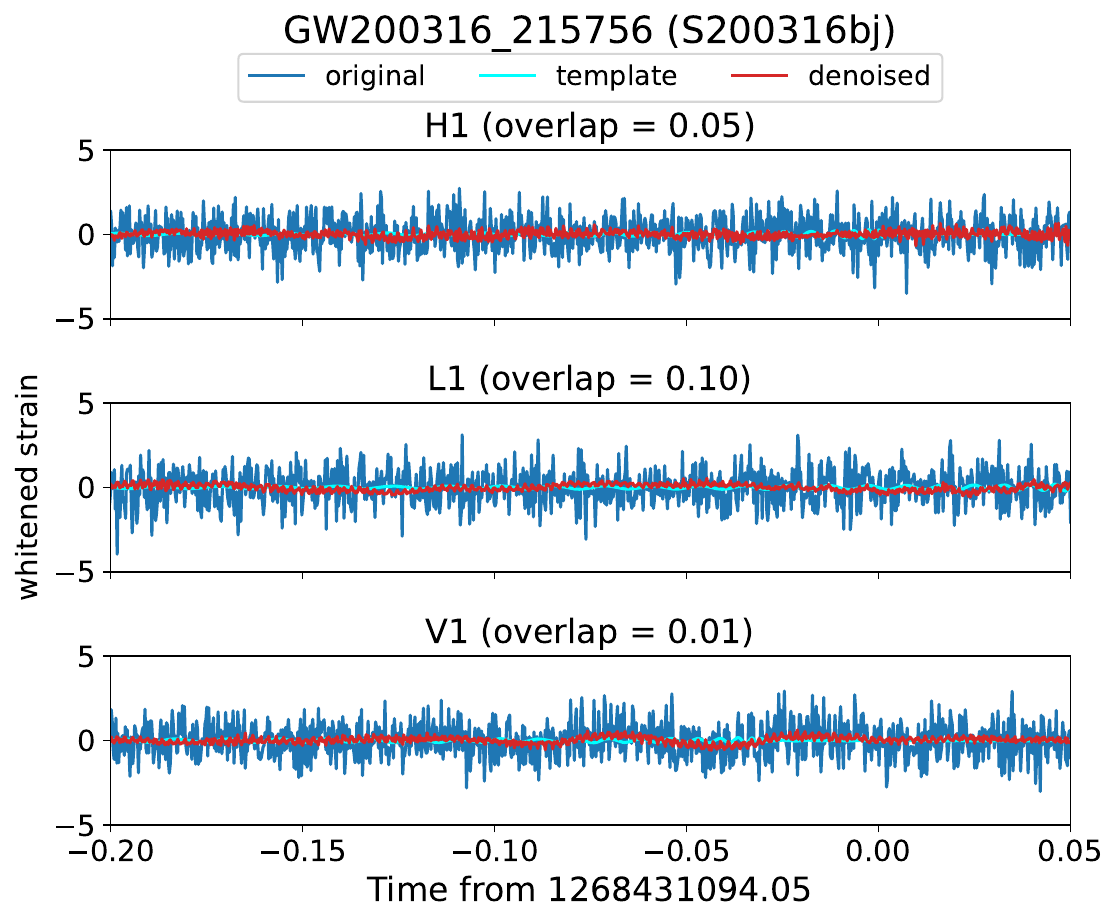}}
    \caption{\label{fig:appendix_43}The $0.25$s windows of the denoised results of the H1, L1 and V1 data around S200311bg and S200316bj.}
\end{figure*}

The denoised results of the event catalog are presented from Figures~\ref{fig:appendix_1} to \ref{fig:appendix_43}. Among all confidently detected events, there are two BNS events: GW170817 and GW190425 (S190425z), an NSBH event: GW191219\_163120 (S191219ax), and the event GW200210\_092254 (S200210ba) that can be either a BBH or an NSBH event~\cite{LIGOScientific:2018mvr, LIGOScientific:2020ibl, 2021arXiv210801045T, KAGRA:2021vkt}. Apart from GW170817 which have shown in Section~\ref{subsec:real_test}, the denoised result of GW190425, GW191219\_163120 and GW200210\_092254 are shown in Figure~\ref{fig:appendix_8_1}, \ref{fig:appendix_32_2} and \ref{fig:appendix_38_1}, respectively. Unfortunately, our model cannot denoise these GW signals due to their low amplitudes.

Other events in the catalog are BBH events. Our model can reveal the structure of most of the loud signals in the H1 and L1 strain data. Moreover, for events GW170608, GW190412 (S190412m), GW190708\_232457 (S190708ap), GW190814 (S190814bv), GW191204\_171526 (S191204r) and GW191216\_213338 (S191216ap), our model can extract the inspiral signal more than 0.1s before the merger event, as shown in Figures~\ref{fig:appendix_2_1}, \ref{fig:appendix_6_1}, \ref{fig:appendix_16_2}, \ref{fig:appendix_21_1}, \ref{fig:appendix_31_1} and \ref{fig:appendix_32_1}, respectively. Also, although our model cannot extract the feature in most of the V1 data due to the large noise background, it can still extract the feature for significant events GW200129p\_065458 (S200129m) and GW200311\_115853 (S200311bg) in Figures~\ref{fig:appendix_35_2} and \ref{fig:appendix_43_1}. Furthermore, although the signal of GW190925\_232845 (S190925ad) is not loud, our model can still extract some of the inspiral features in the V1 data, as shown in Figure~\ref{fig:appendix_25_1}.

Since we only use vanilla matched-filtering to reconstruct waveform templates, sometimes the reconstructed waveform cannot accurately align with the strain data, which may be due to the poor estimation of the PSD. For example, for the event GW200225\_060421 (S200225q) in Figure~\ref{fig:appendix_41_1}, we cannot reconstruct the waveform correctly in the L1 data, but our model can extract the GW features around the event, which leads to the very low overlap value.

Finally, in the event GW200220\_124850 (S200220aw), there are some noise fluctuations around the event in the H1 data, as shown in Figure~\ref{fig:appendix_40_1}. We whiten the data using the PSD estimated from the 512 seconds of data centering at the event time to mitigate the influence of the fluctuation. Our model can extract GW features in the H1 data after the mitigation. Since our model does not intend to separate GW signals from glitches, it also extracts the feature of noise fluctuation, which may affect the result of the following parameter estimation or localization tasks.

\end{document}